\documentclass[final,5p,times,English,twocolumn]{elsarticle}

\usepackage{lineno,hyperref,txfonts, geometry,  graphicx, fleqn}
\usepackage[T1]{fontenc}
\usepackage{amsmath}
\usepackage{diagbox}
\usepackage{longtable}
\usepackage{multirow}
\usepackage{threeparttable}
\usepackage{graphicx}
\usepackage{subfig}
\usepackage{overpic}
\usepackage{color}
\usepackage{makecell}
\usepackage{amssymb}
\usepackage{pifont}

\newcommand{\xmark}{\ding{55}}
\usepackage{amsfonts,amssymb}
\usepackage{mathrsfs}
\usepackage{booktabs}
\graphicspath{{fig/}}
\usepackage{caption}
\usepackage{bm}
\usepackage{upgreek}
\usepackage{float}
\usepackage{enumerate}
\usepackage{pifont}
\usepackage{pdflscape}
\usepackage{array}
\usepackage{appendix}
\usepackage{epstopdf}
\usepackage{overpic}
\usepackage{rotating}
\usepackage{color}
\definecolor{purple}{RGB}{128,0,128}
\usepackage{ulem}
\begin{document}

\begin{frontmatter}

\title{Speaker Recognition Based on Deep Learning: An Overview}
%\tnotetext[t1]{This work was supported in part by the National Key Research and Development Program of China under Grant No. 2018AAA0102200 and in part by the Key Program of National Science Foundation of China (NSFC) Grant No. 61831019 and the NSFC and Israel Science Foundation (ISF) joint research program under Grant No. 61761146001..}

\author{Zhongxin Bai}
\ead{zxbai@mail.nwpu.edu.cn}

\author{Xiao-Lei Zhang}
\ead{xiaolei.zhang@nwpu.edu.cn}

%\author{Jingdong Chen}
%\ead{jingdongchen@ieee.org}

\cortext[cor1]{ Corresponding author: Xiao-Lei Zhang}
\address{Center of Intelligent Acoustics and Immersive Communications (CIAIC) and the School of Marine Science and Technology, \\ Northwestern Polytechnical University, Xi'an, Shaanxi 710072, China.}

\begin{abstract}
Speaker recognition is a task of identifying persons from their voices. Recently, deep learning has dramatically revolutionized speaker recognition. However, there is lack of comprehensive reviews on the exciting progress.
 In this paper, we review several major subtasks of speaker recognition, including speaker verification, identification, diarization, and robust speaker recognition, with a focus on deep-learning-based methods. Because the major advantage of deep learning over conventional methods is its representation ability, which is able to produce highly abstract embedding features from utterances, we first pay close attention to deep-learning-based speaker feature extraction, including the inputs, network structures, temporal pooling strategies, and objective functions respectively, which are the fundamental components of many speaker recognition subtasks. Then, we make an overview of speaker diarization, with an emphasis of recent supervised, end-to-end, and online diarization. Finally, we survey robust speaker recognition from the perspectives of domain adaptation and speech enhancement, which are two major approaches of dealing with domain mismatch and noise problems. Popular and recently released corpora are listed at the end of the paper.
\end{abstract}


\end{frontmatter}

%\linenumbers
\section{Introduction}\label{sec:introduction}

It is known that a speaker's voice contains {{personal traits}} of the speaker, given the unique pronunciation organs and speaking manner of the speaker, e.g. the unique vocal tract shape, larynx size, accent, and  rhythm \cite{kinnunen2010overview}. Therefore, it is possible to identify a speaker from his/her voice automatically via a computer. This technology is termed as \textit{automatic speaker recognition}, which is the core topic of this paper. We do not discuss speaker recognition by humans. Speaker recognition is a fundamental task of speech processing, and finds its wide applications in real-world scenarios. For example, it is used for the voice-based authentication of personal smart devices, such as cellular phones, vehicles, and laptops. It guarantees the transaction security of bank trading and remote payment. It has been widely applied to forensics for investigating a suspect to be guilty or non-guilty \cite{kinnunen2010overview,campbell2009forensic,champod2000inference}, or surveillance and  automatic identity tagging \cite{togneri2011overview}.  It is important in audio-based information retrieval for broadcast news, meeting recordings and telephone calls. It can also serve as a frontend of automatic speech recognition (ASR) for improving the transcription performance of multi-speaker conversations.

\begin{figure*}[t]
  \centering
  \includegraphics[width=18cm]{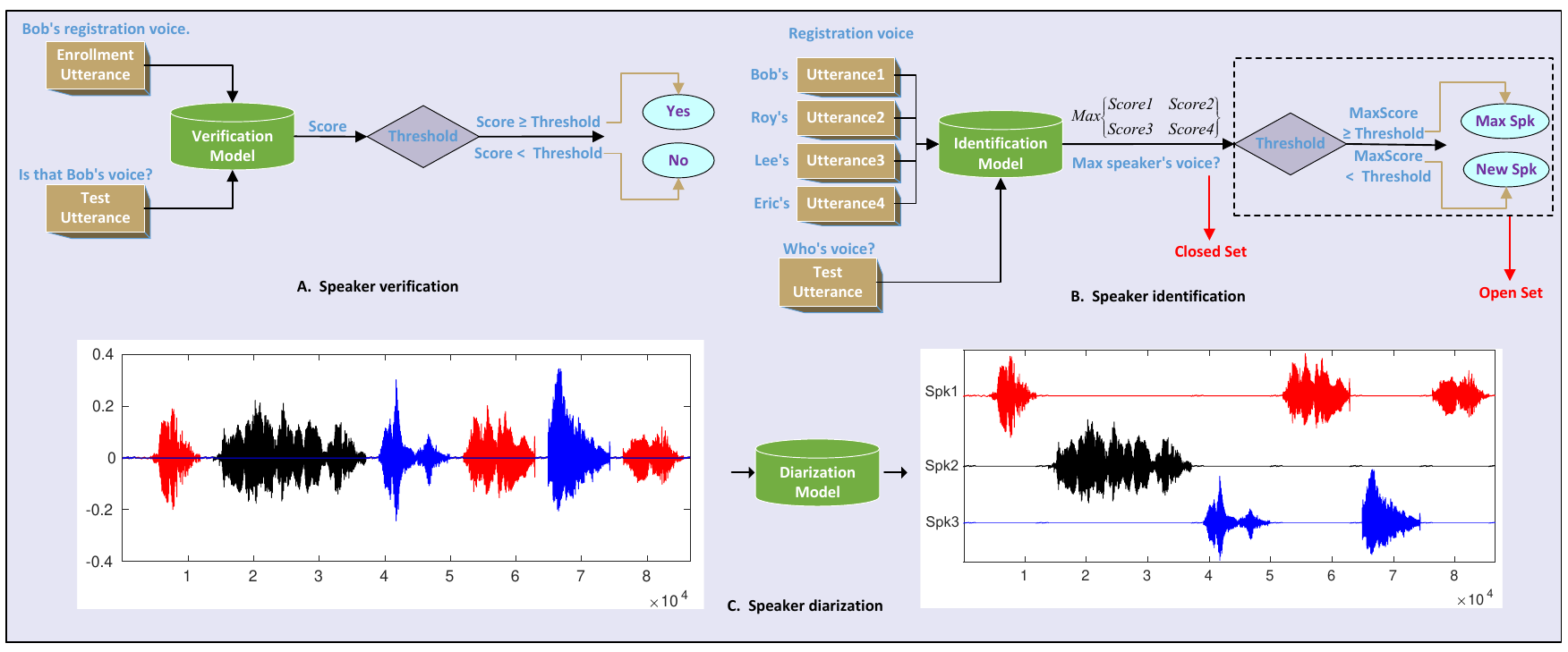}
  \caption{Flowcharts of speaker verification, speaker identification, and speaker diarization. Fig. A describes speaker verification, which is a task of verifying whether a test utterance and an enrollment utterance are uttered by the same speaker via comparing the similarity score of the utterances with a pre-defined threshold. Fig. B describes speaker identification, which is a task of determining the speaker identity of a test utterance from a set of speakers. If the utterance must be produced from the set of the speakers, then it is a closed set identification problem; otherwise, it is an open set problem. Fig. C describes speaker diarization, which addresses the problem of ``who spoke when'', i.e., partitioning a conversation recording into several speech recordings, each of which belongs to a single speaker.}
  \label{fig:veri-iden-diar}
\end{figure*}

The research on speaker recognition can be dated back to at least 1960s \cite{pruzansky1964talker}. In the following forty years, many advanced technologies promoted the development of speaker recognition.
For example, a number of acoustic features (e.g. the linear predictive cepstral coefficients, the perceptual linear prediction coefficient, and the mel-frequency cepstral coefficients) and template models (e.g. vector quantization, and dynamic time warping) have been applied, see \cite{kinnunen2010overview} for the details. Later on,  \cite{reynolds2000speaker} proposed the Gaussian mixture model based universal background model (GMM-UBM), which has been the foundation of speaker recognition for more than ten years since year 2000.
Several representative models based on GMM-UBM have been developed, including the applications of support vector machines \cite{campbell2006support} and joint factor analysis \cite{kenny2007joint}. Among the models, the GMM-UBM/i-vector frontend \cite{dehak2010front} with probabilistic linear discriminant analysis (PLDA) backend \cite{kenny2010bayesian,garcia2011analysis} provided the state-of-the-art performance for several years, until the new era of deep learning based speaker recognition.

Recently, motivated by the powerful feature extraction capability of deep neural networks (DNNs), a lot of deep learning based speaker recognition methods were proposed \cite{lei2014novel,variani2014deep,snyder2018x} right after the great success of deep learning based speech recognition, which significantly boosts the performance of speaker recognition to a new level, even in wild environments \cite{nagrani2017voxceleb,mclaren2016speakers}.

In this survey article, we give a comprehensive overview to the deep learning based speaker recognition methods in terms of the vital subtasks and research topics, including speaker verification, identification, diarization, and robust speaker recognition. By doing the survey, we hope to provide a useful resource for the speaker recognition community. The main contributions of this paper are summarized as follows:
\begin{itemize}
  \item We summarize deep learning based speaker feature extraction techniques for speaker verification and identification, from the aspects of inputs, network structures, temporal pooling strategies, and objective functions which are also the fundamental components of many other speaker recognition subtasks beyond speaker verification and identification.

  \item We make an overview to the deep learning based speaker diarization, with an emphasis of recent supervised,  end-to-end, and online diarization.

   \item We survey robust speaker recognition from the perspectives of domain adaptation and speech enhancement, which are two major approaches to deal with domain mismatch and noise problems.
\end{itemize}

In the last two decades, many excellent overviews on speaker recognition have been published. This paper is fundamentally different from previous overviews. First, this paper focuses on the recently development of deep learning based speaker recognition techniques, while most previous overviews are based on traditional speaker recognition methods \cite{kinnunen2010overview,hansen2015speaker,togneri2011overview,reynolds2002overview,fazel2011overview,wu2015spoofing,anguera2012speaker}.
Although \cite{das2020attacker,irum2019speaker} summarized deep learning based speaker recognition methods in certain aspects, our paper summarizes different subtasks and topics from new perspectives. Specifically, \cite{das2020attacker} presents an overview to the potential threats of adversarial attacks to speaker verification as well as the spoofing countermeasures, which is not the focus of this overview. We provide a broad and comprehensive overview to a wide aspects of speaker verification, speaker diarization, domain adaptation, {{most}} of which have not been mentioned in \cite{irum2019speaker}.

This article is targeted at three categories of readers: The beginners who wish to study speaker recognition, the researchers who want to learn the whole picture of speaker recognition based on deep learning, and the engineers who need to understand or implement specific algorithms for their speaker recognition related products. In addition, we assume that the readers have basic knowledge of speech signal processing, machine leaning and pattern recognition.

The rest of the survey is organized as follows. In Section \ref{sec:overview}, we give a general overview and define some notations. In Sections \ref{sec:DNN/i-vector} to \ref{sec:robust}, we survey the deep learning based speaker recognition methods in various aspects. In Section \ref{sec:data_tool}, we summarize some speaker recognition challenges and  publicly available data. Finally, we conclude this article in Section \ref{sec:discussion_conclusion}.

\section{Overview and scope}
\label{sec:overview}

\begin{figure*}[t]
  \centering
  \includegraphics[width=6.5in]{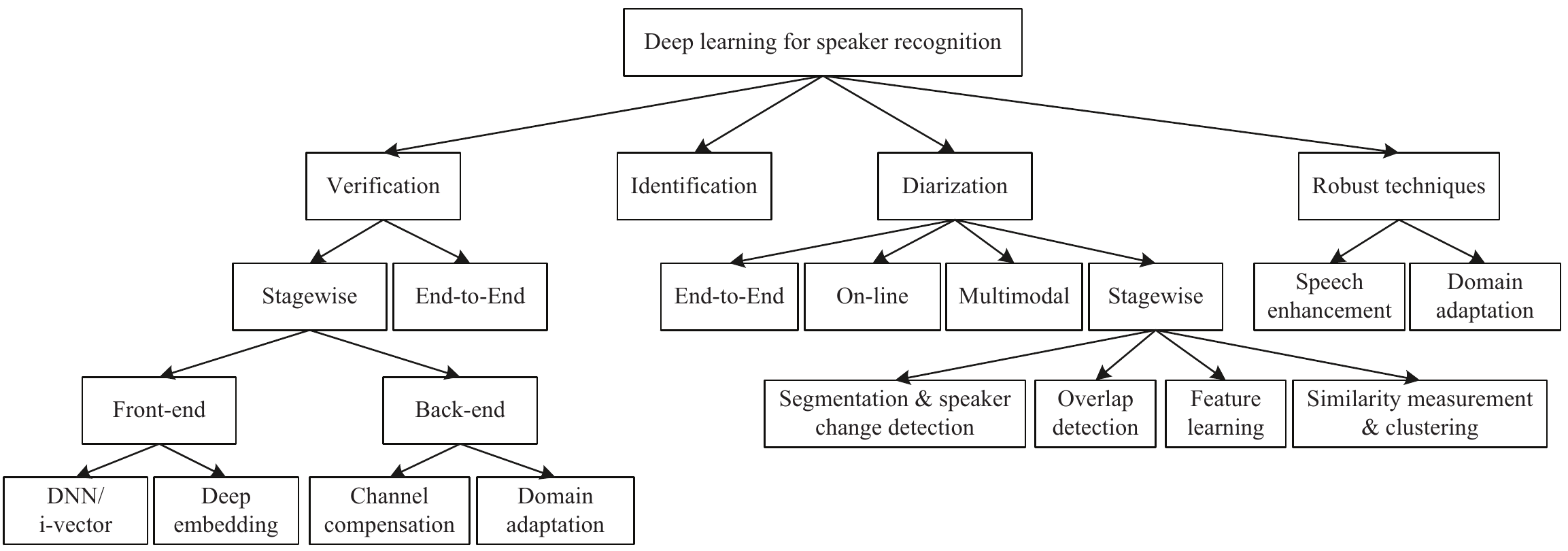}
  \caption{ Overview of deep learning based speaker recognition.}
  \label{fig:overview}
\end{figure*}

This overview summarizes four major research branches of speaker recognition, which are speaker verification, identification, diarization, and robust speaker recognition respectively. The  flowcharts of the first three branches are described in Fig. \ref{fig:veri-iden-diar}, while robust speaker recognition deals with the challenges of noise and domain mismatch problems. The contents of the overview are organized in Fig. \ref{fig:overview}, which are described briefly as follows.

\textit{Speaker verification} aims at verifying whether an utterance is pronounced by a hypothesized speaker based on his/her pre-recorded utterances. Speaker verification algorithms can be categorized into \textit{stage-wise} and \textit{end-to-end}  ones.  A stage-wise speaker verification system usually consists of a front-end for the extraction of speaker features and a back-end for the similarity calculation of speaker features.
 The front-end transforms an utterance in time domain or time-frequency domain into a high-dimensional feature vector. It accounts for the recent advantage of the deep learning based speaker recognition.
We survey the research on the front-end comprehensively in Sections \ref{sec:DNN/i-vector} to \ref{sec:embedding_objectives}.
The back-end first calculates a similarity score between enrollment and test speaker features and then compare the score with a threshold:
\begin{equation}\label{eq:verifi_def}
    f(\mathbf{x}^e, \mathbf{x}^t;\mathbf{w}) \mathop{\gtrless}\limits_{H_1}^{H_0} \xi
\end{equation}
where $f(\cdot)$ denotes a function for calculating the similarity, $\mathbf{w}$ stands for the parameters of the back-end, $\mathbf{x}^e$ and $\mathbf{x}^t$ are the enrollment and test speaker features respectively, $\xi$ is the threshold, $H_0$ represents the hypothesis of $\mathbf{x}^e$ and $\mathbf{x}^t$ belonging to the same speaker, and $H_1$ is the opposite hypothesis of $H_0$. One of the major responsibilities of the back-end is to compensate the channel variability and reduce interferences, e.g. language mismatch.
Because most back-ends aim at alleviating the interferences, which belongs to the problem of robust speaker recognition, we put the overview of the back-ends in Section \ref{sec:robust}.

In contrast to the stage-wise techniques, {{end-to-end}} speaker verification takes a pair of speech utterances as the input, and produces their similarity score directly.
Because a fundamental difference between the end-to-end speaker verification and the deep embedding techniques in the stage-wise speaker verification is the loss function, we mainly summarize the loss functions of the end-to-end speaker verification in Section \ref{sec:end2end_loss}.

\textit{Speaker identification} aims at detecting the speaker identity of a test utterance $\mathbf{x}^t$ from an enrollment database $\{\mathbf{x}_k^e|k=1,2,\cdots,K\}$  by\footnote{Although some work used all speakers in a given database for both training and test which is essentially regarded as a close-set speaker classification problem \cite{nagrani2017voxceleb}, most real world speaker recognition systems must be able to ``enroll'' and ``test'' new speakers dynamically.}:
\begin{equation}
    \label{eq:ident_def}
    k^{*} = \arg \max_{k} \{f(\mathbf{x}^e_1, \mathbf{x}^t; \mathbf{w}),  f(\mathbf{x}^e_2, \mathbf{x}^t; \mathbf{w}), \cdots, f(\mathbf{x}^e_K, \mathbf{x}^t; \mathbf{w})\}
\end{equation}
where $K>1$ denotes the number of the enrollment speakers. If $\mathbf{x}^t$ can never be out of the $K$ registered speakers, then the speaker identification problem is a {{closed set problem}}; otherwise, it is an open set problem. Comparing \eqref{eq:verifi_def} with \eqref{eq:ident_def}, we see that speaker verification is a special case of the open set speaker identification problem with $K=1$, therefore, it is possible that the fundamental techniques of speaker identification and verification are similar, as what we have observed in \cite{yadav2018learning,wang2019centroid,ji2018end,9053640,9053152}. Taking this point into consideration, we make a joint overview to speaker verification and identification with an emphasis on the former.

\textit{Speaker diarization} addresses the problem of ``who spoke when'',
which is a process of partitioning a conversation recording into several speech recordings, each of which belongs to a single speaker.
As shown in Fig. \ref{fig:overview}, a conventional framework of speaker diarization is stage-wise with multiple modules.
 Although the stage-wise speaker verification and diarization share some common modules, e.g. voice activity detection and speaker feature extraction, they have many differences. First, speaker verification assumes that each utterance belongs to a single speaker, while the number of speakers of a conversation in speaker diarization changes case by case. Moreover, speaker verification has an explicit registration/enrollment procedure, while speaker diarization intends to detect speakers on-the-fly without an enrollment procedure. At last, overlapped speech is one of the biggest challenges of speaker diarization, while speaker verification usually assumes that the enrollment or test utterance contains a single speaker only. Therefore, we focus on reviewing the work on the above distinguished properties of the stage-wise speaker diarization in Section \ref{sec:diarization}.
Recently, end-to-end speaker diarization, which outputs the diarization result directly, attracted much attention. Online speaker diarization, which meets the requirement of real-world applications, is also an emerging direction. Furthermore, multimodal speaker diarization, which integrates speech with video or text signals, was also studied extensively.
We review the aforementioned end-to-end, online, and multimodal speaker diarization techniques in Section \ref{sec:diarization}.

Besides, speech is easily contaminated by additive noise, reverberation, channel distortions. {{Therefore,}} robust speaker recognition is also one of the main topics. It mainly includes speech enhancement and domain adaptation techniques, which will be summarized in detail in Section \ref{sec:robust}. At last, we survey benchmark corpora in Section \ref{sec:data_tool}.

To summarize, the aforementioned contents will be organized as listed in Table \ref{tab:contens}. The notations are summarized in Table \ref{tab:notations}.

\renewcommand\arraystretch{1.5}
\begin{table}[t]
  \footnotesize
  \centering
  \caption{Organization of the contents of this paper.}\label{tab:contens}
  \begin{tabular}{m{1.5cm}<{\centering} m{5.5cm}}
  \hline
  \hline
  \textbf{ Sections}                    & \textbf{Contents}                \\
  \hline
  \ref{sec:introduction}, \ref{sec:overview}& Introduction and brief overview.  \\
  \ref{sec:DNN/i-vector}, \ref{sec:embedding}, \ref{sec:embedding_input_network}, \ref{sec:embedding_pooling},  \ref{sec:embedding_objectives} & Speaker feature extraction. \\
  \ref{sec:end2end_loss}    & The loss functions of the end-to-end speaker verification. \\

  \ref{sec:diarization}     & Speaker diarization. \\
  \ref{sec:robust}          & Robust speaker recognition.               \\
  \ref{sec:data_tool}       & Benchmark corpora. \\
  \ref{sec:discussion_conclusion} & Conclusions and discussions.\\
  \hline
  \hline
  \end{tabular}
\end{table}

\renewcommand\arraystretch{1.5}
\begin{table}[t]
  \footnotesize
  \centering
  \caption{Summary of the notations in this paper.}
  \label{tab:notations}
  \scalebox{1}{
  \begin{tabular}{p{1.5cm} p{6.2cm}}
  \hline
  \hline
  \textbf{Notation} & \textbf{Description}\\
  \hline
   $\mathbb{R}$ & Set of real numbers \\
   $\mathbb{R}^d$ & Set of $d$-dimensional real-valued vectors\\
   $\mathbb{R}^{d_1 \times d_2}$ & Set of $d_1 \times d_2$ real-valued matrices\\

   \hline
   $\mathcal{Y} $ & Set of acoustic features \\
   $\mathcal{H} $ & Set of the last  frame-level hidden layer's outputs \\
   $\mathcal{E}$  & Set of embeddings       \\
   $\mathcal{X} $ & Set of the inputs to  loss functions \\
   $\mathcal{L} $ & The symbol of loss functions  \\

   \hline
   $\mathbf{y}$   &  A frame acoustic feature \\
   $\mathbf{h}$   &  A hidden feature of the last frame-level layer's output \\
   $\mathbf{e}$   &  An embedding feature of the embedding layer's  output\\
   $\mathbf{x}$   &  An input feature to loss functions  \\
   $\mathbf{u}$   &  An output of the temporal pooling layer \\
   \hline
   $t,\, T$       &  Index and total number of frames in an utterance \\
   $i,\, I$       &  Index and total number of the utterance \\
   $j,\, J$       &  Index and total number of speakers in the training set\\
   \hline
   $\|\cdot\|$        & The $\ell_2$ norm                   \\
   $\odot$            & The Hadamard product             \\
   $\delta(\cdot)$    & Indicator function               \\
   $(\cdot)^T$  & The transform of matrix or vector\\
  \hline
  \hline
  \end{tabular}
  }
\end{table}

\section{Speaker feature extraction with DNN/i-vector}\label{sec:DNN/i-vector}
In this section, we first introduce two main streams of the deep learning based improvement to the i-vector framework in Section \ref{sec:GMM-UBM-ivector}, and then comprehensively review the two streams in Sections \ref{sec:DNN-UBM} and \ref{sec:DNN-BNF} respectively. Finally, we make some discussions to the DNN/i-vector in Section \ref{sec:discuss_DNN/i-vector}.
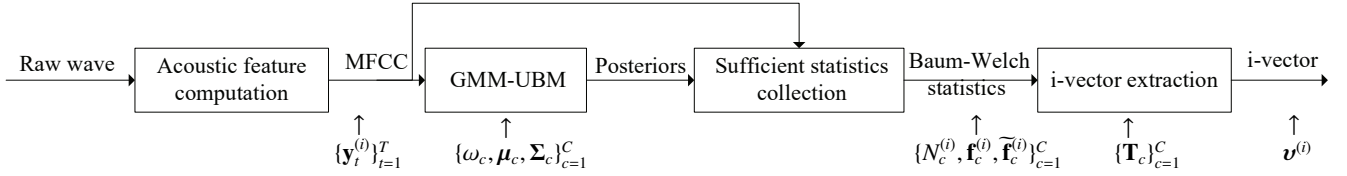
\begin{figure*}[!ht]
  \centering
  \includegraphics[width=7in]{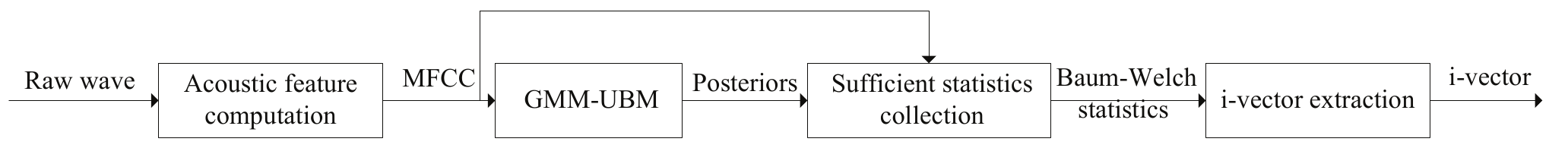}
  \put(-318,-5){\small\bfseries\color{black}{$\uparrow$}}
  \put(-335,-15){\small\bfseries\color{black}{$\{\omega_c,  \bm{\mu}_c,  \bm{\Sigma}_c\}_{c=1}^{C}$}}
  \put(-85,-5){\small\bfseries\color{black}{$\uparrow$}}
  \put(-88,-15){\small\bfseries\color{black}{$\{\mathbf{T}_c\}_{c=1}^{C}$}}
  \put(-373,-4){\small\bfseries\color{black}{$\uparrow$}}
  \put(-380,-15){\small\bfseries\color{black}{$\{\mathbf{y}_t^{(i)}\}_{t=1}^T$}}
  \put(-23,-4){\small\bfseries\color{black}{$\uparrow$}}
  \put(-25,-15){\small\bfseries\color{black}{$\bm{\upsilon}^{(i)}$}}
  \put(-142,-4){\small\bfseries\color{black}{$\uparrow$}}
  \put(-163,-15){\small\bfseries\color{black}{$\{N_c^{(i)},\mathbf{f}_c^{(i)}, \mathbf{\widetilde{f}}_c^{(i)} \}_{c=1}^C$}}
  \caption{The traditional GMM/i-vector framework. The term MFCC denotes Mel-frequency cepstral coefficent.}
  \label{fig:IvectorFramework}
\end{figure*}

\subsection{From GMM/i-vector to DNN/i-vector}
\label{sec:GMM-UBM-ivector}

The performance of the conventional GMM-UBM based speaker recognition is largely affected by the speaker and channel variations of utterances. To address this issue,  \cite{dehak2010front} proposed to reduce the high-dimensional GMM-UBM supervectors into low-dimensional vectors, named \textit{i-vectors} by factor analysis. The GMM/i-vector system eliminates the within-speaker and channel variabilities effectively, which leads to significant performance improvement.

The GMM/i-vector system is shown in Fig. \ref{fig:IvectorFramework}. We assume that  $\mathcal{Y}=\{\mathbf{y}_t^{(i)}  \in \mathbb{R}^{d_1}|t=1,2,\cdots,T\}$ represents the $i$th ($i=1,2,\cdots,I$ ) utterance of $T$ successive Mel-frequency cepstral coefficent (MFCC) frames, and $\Omega =\{\omega_c \in \mathbb{R}, \bm{\mu}_c \in \mathbb{R}^{d_1},\bm{\Sigma}_c \in \mathbb{R}^{d_1 \times d_1}| c=1,2,\cdots, C\}$ ($\omega_c \geq 0$ for all $c$, and $\sum_{c=1}^C \omega_c = 1$) denotes a GMM-UBM model where $C$ is the total number of components and $\omega_c$, $\bm{\mu}_c$ and  $\bm{\Sigma}_c$ are the weight, mean, and
covariance matrix of the $c$th component respectively. Then, $\mathbf{y}_t^{(i)}$ is assumed to be generated by the following distribution \cite{lei2014novel, snyder2020x}:
\begin{equation}
 \mathbf{y}_t^{(i)} \sim \sum_{c=1}^{C} \omega_c N(\bm{\mu}_c+\mathbf{T}_c \bm{\upsilon}^{(i)}, \bm{\Sigma}_c )
\end{equation}
where  $\{\mathbf{T}_c\}_{c=1}^{C}$  is a so called  total variability subspace, $\bm{\upsilon}^{(i)}$ is a segment-specific standard normal-distributed latent vector. The i-vector used to represent the speech signal is the maximum a posterior (MAP) point estimate of the latent vector $\bm{\upsilon}^{(i)}$, and it can be  regarded as a kind of ``\textit{speaker embedding}''\footnote{In this paper, the  'embedding' denotes the problem of learning a vector space where speakers are ``embedded''. The i-vectors,  d-vectors (introduced in Section \ref{sec:d-vector}), and  x-vectors (introduced in Section \ref{sec:x-vector}) are different embedding models for learning the vector spaces.}.

Given a speech segment, the  following sufficient statistics can be accumulated from the GMM-UBM:
\begin{equation}
\label{eq:BW0}
   N_c^{(i)}= \sum_{t=1}^{T}  p(c|\mathbf{y}_t^{(i)})
\end{equation}
\begin{equation}
\label{eq:BW1}
  \mathbf{f}_c^{(i)}= \sum_{t=1}^{T}  p(c|\mathbf{y}_t^{(i)})\mathbf{y}_t^{(i)}
\end{equation}
\begin{equation}
\label{eq:BW2}
  \mathbf{S}_c^{(i)}= \sum_{t=1}^{T}  p(c|\mathbf{y}_t^{(i)})\mathbf{y}_t^{(i)}(\mathbf{y}_t^{(i)})^T
\end{equation}
where $ p(c|\mathbf{y}_t^{(i)})= \frac{\omega_c N(\mathbf{y}_t^{(i)};\bm{\mu}_c,\bm{\Sigma}_c)}{\sum_{c^\prime=1}^{C}\omega_c N(\mathbf{y}_t^{(i)};\bm{\mu}_{c^\prime},\bm{\Sigma}_{c^\prime})} $ denotes the posterior probability of  $\mathbf{y}_t^{(i)}$ against the $c$th Gaussian component. These sufficient statistics are all that are needed to train the subspace $\{\mathbf{T}_c\}_{c=1}^{C}$ and extract the i-vector $\bm{\upsilon}^{(i)}$  \cite{lei2014novel}.  See \cite{dehak2010front,kenny2008study} for the details of training $\mathbf{T}_c$ and estimating the i-vectors.

Motivated by the success of deep learning for speech recognition, many efforts have been made to replace the GMM-UBM module of the GMM/i-vector system by DNN, which can be categorized to two main streams---DNN-UBM/i-vector and DNN based bottleneck feature (DNN-BNF)/i-vector. The two main streams will be presented in detail in the following two subsections, with selected references summarized in Table \ref{tab:DNN_ivector}.

\begin{table}[t]
  \centering
  \footnotesize
  \caption{Two main streams of the DNN/i-vector techniques. }
  \label{tab:DNN_ivector}
  \begin{tabular}{m{2.3cm}m{5.3cm}}
    \hline
    \hline
    \textbf{Approaches} & \textbf{References}\\
    \hline
      DNN-UBM/i-vector                  &\cite{lei2014novel} \cite{dey2016deep} \cite{zeinali2016deep} \cite{lei2014deep} \cite{dey2017exploiting}  \cite{zeinali2017text} \cite{richardson2015deep}  \cite{kenny2014deep} \cite{snyder2015time}  \cite{mclaren2015advances} \cite{chen2015phone}  \cite{richardson2015unified} \cite{sadjadi2016ibm} \cite{mclaren2014application} \cite{garcia2015insights} \cite{zheng2015exploring} \\
  %  \hline
       DNN-BNF/i-vector                 & \cite{ghalehjegh2015deep} \cite{zeinali2016deep} \cite{richardson2015deep} \cite{mclaren2015advances} \cite{richardson2015unified} \cite{mclaren2016exploring} \cite{lozano2016analysis} \cite{do2013augmenting}   \cite{sarkar2014combination}      \\
    \hline
    \hline
  \end{tabular}
\end{table}

\begin{figure}[t]
  \centering
  \includegraphics[width=2.5in]{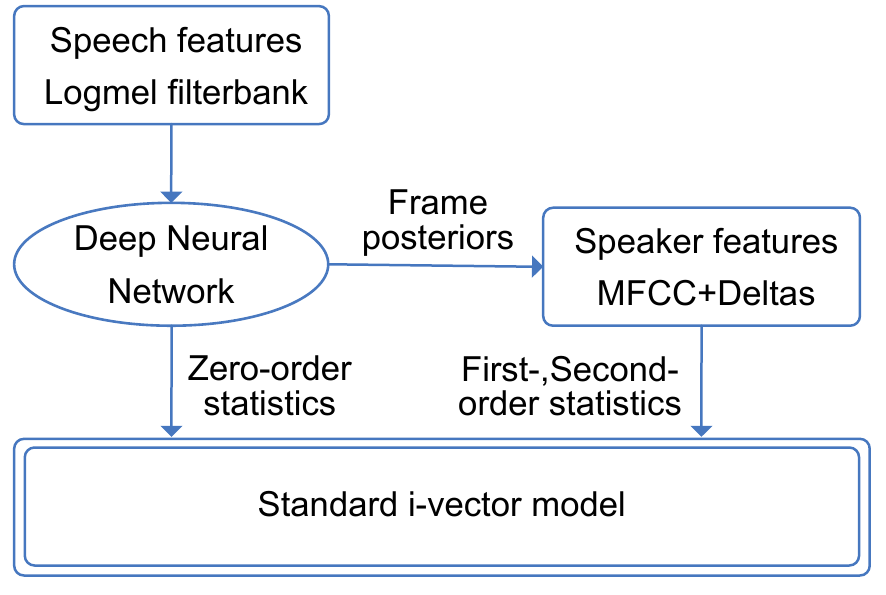}
  \caption{DNN-UBM/i-vector (from \cite{lei2014novel}). The posteriors are produced from the DNN acoustic model of an automatic speech recognition (ASR) system that is trained with, e.g. Logmel filterbank features. On the contrary, the sufficient statistics are computed from a speaker verification (SV) system that is trained with, e.g. MFCC which is not necessarily the same as the features for ASR. That is to say, one does not have to find a feature that works well for both ASR and SV in this framework.}
  \label{fig:DNN-ASR/i-vector}
\end{figure}

\subsection{DNN-UBM/i-vector}
\label{sec:DNN-UBM}
From \eqref{eq:BW0}, \eqref{eq:BW1}, and \eqref{eq:BW2}, one can see that only the posteriors of speech frames are needed to collect sufficient statistics for producing the i-vectors. Thus, we can use any probabilistic models beyond GMM-UBM to produce the posteriors theoretically \cite{lei2014novel}. Motivated by this insight,  \cite{lei2014novel} proposed the DNN-UBM/i-vector framework (Fig.\ref{fig:DNN-ASR/i-vector}) which takes a DNN acoustic model trained for ASR, denoted as DNN-UBM, to generate the posterior probabilities instead of GMM-UBM.

Specifically, DNN-UBM uses a set of senones  $\mathcal{Q}=\{Q_c|c=1,2,\cdots,C\}$, e.g., the tied-triphone states, to mimic  the mixture components of the GMM-UBM. It first trains a DNN-based ASR acoustic model to align each training frame with a senone, and then generates the posterior probabilities of each frame over the senones from the softmax output layer of the DNN acoustic model. The posteriors can be directly applied to \eqref{eq:BW0}, \eqref{eq:BW1} and \eqref{eq:BW2} to extract the DNN-UBM based i-vector. Due to the strong representation ability of DNN over GMM, DNN-UBM/i-vector yields 30\% relative equal error rate (EER) reduction over GMM/i-vector on the telephone condition of  the 2012 NIST speaker recognition evaluation (SRE) \cite{lei2014novel}. Later on, the authors in  \cite{lei2014deep,mclaren2015advances,mclaren2014application} further analyzed the performance of the DNN-UBM/i-vector in microphone and noisy conditions.

A lot of further studies bloomed the DNN-UBM/i-vector related techniques. For example, \cite{richardson2015deep,richardson2015unified} proposed to use a single ASR-DNN for  both the speaker and language recognition tasks simultaneously. Additionally, \cite{snyder2015time} employed a time delay deep neural network (TDNN), which was originally applied to speech recognition, to compute the posteriors. It achieved the state-of-the-art performance on the NIST SRE10 corpus at the time. As a third instance, \cite{zheng2015exploring} replaced the feedforward DNN by a long short-term memory (LSTM) recurrent neural network (RNN). The last but not all, \cite{garcia2015insights} studied a number of open issues relating to performance, computational complexity, and applicability of different types of DNNs.

The advantage of the DNN acoustic model may be brought by its strong ability in modeling content-related phonetic states explicitly, which not only generates highly compact representation of data but also provides precise frame alignment. This advantage is particularly apparent in text-dependent speaker verification \cite{chen2015phone,dey2016deep,dey2017exploiting,zeinali2016deep,zeinali2017text}.
However, this comes at the cost of greatly increased computational complexity over the traditional GMM-UBM/i-vector systems \cite{snyder2015time, snyder2017deep}, since that a DNN usually has more parameters than GMM. In addition, the training of the DNN based acoustic model  requires a large number of labeled training data.

To overcome the  computational complexity, a supervised GMM-UBM was also investigated based on the DNN acoustic model \cite{snyder2015time}. In specific, a  GMM  is obtained by:
\begin{equation}\label{eq:dnn-asr-poster}
  \begin{split}
      \gamma_{ct}^{(i)} = & p(c|\mathbf{\overline{y}}_t^{(i)}) \\
      \omega_{c}  =   & \sum_{i,t}\gamma_{ct}^{(i)} \\
      \bm{\mu}_c = & \frac{\sum_{i,t}\gamma_{ct}^{(i)}\mathbf{y}_t^{(i)}}{\sum_{i,t}\gamma_{ct}^{(i)}} \\
      \bm{\Sigma}_c = & \frac{\sum_{i,t}\gamma_{ct}^{(i)}\mathbf{y}_t^{(i)}(\mathbf{y}_t^{(i)})^T}{\sum_{i,t}\gamma_{ct}^{(i)}}-\bm{\mu}_c\bm{\mu}_c^T\\
  \end{split}
\end{equation}
where $\mathbf{\overline{y}}_t^{(i)}$ and $\mathbf{y}_t^{(i)}$ denote the acoustic features for ASR and speaker recognition respectively, and $p(c|\mathbf{\overline{y}}_t^{(i)})$ is the posterior probability corresponding  to the $c$th senones. By this way, the
supervised-GMM maintains the training computational complexity of the traditional unsupervised-GMM, with a 20\% relative EER reduction on the NIST
SRE10 corpus \cite{snyder2015time}. Similar idea was also studied in \cite{lei2014novel}, though no performance improvement over the baseline is observed. Although the supervised-GMM reduces the training computational complexity, training the DNN acoustic model still needs a large amount of labeled training data.

\subsection{DNN-BNF/i-vector}
\label{sec:DNN-BNF}

\begin{figure}[t]
  \centering
  \includegraphics[width=3.2in]{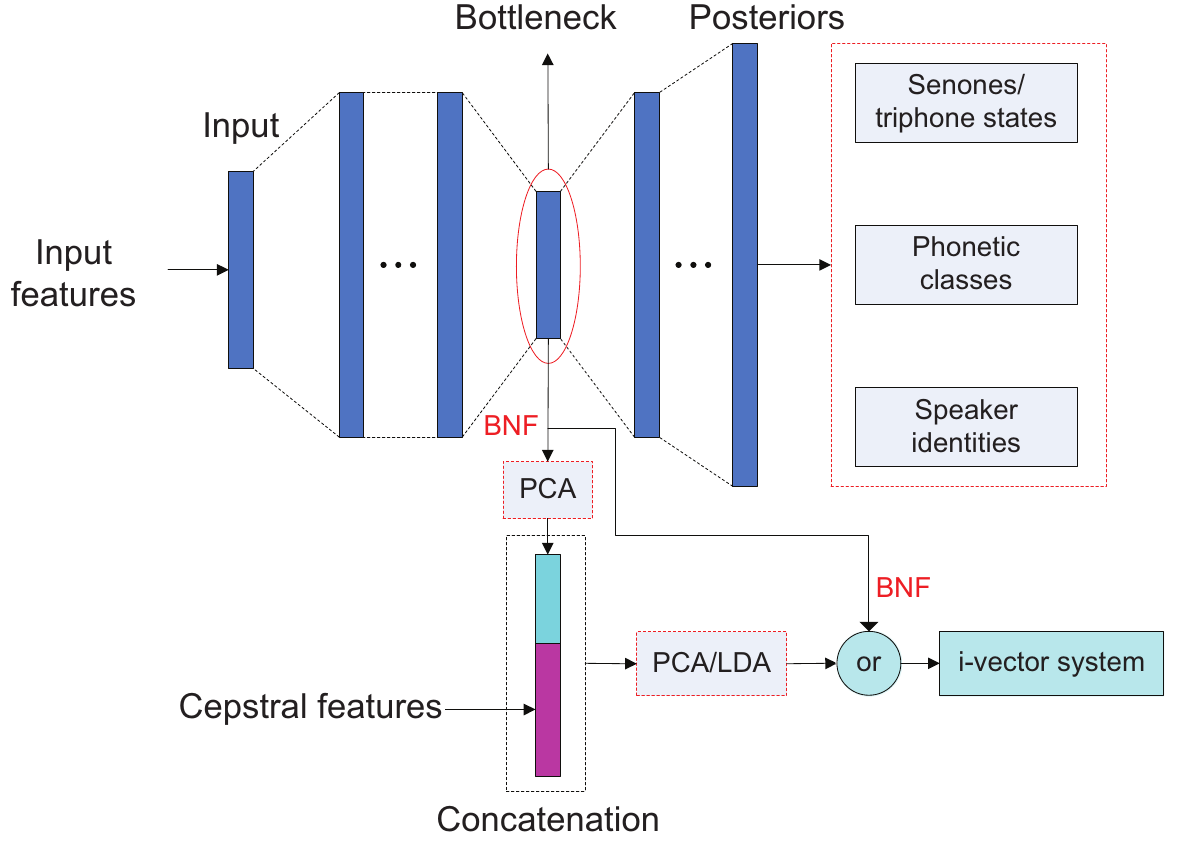}
  \caption{The DNN-BNF/i-vector framework. The framework is a summarization of related work. The term ``PCA'', ``LDA'', and ``BNF'' are short for principal components analysis, linear discriminant analysis, and bottleneck feature respectively.}
  \label{fig:DNN-BNF/i-vector}
\end{figure}

The fundamental idea of DNN-BNF/i-vector is to extract a compact feature from the bottleneck layer of a DNN as the input of the factor analysis, where the bottleneck layer is a special hidden layer of the DNN that has much less hidden units than the other hidden layers. In practice, DNN-BNF/i-vector has many variants, as we have summarized in Fig. \ref{fig:DNN-BNF/i-vector}. Like DNN-UBM/i-vector, the deep model in DNN-BNF/i-vector is mainly trained to discriminate senones \cite{zeinali2016deep,mclaren2015advances,mclaren2016exploring,lozano2016analysis,richardson2015deep,richardson2015unified} or phonemes \cite{do2013augmenting,sarkar2014combination}.

 The input of the factor analysis can be either the bottleneck feature (BNF) produced from the bottleneck layer, a concatenation of BNF with other acoustic feature \cite{do2013augmenting,sarkar2014combination}, or a post-processed feature by principal components analysis (PCA) or linear discriminant analysis (LDA) \cite{do2013augmenting,sarkar2014combination}. One can find that no matter whether we apply BNF alone \cite{richardson2015unified} or concatenate it with other acoustic features \cite{mclaren2015advances}, DNN-BNF/i-vector can significantly outperform the conventional GMM/i-vector, which indicates the effectiveness of the framework \cite{richardson2015deep}.

However, it is unclear why a deep model trained to discriminate phonemes or senones can produce speaker-sensitive BNF. To address this issue, {{the authors of}} \cite{mclaren2016exploring} assumed that speaker information is traded for dense phonetic information when the bottleneck layer moves toward the DNN output layer. Under this hypothesis, they experimentally analyzed the role of BNF by placing the bottleneck layer at different depths of the DNN. They found that, if the training and test conditions match, the closer the bottleneck layer is to the output layer, the better the performance is; otherwise, the bottleneck layer should be placed around the middle of the DNN.
{{The authors of}} \cite{lozano2016analysis} explored whether weakening the accuracy of the acoustic model on speech recognition yields better BNF for speaker recognition. They analyzed the speaker recognition performance in different respects of the acoustic model, including
under-trained DNN, different inputs, and different feature normalization strategies. Results indicate that high speech recognition performance in terms of phonetic accuracy does not necessarily imply {{increased speaker recognition accuracy}}.
In addition, {{\cite{ghalehjegh2015deep}}} proposed to take speaker identity as the training target, under the conjecture that this training target should be able to improve the robustness of the phonetic variability of BNF.

\subsection{Discussion to the DNN/i-vector}
\label{sec:discuss_DNN/i-vector}

\renewcommand\arraystretch{1.5}
\begin{table}[t]
  \centering
  \footnotesize
  \caption{Comparison results between DNN/i-vector and GMM/i-vector. Each row denotes a comparison. The last three columns list the EER  of the main models, the EER  of the baselines, and the  relative EER reductions, respectively. \textbf{The results across rows are not comparable}, since they are collected from different references, and their comparisons are not apple-to-apple comparisons.}
  \label{tab:experiments_DNN/i-vector}
  \scalebox{0.75}{
  \begin{tabular}{m{2.45cm} m{1.45cm} m{2.5cm}<{\centering} m{0.8cm}<{\centering} m{0.8cm}<{\centering} m{1.2cm}<{\centering}}
    \hline
    \hline
    \multicolumn{2}{c}{Comparisons}   & \multirow{2}*{Test dataset [condition]} &\multicolumn{3}{c}{EER}\\
    \cline{1-2}    \cline{4-6}
    Main models         & Baselines           &       & Main   & Baseline & Relative reduction \\
    \hline
    DNN-UBM \cite{lei2014novel}  & GMM-UBM    &  NIST SRE12 C2       &  1.39\%    & 1.81\%     & \textbf{23\%}\\
    DNN-UBM \cite{lei2014novel}  & GMM-UBM    &  NIST SRE12 C5       &  1.92\%    & 2.55\%     & \textbf{25\%}\\
    \hline
    TDNN-UBM\cite{snyder2015time}& GMM-UBM    &  NIST SRE10 C5       &  1.20\%    & 2.42\%     & \textbf{50\%}\\
    Sup-GMM-UBM\cite{snyder2015time}& GMM-UBM &  NIST SRE10 C5       &  1.94\%    & 2.42\%     & \textbf{20\%}\\
    \hline
    BNF \cite{richardson2015deep}& MFCC    &In-domain DAC13   & 2.00\%  &2.71\%  & \textbf{26\%}\\
    BNF \cite{richardson2015deep}& MFCC    &Out-domain DAC13  & 2.79\%  &6.18\%  & \textbf{55\%}\\
    \hline
    \hline
  \end{tabular}}
\end{table}

It is known that a major difference between DNN-UBM and GMM-UBM is that DNN-UBM is a discriminant model, while GMM-UBM is a generative one. {{DNN-UBM is  more powerful than GMM-UBM in modeling a complicated data distribution \cite{lei2014novel, snyder2015time}}}. Moreover, the DNN acoustic model is trained to align each speech frame to its corresponding senone in a supervised fashion. Its output nodes have a clear physical explanation. It mines the pronunciation characteristics of speakers. On the contrary, GMM-UBM is trained by the expectation-maximum algorithm in an unsupervised manner. Its mixtures have no inherent meaning.
Although DNN-UBM/i-vector needs labeled training data and heavier computation power than GMM-UBM, it does yield excellent performance. In addition,  many corpora are also developed for the demand of training strong DNN, which will be reviewed in Section \ref{sec:data_tool}.

{{To demonstrate general performance differences of DNN/i-vector and GMM/i-vector,}} some carefully selected experimental results from literatures are listed in Table \ref{tab:experiments_DNN/i-vector}.  Compared to the GMM-UBM/i-vector baseline, one can find that DNN-UBM/i-vector achieves more than 20\% relative EER reduction over GMM-UBM/i-vector. In addition, the supervised GMM-UBM in \eqref{eq:dnn-asr-poster} can also get 20\% relative improvement according to the fourth row. Finally, from the last two rows, one can see that, when taking DNN-BNF and MFCC as the input features of the GMM-UBM/i-vector respectively, the former achieves better performance than the latter.

It should be note that, as far as we know, different test conditions may yield slightly different conclusions from those in Table \ref{tab:experiments_DNN/i-vector}. However, to our knowledge, the results in the table can be a representative of the research trend.

\section{Speaker feature extraction with deep embedding}
\label{sec:embedding}
 In this section, we first introduce two representative deep embeddings---d-vector and x-vector in Section \ref{subsec:twoseminal} with some discussions in Section \ref{subsec:discussion}, and then identify their key components in Section \ref{subsec:four}, which provide a taxonomy to existing algorithms.

\subsection{Two seminal work of deep embeddings}\label{subsec:twoseminal}
\subsubsection{Frame-level embedding --- d-vector}
\label{sec:d-vector}
\begin{figure}[t]
  \centering
  \includegraphics[width=2.75in]{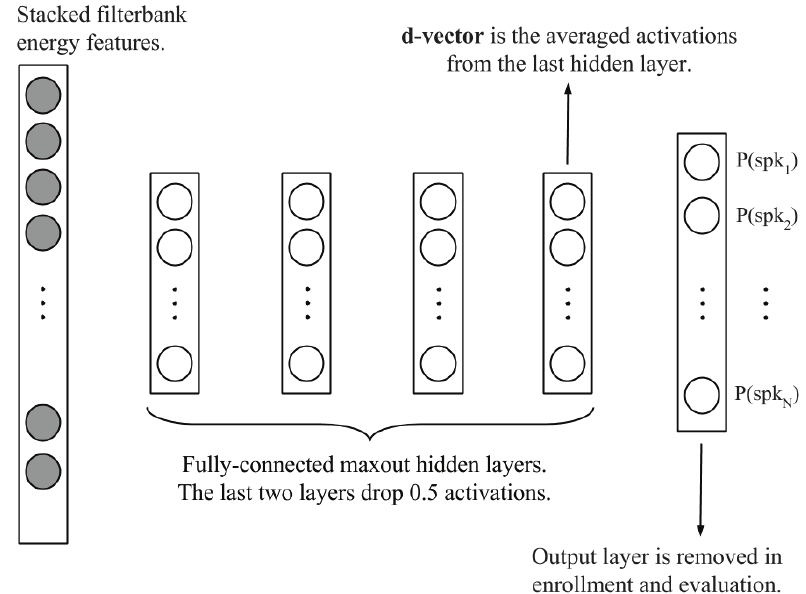}
  \caption{Diagram of the d-vector framework (from \cite{variani2014deep}).  }
  \label{fig:d-vector}
\end{figure}

D-vector is one of the earliest DNN-based embeddings \cite{variani2014deep}. The core idea of d-vector is to assign the ground-truth speaker identity of a training utterance as the labels of the training frames belonging to the utterance in the training stage, which transforms the model training as a classification problem. As shown in Fig. \ref{fig:d-vector}, d-vector expands each training frame with its context, and employs a \textit{maxout} DNN to classify the frames of a training utterance to the speaker identity of the utterance, where the DNN takes softmax as the output layer to minimize the cross-entropy loss between the ground-truth labels of the frames and the network output.

In the test stage, d-vector takes the output activation of each frame from the last hidden layer of the DNN as the deep embedding feature of the frame, and averages the deep embedding features of all frames of an utterance as a new compact representation of the utterance, named \textit{d-vector}.
An underlying hypothesis of d-vector is that the compact representation space produced from a development set may generalize well to unseen speakers in the test stage.

\subsubsection{Segment-level embedding --- x-vector}
\label{sec:x-vector}
\begin{figure}[t]
  \centering
  \includegraphics[width=2.85in]{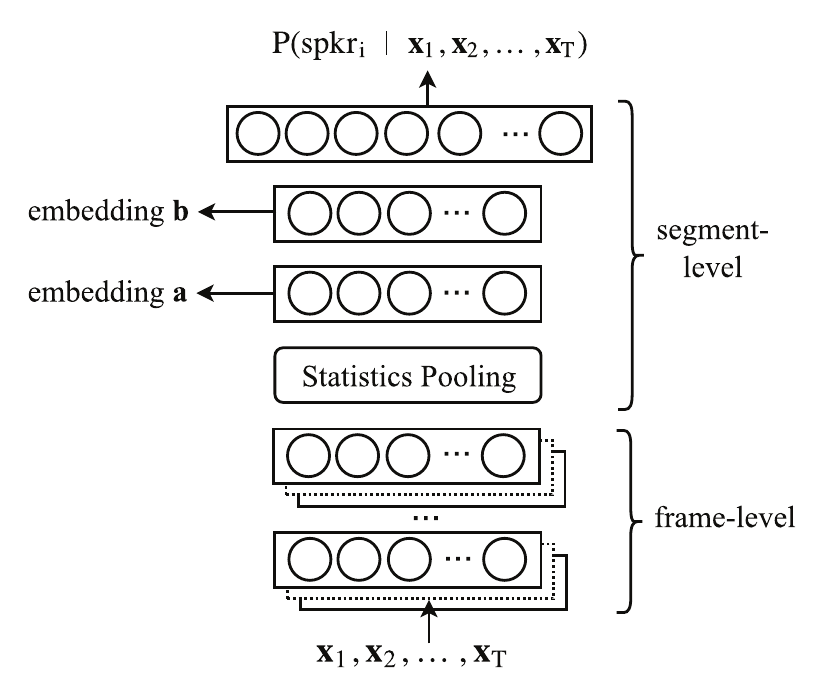}
  \caption{Diagram of the DNN model for extracting x-vectors (from \cite{snyder2017deep}). Note that segment-level embeddings (e.g., $\mathbf{a}$ or $\mathbf{b}$) can be extracted from any layer of the network after the statistics pooling layer \cite{snyder2017deep}. \cite{snyder2018x} where the name ``x-vector'' comes from uses the embedding \textbf{a} as the speaker feature. }
  \label{fig:x-vector}
\end{figure}

X-vector \cite{snyder2017deep,snyder2018x} is an important evolution of d-vector that evolves speaker recognition from frame-by-frame speaker labels to utterance-level speaker labels with an aggregation process. The network structure of x-vector is shown in Fig. \ref{fig:x-vector}.
It first extracts frame-level embeddings of speech frames by time-delay layers, then concatenates the mean and standard deviation of the frame-level embeddings of an utterance as a segment-level (a.k.a., utterance-level) feature by a statistical pooling layer, and finally classifies the segment-level feature to its speaker by a standard feedforward network. The time-delay layers, statistical pooling layer, and feedforward network are jointly trained.
 \textit{X-vector} is defined as the segment-level speaker embedding produced from the second to last hidden layer of the feedforward network, i.e. the variable $\mathbf{a}$ in Fig. \ref{fig:x-vector}.

{{The authors in \cite{snyder2018x}}} found that data augmentation is important in improving the performance of x-vector. We will introduce the data augmentation techniques in Section \ref{sec:data_augmentation}.

\subsection{Discussion to the speaker embedding}\label{subsec:discussion}
\renewcommand\arraystretch{1.5}
\begin{table*}[t]
  \footnotesize
  \centering
  \caption{Characteristics of different speaker embeddings and their favorite back-ends. }
  \label{tab:comparsion_speaker_embedding}
  \begin{tabular}{m{2.5cm}|  m{3.2cm}<{\centering} m{3.2cm}<{\centering} m{2.1cm}<{\centering}  m{1.2cm}<{\centering} m{2.5cm}<{\centering}}
  \hline  \hline
  Model name                & Type & Training strategy & Label for model training     & Back-end name        &  Label for back-end training    \\
  \hline
  GMM-UBM/i-vector      & Generative/Generative        & Unsupervised/Unsupervised    & \xmark / \xmark          & PLDA  &Speaker identity\\
  DNN-UBM/i-vector      & Discriminative/Generative    & Supervised /Unsupervised     & Phonetic labels/\xmark & PLDA  &Speaker identity\\
  DNN-BNF/i-vector      & Discriminative/Generative   & Supervised/Unsupervised      & Phonetic labels /\xmark & PLDA  &Speaker identity \\
  D-vector              & Discriminative               & Supervised & Speaker identity                         & Cosine& \xmark \\
  X-vector              & Discriminative               & Supervised & Speaker identity                         & PLDA  &Speaker identity \\
  \hline
  \hline
  \end{tabular}
\end{table*}
Similar to the i-vector, the d-vector and x-vector are also a kind of speaker embedding, which discriminatively embeds speakers into a vector space by using DNNs. We call this type of speaker embedding as {{\textit{deep speaker embedding}, or \textit{deep embedding} for short}}. The main characteristics between different speaker embeddings are summarized in Table \ref{tab:comparsion_speaker_embedding}. Compared to the traditional GMM-UBM/i-vector, the deep embedding is a discriminant model and trained in a supervised fashion. Compared to DNN-UBM, its training data does not need phonetic-level labels. Therefore, the training of the deep embedding is much simpler than that of DNN-UBM and DNN-BNF. In addition, the deep embedding is a new framework, while DNN-UBM/i-vector and DNN-BNF/i-vector are hybrid ones.

\renewcommand\arraystretch{1.5}
\begin{table*}[t]
  \centering
  \footnotesize
  \caption{Selected results on deep embedding in literature. Each row represents a comparison. The results across rows are not comparable.}
  \label{tab:experiments_deep_embedding}
  \scalebox{0.9}{
  \begin{tabular}{m{3.5cm}<{\centering} m{2.45cm}<{\centering} m{5cm}<{\centering} m{2cm}<{\centering} m{2cm}<{\centering} m{1.8cm}<{\centering}}
    \hline
    \hline
    \multicolumn{2}{c}{Comparison methods}   & \multirow{2}*{Test dataset [condition]}  &\multicolumn{3}{c}{EER}\\
    \cline{1-2}    \cline{4-6}
    Deep embedding         & Baseline          &       & Deep embedding    & Baseline & Relative reduction  \\
    \hline
    d-vector \cite{variani2014deep}  & GMM-UBM/i-vector    &  Google data     &  4.54\%  & 2.83\%   & \textbf{-37\%}  \\
    d-vector+i-vector \cite{variani2014deep}  & GMM-UBM/i-vector    & Google data [clean , noisy]    &  -----  & -----   & \textbf{[14\% , 25\%]}  \\
    \hline
     embedding \textbf{a+b} (in Fig.\ref{fig:x-vector}) \cite{snyder2017deep} & GMM-UBM/i-vector &  NIST SRE10 [10s-10s , 60s ] & [7.9\% , 2.9\%] &[11.0\% , 2.3\% ] &  \textbf{[28\% , -21\% ]} \\
    embedding \textbf{a+b} (in Fig.\ref{fig:x-vector}) \cite{snyder2017deep} & GMM-UBM/i-vector &  NIST SRE16 [Cantonese , Tagalog ] & [6.5\% , 16.3\% ] &[8.3\% , 17.6\% ] &  \textbf{[22\% , 7\%]} \\
   \hline
   x-vector (embedding \textbf{a}) \cite{snyder2018x}   & GMM-UBM/i-vector & SITW Core [ PLDA and extractor aug. , Incl. VoxCeleb] & [ 6.00\% , 4.16\% ]      &   [ 8.04\% , 7.45\%]  & \textbf{[25\%, 44\%]}    \\
   x-vector (embedding \textbf{a}) \cite{snyder2018x}   & GMM-UBM/i-vector & SRE16 Cantonese [ PLDA and extractor aug. , Incl. VoxCeleb] & [ 5.86\%, 5.71\% ]      &   [ 8.95\% , 9.23\%]  & \textbf{[ 34\%, 38\%]}    \\
   \hline
   \hline
  \end{tabular}}
\end{table*}

Some experimental results on deep embedding are listed in Table \ref{tab:experiments_deep_embedding}. From the table, one can find that, the d-vector alone yields higher EER than the i-vector. When fusing the d-vector and i-vector, the combined system achieves 14\% and 25\% relative EER reduction in clean and noisy test conditions respectively over the i-vector. The ``embedding a+b'' model, which is the predecessor of the x-vector, achieves lower EER than the GMM-UBM/i-vector baseline on the 10-second short utterances of NIST SRE10, and higher EER than the latter on the 60-second long utterances of NIST SRE10. With enlarged training data and data augmentation, the x-vector achieves significant performance improvement over the GMM-UBM/i-vector.

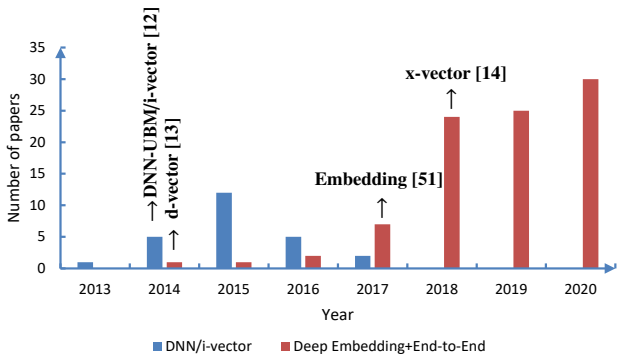
\begin{figure}[t]
  \centering
  \includegraphics[width=3.2in]{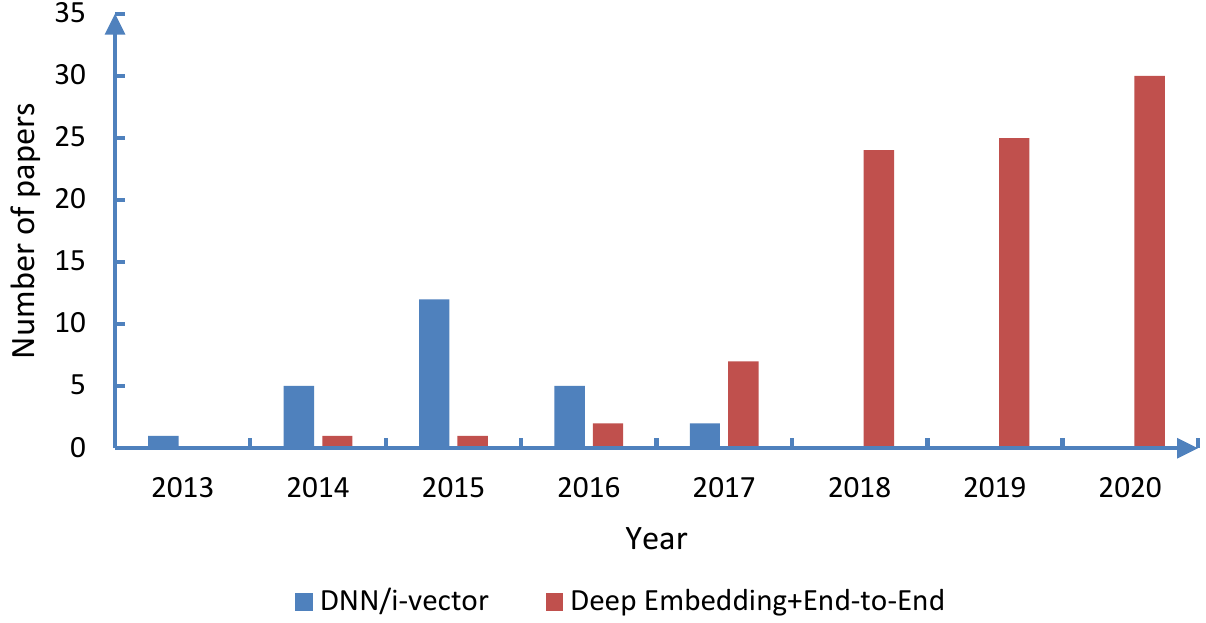}
  \put(-170,52){\small\bfseries\color{black}{\begin{turn}{90} \scriptsize d-vector \cite{variani2014deep} \end{turn}}}
  \put(-169,42){\small\bfseries\color{black}{$\uparrow$}}
  \put(-178,60){\small\bfseries\color{black}{\begin{turn}{90}\scriptsize DNN-UBM/i-vector \cite{lei2014novel}\end{turn}}}
  \put(-177,53){\small\bfseries\color{black}{$\uparrow$}}
  \put(-80,105){\small\bfseries\color{black}{\scriptsize x-vector \cite{snyder2018x} }}
  \put(-114,65){\small\bfseries\color{black}{\scriptsize Embedding \cite{snyder2017deep}}}
  \put(-91,54){\small\bfseries\color{black}{$\uparrow$}}
  \put(-65,95){\small\bfseries\color{black}{$\uparrow$}}
  \caption{Statistics of the published papers on DNN/i-vector and deep embedding cited by this article.}
  \label{fig:Embedding_time_line}
\end{figure}

Fig. \ref{fig:Embedding_time_line} shows the number of the related papers. {{We observe the following phenomena.}} First, the d-vector and DNN-UBM/i-vector was proposed both in 2014, where the former achieved better performance at the time. Second, the research on DNN/i-vector was mainly conducted in the first few years after its appearance, and then became less studied. Third, the research on deep embedding becomes bloom along with its performance improvement after that the x-vector achieved the state-of-the-art performance. At present, the deep embedding is the trend of speaker recognition, which has been developed in several aspects as summarized in the following subsection.

\subsection{Four key components of deep embedding}\label{subsec:four}
Motivated by the seminal work d-vector and x-vector, many deep embedding techniques were proposed, most of which are composed of four key components---network input, network structure, temporal pooling, and training objective. These components include but not limited to the  following contents:
\begin{itemize}
  \item \textit{Network inputs and structures}:
  The network input can be categorized into two classes---raw wave signals in time domain and acoustic features in time-frequency domain, including spectrogram, Mel-filterbanks (f-bank), and MFCC. The network structure is diverse, which is rooted essentially at DNN,  RNN/LSTM, and CNN. Because the network input and structure were jointly designed case by case in practice, we will jointly summarize them in Section  \ref{sec:embedding_input_network}.

  \item \textit{Temporal pooling}: Temporal pooling represents the transition layer of a neural network that transforms frame-level embedding features to utterance-level embedding features. The temporal pooling strategies consist of two classes---statistical pooling and learning based pooling. We will introduce them in Section \ref{sec:embedding_pooling}.

  \item \textit{Objective functions}: Objective functions affect the effectiveness of speaker recognition much. Both d-vector and x-vector adopt softmax as the output layer and take the cross-entropy minimization as the objective function, which may not be optimal. Recently, many objectives were designed to further improve the performance. We will survey the objective functions in Section \ref{sec:embedding_objectives}.
\end{itemize}

\section{Deep embedding: network  structures and  inputs}
\label{sec:embedding_input_network}

\begin{table*}[t]
  \centering
  \footnotesize
  \caption{A brief summary of the inputs and neural network structures in deep speaker feature extraction.}
  \label{tab:networks_inputs_summary}
  \scalebox{1}{
  \begin{tabular}{m{2cm} <{\centering}|  m{6cm} |  m{2cm} <{\centering} | m{5.5cm} }
  \hline
  \hline
  \textbf{Inputs}&\makecell[c]{ \textbf{CNN  } }& \makecell[c]{\textbf{LSTM  } } & \makecell[c]{\textbf{Hybrid structures}}    \\
  \hline
  \textbf{Wave} &Others \cite{muckenhirn2018towards, ravanelli2018speaker}. &------ &CNN-LSTM \cite{jung2018avoiding, jung2018complete}; CNN-GRU \cite{jung2019rawnet, jung2019short}.\\

  \textbf{Spectrogram}& ResNet \cite{yu2019ensemble, chung2018voxceleb2, xie2019utterance, 9054440};  VGGNet \cite{nagrani2017voxceleb,
  yadav2018learning}; Inception-resnet-v1 \cite{zhang2017end, zhang2018text}. &------ & CNN-GRU \cite{zhang2019seq2seq} \\

  \textbf{F-bank}       & TDNN \cite{snyder2018x,9053198,Zhu2020Orthogonality,9053209}; ResNet \cite{9053217,li2017deep,kim2019deep,Garcia-Romero2020MagNetO}; VGGNet \cite{bhattacharya2017deep}; Inception-resnet-v1 \cite{zhang2018text,li2018deep,li2019boundary}; Others \cite{torfi2018text,li2017deepspeaker}.    &   \cite{rahman2018attention,wan2018generalized,heigold2016end}. &BLSTM-ResNet \cite{9053767}, TDNN-LSTM \cite{tang2019deep}                 \\

  \textbf{MFCC}          &TDNN \cite{snyder2019speaker,snyder2017deep,liu2018speaker,zhu2018self,okabe2018attentive,bai2020partial,li2019gaussian,
  xiang2019margin,9053209,villalba2019state,9054350,Li2020Bayesian}; ResNet \cite{zhou2019deep}; Others \cite{gao2018improved,jiang2019effective}.     &  ------       &     TDNN-LSTM \cite{chen2019speaker}               \\
  \hline
  \hline
  \end{tabular}
  }
\end{table*}

Although deep neural networks can be divided roughly into DNN, CNN, and RNN/LSTM structures, the network structure and input for speaker recognition are quite flexible. Each component of a network has many candidates. For example, the hidden layer of a neural network may be a standard convolutional layer \cite{bhattacharya2017deep}, a dilated convolution layer \cite{gao2018improved}, a LSTM layer \cite{jung2018avoiding}, a gated recurrent unit (GRU) layer \cite{jung2019rawnet}, a multi-head attention layer, a fully-connected layer, and even a combination of these  different layers \cite{jung2018avoiding,jung2019rawnet}, etc. The activation functions can be Sigmoid, Rectified Linear Unit (ReLU), Leaky ReLU, or  Parametric Rectified Linear Unit (PReLU) etc. Besides, the topology of a network and connection mode between layers are all variables. Even the number of layers and number of hidden units at a layer can also affect the performance. To prevent enumerating the networks case by case, here we first review some commonly used networks for the speaker feature extraction, and then briefly review their inputs.

\textit{Time delay neural network} (TDNN) \cite{snyder2017deep}: TDNN takes a one-dimensional convolution structure along the time axis as a feature extractor \cite{peddinti2015time}. It is adopted by the well known x-vector, as shown in Fig. \ref{fig:x-vector}. Due to the success of the x-vector \cite{snyder2018x,snyder2019speaker}, TDNN becomes one of the most popular structures for speaker recognition.
For example, \cite{liu2018speaker} introduced phonetic information to the TDNN architecture based embedding extractor.
\cite{stafylakis2019self} trained a TDNN embedding extractor without speaker labels via self-supervised training. \cite{9053198,Zhu2020Orthogonality}  explored the effectiveness of the orthogonality regularization by TDNN. Generally, TDNN has been frequently used as a framework to study other key components of the deep embedding models, such as the temporal pooling layers \cite{zhu2018self, okabe2018attentive} and objective functions \cite{bai2020partial,li2019gaussian,xiang2019margin}.

The TDNN structure has also been intensively improved. For instance, an extended TDNN architecture (E-TDNN) was introduced in \cite{snyder2019speaker}, which greatly outperforms the  x-vector baseline \cite{snyder2018x}. It adopts a slightly wider temporal context than TDNN, and interleaves affine layers in between the convolutional layers \cite{garcia2020jhu}. \cite{povey2018semi} developed a factorized TDNN (F-TDNN) to reduce the number of parameters. It factorizes the weight matrix of each TDNN layer into the product of two low-rank matrices. It further constrains the first low-rank matrix to be semi-orthogonal under the assumption that the semi-orthogonal constraint prevents information loss. The application of F-TDNN to deep embedding was also investigated \cite{villalba2019state, garcia2020jhu,  snyder2019jhu}. Some other parameter reduction works can be found in \cite{Yu2020,Georges2020Compact}. Recently, \cite{9054350} integrated TDNN with statistics pooling at each layer for compensating the variation of temporal context in the frame-level transforms. Similarly, \cite{tang2019deep,chen2019speaker} inserted LSTM layers into TDNN to capture the temporal information for remedying the weakness of TDNN whose time delay layers focus on local patterns only. \cite{Li2020Bayesian} alleviated the  mismatch problem between training and evaluation by incorporating  Bayesian neural networks into TDNN.

\textit{Residual networks} (ResNet) \cite{he2016deep}: it is another popular structure in speaker embedding. Its trunk architecture is a 2-dimensional CNN with convolutions in both the time and frequency domains. Some work directly used the standard ResNet as their speaker  feature extractors \cite{9053217,yu2019ensemble,chung2018voxceleb2,li2017deep}. Some other work employed ResNet as a backbone and modified it for specific purposes or applications \cite{xie2019utterance,zhou2019deep,9053767,9054440,kim2019deep,Garcia-Romero2020MagNetO}.
For example, to reduce the number of parameters, \cite{xie2019utterance} modified the standard ResNet-34 to a \textit{thin} ResNet by cutting down the number of channels in each residual block.  {{The authors in \cite{9053767}}} combined bi-directional LSTM (BLSTM) and ResNet into a unified architecture, where the BLSTM is used to model long temporal contexts. {{The authors in \cite{zhou2019deep}}} incorporated a so-called ``squeeze-and-excitation'' block into ResNet.

\textit{Raw wave neural networks} \cite{muckenhirn2018towards, ravanelli2018speaker,  jung2018avoiding, jung2018complete,jung2019rawnet, jung2019short, Jung2020Improved, LinWeiwei2020}: some work takes raw waves in the time domain as the input, which aims to extract learnable acoustic features instead of handcrafted features. For example, \cite{muckenhirn2018towards} applied CNN to capture raw speech signal. The experimental results indicate that the filters of the first convolution layer give emphasis to speaker information in low frequency regions. {{The authors in \cite{ravanelli2018speaker}}} believed that the first layer is critical for the waveform-based CNNs, since it not only deals with high-dimensional inputs, but suffers more from the gradient vanishing problem than the other layers. Therefore, they proposed a SincNet architecture based on parametrized sinc functions, where only low and high cutoff frequencies of band-pass filters  are learned from data \cite{ravanelli2018speaker}.
In \cite{jung2018complete}, {{the authors}} thought that  the difficulty of  processing raw audio signals by DNN is  mainly caused by the fluctuating scales of the signals. To stabilize the scales, they employed a convolutional layer, named pre-emphasis layer, to mimic the well-known signal  pre-emphasis technique $p(t)=s(t)-\alpha s(t-1)$. They also made several improvements to the original  raw wave  network \cite{jung2019rawnet, jung2019short, Jung2020Improved} which results in excellent performance. \cite{LinWeiwei2020} designed a \textit{Wav2Spk} architecture to learn speaker embeddings from waveforms, where the traditional MFCC extraction, voice activity detection, and cepstral mean and variance normalization are replaced by a feature encoder, a temporal gating unit and an instance normalization
scheme respectively. Wav2Spk performs better than the convention x-vector network.

\textit{Other neural networks}: in addition to TDNN and ResNet, many other well-known neural network architectures have also been applied to speaker recognition, including VGGNet \cite{bhattacharya2017deep,yadav2018learning,nagrani2017voxceleb}, Inception-resnet-v1 \cite{zhang2017end, zhang2018text, li2018deep, li2019boundary}, BERT \cite{Ling2020BERTphone}, and Transformer \cite{Safari2020}.
Besides, recurrent neural networks, such as LSTM and gated recurrent units, are often used for text-dependent speaker verification \cite{rahman2018attention,wan2018generalized,heigold2016end}. The CNN models can also be improved by inserting LSTM or gated recurrent units into the backbone networks \cite{jung2018avoiding,jung2018complete,jung2019rawnet,jung2019short,zhang2019seq2seq,9053767,tang2019deep,chen2019speaker}.
Finally, apart from the above handcrafted neural architectures, neural architecture search was also recently applied to speaker recognition \cite{Ding2020AutoSpeech,Qu2020Evolutionary}.

\textit{Neural network inputs}: Table \ref{tab:networks_inputs_summary} provides a summary to the common inputs and neural networks   for the deep embedding based speaker feature extraction. From the table, one can see that CNN-based neural networks and f-bank/MFCC acoustic features are becoming popular, while some 2-dimensional convolution structures, e.g. ResNet, use spectrogram as the input feature.
In addition to the above common inputs, such as MFCC, spectrum and mel-filterbanks, \cite{Liu2020Comparative} recently presented an extensive re-assessment of 14 acoustic feature extractors. They found that the acoustic features equipped with the techniques of spectral centroids, group delay function, and integrated noise suppression provide promising alternatives to MFCC.

\subsection{Discussion to the networks}
\renewcommand\arraystretch{1.5}
\begin{table}[t]
  \centering
  \footnotesize
  \caption{Selected examples of the effect of neural network structures on performance (from \cite{villalba2020state}).}
  \label{tab:experiments_network}
  \scalebox{0.8}{
  \begin{tabular}{m{1.8cm} m{1.35cm} m{2.5cm}<{\centering} m{0.8cm}<{\centering} m{0.8cm}<{\centering} m{1.2cm}<{\centering}}
    \hline
    \hline
    \multicolumn{2}{c}{Comparison methods}   & \multirow{2}*{Test dataset [condition]} &\multicolumn{3}{c}{EER}\\
    \cline{1-2}    \cline{4-6}
    Main models         & Baselines           &        & Main   & Baseline & Relative reduction  \\
    \hline
    E-TDNN(10M)    & TDNN(8.5M)    &  SITW EVAL CORE (\textit{16 kHz systems})     &  2.74\%    & 3.40\%     & \textbf{19\%}\\
    F-TDNN(9M)     & TDNN(8.5M)    &  SITW EVAL CORE (\textit{16 kHz systems})     &  2.39\%    & 3.40\%     & \textbf{30\%}\\
    F-TDNN(17M)    & TDNN(8.5M)    &  SITW EVAL CORE (\textit{16 kHz systems})     &  1.89\%    & 3.40\%     & \textbf{44\%}\\
    ResNet(8M)     & TDNN(8.5M)    &  SITW EVAL CORE (\textit{16 kHz systems})     &  3.01\%    & 3.40\%     & \textbf{11\%}\\
    \hline
    \hline
  \end{tabular}}
\end{table}
The network structure plays a key role on performance. For example, as shown in Table \ref{tab:experiments_network}, E-TDNN and F-TDNN significantly reduced EER on the SITW dataset \cite{mclaren2016speakers}, where F-TDNN achieves  more than 40\% relative EER reduction over the original TDNN. Although this promotion is not consistent across all datasets \cite{villalba2020state}, it demonstrates the importance of the network structure on performance.

For the acoustic features, the delta and double-delta features are helpful in statistical model based speaker recognition, e.g. the GMM-UBM/i-vector. However, they are not very effective in convolution and time-delay neural networks. This may be caused by that, the statistical model needs the delta and double-delta operations to capture the time dependency between frames, while the neural networks are able to achieve this goal intrinsically.

Although the deep embedding networks have achieved a great success, {{in our view}}, the following aspects can be further studied. First, the raw wave networks did not attract much attention. The mainstream of speaker recognition still adopts handcrafted features, which may lose useful information, e.g. the phase information, and finally may result in suboptimal performance as what we have observed in speech separation.
Second, the model size and inference efficiency, which is important for the devices with limited computation source, e.g. edge or mobile devices, have not been fully studied. The topic was just recently investigated in \cite{Safari2020, Georges2020Compact, nunes2020mobilenet1d}.

\section{Deep embedding: Temporal pooling layers}
\label{sec:embedding_pooling}
As shown in Fig. \ref{fig:x-vector}, the temporal pooling layer is a bridge between the frame-level and utterance-level hidden layers. Given a speech segment, we assume that the input and output of the temporal pooling layer are $\mathcal{H}=\{\mathbf{h}_t \in \mathbb{R}^{d_2}| t=1,2,\cdots, T\}$ and $\mathbf{u}$, respectively, where $\mathbf{h}_t$ denotes the $t$th frame-level speaker feature produced from the frame-level hidden layers. In this section, we introduce a number of temporal pooling functions.

\subsection{Average pooling}
\label{sec:average_pooling}
 Average pooling \cite{zhang2017end,yadav2018learning,li2017deep,li2018deep} is the   most common pooling function:
\begin{equation}\label{eq:avg_pooling}
  \mathbf{u}=\frac{1}{T}\sum_{t=1}^{T} \mathbf{h}_t
\end{equation}

\subsection{Statistics  pooling}
Statistics pooling \cite{snyder2017deep,snyder2018x} calculates both the statistic mean  $\mathbf{m}$ and  standard deviation $\mathbf{d}$ of  $\mathcal{H}$:
\begin{equation}\label{eq:sta_pooling_avg}
   \mathbf{m} = \frac{1}{T}\sum_{t=1}^{T} \mathbf{h}_t
\end{equation}
\begin{equation}\label{eq:sta_pooling_dev}
   \mathbf{d} = \sqrt{\frac{1}{T}\sum_{t=1}^{T}\mathbf{h}_t \odot \mathbf{h}_t - \mathbf{m} \odot \mathbf{m} }  \\
\end{equation}
where $\odot$  denotes the Hadamard product. The output of the statistics pooling layer is a concatenation of $\mathbf{m}$  and $\mathbf{d}$, i.e. $\mathbf{u} = [\mathbf{m}^T,  \mathbf{d}^T]^T$.

\subsection{Self-attention-based pooling}
Obviously, \eqref{eq:avg_pooling}, \eqref{eq:sta_pooling_avg}, and \eqref{eq:sta_pooling_dev} assume that all elements of $\mathcal{H}$ contribute equally to $\mathbf{u}$. However, the assumption may not be true, since that the frames may not provide equal speaker-discriminative information.  To address this issue, many works applied self attention mechanisms for weighted statistics pooling layers.
Specifically, the attention can be broadly interpreted as a vector of importance weights\footnote{https://lilianweng.github.io/lil-log/2018/06/24/attention-attention.html}, which allows a neural network to focus on a specific portion of its input. Further more, the self attention  computes attentive weights within a single sequence.

In the following two subsections, we first present a general self attention framework which produces weighted means and standard deviations of the input from a self-attentive scoring function in Section \ref{subsubsec:g_att}, and then list a number of specific self-attention-based pooling methods under the framework in Section \ref{subsubsec:att_pool}.

\subsubsection{A self attention pooling framework}\label{subsubsec:g_att}
Without loss of generality, self-attentive scoring is defined as:
\begin{equation}
  \big\{f_{\rm Att}^{(k)}(\cdot)|k=1,2,\cdots, K \big\}
\end{equation}
where  $f_{\rm Att}^{(k)}(\cdot)$ is usually referred as one-head, and $K$ is the total number of heads. If $K \ge 2$,
the self attention mechanism  is usually called multi-head self attention which allows the model to jointly attend to information from different representation subspaces \cite{vaswani2017attention}; otherwise, it degenerates into a single-head one.
Although $f_{\rm Att}^{(k)}(\cdot)$ has many different implementations, many of the implementations share similar forms with the structured self-attentive function \cite{lin2017structured} which obtains the importance weights by:
\begin{equation}\label{eq:multi_head_atten}
    f_{\rm Att}^{(k)}(\mathbf{h}_t) = \mathbf{v}^{(k)^T} {\rm tanh}(\mathbf{W}^{(k)}\mathbf{h}_t+\mathbf{g}^{(k)})+b^{(k)},\quad  k=1,2, \cdots, K
\end{equation}
where  $\mathbf{W}^{(k)} \in \mathbb{R}^{d_3 \times d_2}$,   $\mathbf{g}^{(k)} \in \mathbb{R}^{d_3}$,  $\mathbf{v}^{(k)} \in \mathbb{R}^{d_3}$ and  $b^{(k)} \in \mathbb{R}$ are learnable parameters of the $k$th scoring function. Suppose
   $ s_t^{(k)} = f_{\rm Att}^{(k)}(\mathbf{h}_t)$, $\quad  k=1,2, \cdots, K$, then the importance weights for the frame-level feature $\mathbf{h}_t$ are obtained by normalizing $s_t^{(k)}$ with a softmax function:
\begin{equation}\label{eq:multi_head_weights}
      \alpha_t^{(k)} = \frac{{\rm exp}\big(s_t^{(k)}\big)}{\sum_{t^\prime}^{T} {\rm exp}\big(s_{t^\prime}^{(k)} \big)}, \quad  k=1,2, \cdots, K
\end{equation}
where the normalization guarantees that the weights satisfy $0 \leq \alpha_t^{(k)} \leq 1$ and $\sum_{t=1}^{T}\alpha_t^{(k)} =1$.
Finally, the weighted mean and standard deviation produced from the $k$th self-attentive scoring function can be derived as follows:

\begin{equation}\label{eq:atten_pooling_avg}
 \widetilde{\mathbf{m}}^{(k)} = \sum_{t=1}^{T} \alpha_t^{(k)} \mathbf{h}_t, \quad  k=1,2, \cdots, K
\end{equation}

\begin{equation}\label{eq:atten_pooling_dev}
   \widetilde{\mathbf{d}}^{(k)}= \sqrt{\sum_{t=1}^{T} \alpha_t^{(k)} \mathbf{h}_t \odot \mathbf{h}_t -  \mathbf{\widetilde{m}}^{(k)} \odot  \mathbf{\widetilde{m}}^{(k)}},  \quad  k=1,2, \cdots, K
\end{equation}
Finally, $\widetilde{\mathbf{m}}^{(k)}$ and $\widetilde{\mathbf{d}}^{(k)}$ are used to calculate an utterance-level representation as  described in the following subsection.

\subsubsection{Attention pooling methods}\label{subsubsec:att_pool}
Under the above attention framework, this subsection categorizes existing self-attention based pooling layers into the following six classes, where all methods take \eqref{eq:multi_head_atten} as the self-attentive scoring function and take \eqref{eq:multi_head_weights} as the normalization function, unless otherwise stated.

\begin{itemize}
 \item   \textbf{Single-head attentive average pooling} \cite{bhattacharya2017deep, bhattacharya2018deeply, rahman2018attention}:  \cite{bhattacharya2017deep} takes a fully-connected layer as $f_{\rm Att}(\cdot)$. \cite{bhattacharya2018deeply} adopts the cosine function to compute attention scores:
     \begin{equation}\label{eq:saap}
       s_t = f_{\rm Att}(\mathbf{h}_t,\mathbf{r}) = \frac{\mathbf{h}_t^T\mathbf{r}}{\|\mathbf{h}_t\|_2\|\mathbf{r}\|_2}
     \end{equation}
     where $\mathbf{r}$ is a nonlinearly transformed i-vector from  the same utterance as $\mathbf{h}$. Obviously, the attention weights in \eqref{eq:saap} are determined by both the frame-level $\mathbf{h}_t$ and the utterance-level information $\mathbf{r}$. In \cite{rahman2018attention}, several attentive functions similar to \eqref{eq:multi_head_atten} are investigated.

     The output of the single-head attentive average pooling is set to the weighted mean:
  \begin{equation}
    \mathbf{u}= \mathbf{\widetilde{m}}^{(1)}.
  \end{equation}

  \item  \textbf{Single-head attentive statistics pooling} \cite{okabe2018attentive}: It uses a single-head attention function, i.e. $K=1$. Its output is a concatenation of both the weighted mean and weighted standard deviation:
    \begin{equation}\label{eq:Single-head attentive statistics pooling}
        \mathbf{u} = [\widetilde{\mathbf{m}}^{(1)^T}, \widetilde{\mathbf{d}}^{(1)^T}]^T.
    \end{equation}
  \item \textbf{Single-head Baum-Welch statistics attention mechanism based statistics pooling} \cite{9054151}:
  To overcome the weakness of \eqref{eq:multi_head_atten} which cannot fully mine the inner relationship between an utterance and its frames, \cite{9054151} integrated the Baum-Welch statistics into the attention mechanism:
    \begin{equation}
    \label{eq:multi_head_score}
        s_t = \mathbf{v}^T {\rm tanh}(\mathbf{K}\mathbf{q}_t+\mathbf{g})
    \end{equation}
    where $\mathbf{K}$ is named the key matrix and $\mathbf{q}_t$ is a query vector calculated by:
  \begin{equation}
    \mathbf{q}_t=f(\mathbf{h}_t^{(-1)})
  \end{equation}
where $f(\cdot)$ is a nonlinear function, and $\mathbf{h}_t^{(-1)}$ denotes the output of a penultimate frame-level hidden layer.
  The key matrix $ \mathbf{K}$ is calculated from the Baum-Welch statistics. Specifically, \cite{9054151} first calculates the normalized first order statistics $\mathbf{f}_c$ from the $c$th component of a GMM-UBM model $\Omega$ (see \eqref{eq:BW1}), and then conducts the following nonlinear transform:
  \begin{equation}
   \mathbf{f}_c^\prime=\mathbf{V}_2 {\rm tanh}(\mathbf{V}_1\mathbf{f}_c+\mathbf{g}),\quad \forall c = 1,\ldots,C
  \end{equation}
  where $\mathbf{V}_1$, $\mathbf{V}_2$ and $\mathbf{g}$ are the parameters of DNN.
  Finally, it concatenates $\mathbf{F}^\prime=[\mathbf{f}_1^{\prime},\mathbf{f}_2^{\prime}, \cdots, \mathbf{f}_C^{\prime}]$ and the trainable  matrix $\mathbf{W}$ as the key matrix:
\begin{equation}
   \mathbf{K}=[\mathbf{F}^\prime, \mathbf{W}]^T
\end{equation}
After obtaining $s_t$, $\mathbf{u}$ is obtained in the same way as \eqref{eq:Single-head attentive statistics pooling}.

\item \textbf{Global multi-head attentive average pooling}:
  It first applies a $K$-head ($K \geq 2$) attention function to $\mathcal{H}$ by \eqref{eq:multi_head_atten}. Then, the attentive weights and weighted means are calculated by \eqref{eq:multi_head_weights} and \eqref{eq:atten_pooling_avg} respectively \cite{9053217}.
  Finally, the output of the pooling layer $\mathbf{u}$ is the concatenation of the weighted means:
  \begin{equation}\label{eq:Global_multi_head}
    \mathbf{u} = [\widetilde{\mathbf{m}}^{(1)^T}, \widetilde{\mathbf{m}}^{(2)^T}, \cdots, \widetilde{\mathbf{m}}^{(K)^T}]^T
  \end{equation}
  It can be seen that $\mathbf{u} \in \mathbb{R}^{Kd_2}$. Similar ideas can also be found in \cite{zhu2018self,zhou2019cnn}.
 \cite{zhu2018self} also added an additional penalty term into the objective function to enlarge the diversity between the heads.

\item \textbf{Sub-vectors based multi-head attentive average pooling} \cite{safari2019self}:
It first splits $\mathbf{h}_t$ into $K$ ($K \geq 2$) non-overlapping homogeneous sub-vectors $\mathbf{h}_t= [\mathbf{h}_t^{(1)^T},\mathbf{h}_t^{(2)^T},\cdots,\mathbf{h}_t^{(K)^T} ]^T$, where $\mathbf{h}_t^{(k)} \in \mathbb{R}^{d_2/K}$.  Then, it applies single-head attention to each of the sub-vectors
$\mathcal{H}^{(k)}=\{\mathbf{h}_t^{(k)}\in \mathbb{R}^{d_2/K}| t=1,2, \cdots, T\}$. Finally, it obtains the sub-pooling outputs by:
\begin{equation}
  \mathbf{u}^{(k)}=\sum_{t=1}^{T} \alpha_t^{(k)} \mathbf{h}_t^{(k)}, \quad k=1,2, \cdots, K
\end{equation}
 It can be seen that $\mathbf{u}^{(k)} \in \mathbb{R}^{d_2/K}$.
The output of the pooling layer is a concatenation of the sub-pooling outputs:
\begin{equation}
  \mathbf{u} = [\mathbf{u}^{(1)^T},\mathbf{u}^{(2)^T},\cdots,\mathbf{u}^{(K)^T}]^T.
\end{equation}

  \item \textbf{Multi-resolution multi-head attentive average pooling} \cite{9053217}: Because the speaker characteristics are obtained through the aggregation of the attentive weights reweighted frame-level features,  \cite{9053217} proposed to control the resolution of the attentive weights with a temperature parameter. They modify the softmax function as:
      \begin{equation}
      \label{eq:softmax_temperature}
        \alpha_t=\frac{{\rm exp}(s_t/E)}{\sum_{t^\prime=1}^{T}{\rm exp}(s_{t^\prime}/E)}
      \end{equation}
      where $E$ is the temperature parameter. It is obvious that increasing $E$ makes the distribution of $ \alpha_t$ less sharp, i.e. lower resolution. By incorporating the above intuition, the weighting equation \eqref{eq:multi_head_weights} is changed to:
      \begin{equation}\label{eq:softmax_temperature_k}
           \alpha_{t}^{(k)} = \frac{{\rm exp}\big(s_t^{(k)}/E_k\big)}{\sum_{t^\prime}^{T} {\rm exp}\big(s_{t^\prime}^{(k)}/E_k\big)}
      \end{equation}
     where $E_k\geq 1$  is a temperature hyperparameter of the $k$th head. Finally, the output $\mathbf{u}$ is calculated in a similar way with that of the  global multi-head attentive average pooling except that $\alpha_t^{(k)}$ is replaced by \eqref{eq:softmax_temperature_k}.
\end{itemize}

It is clear that the above attentive pooling methods all employ scalar attention weights for each frame-level vector. \cite{Wu2020} further proposed a vector-based attentive pooling method, which adopts vectorial attention weights for each frame-level vector.

\subsection{NetVLAD $\&$ GhostVLAD pooling}
In \cite{xie2019utterance}, the authors applied a dictionary-based NetVLAD layer to aggregate features across time, which can  be intuitively regarded as trainable discriminative clustering: every frame-level descriptor will be softly assigned to different clusters, making the residuals encoded as the output feature \cite{xie2019utterance}.
\begin{figure}[t]
  \centering
  \includegraphics[width=3.3in]{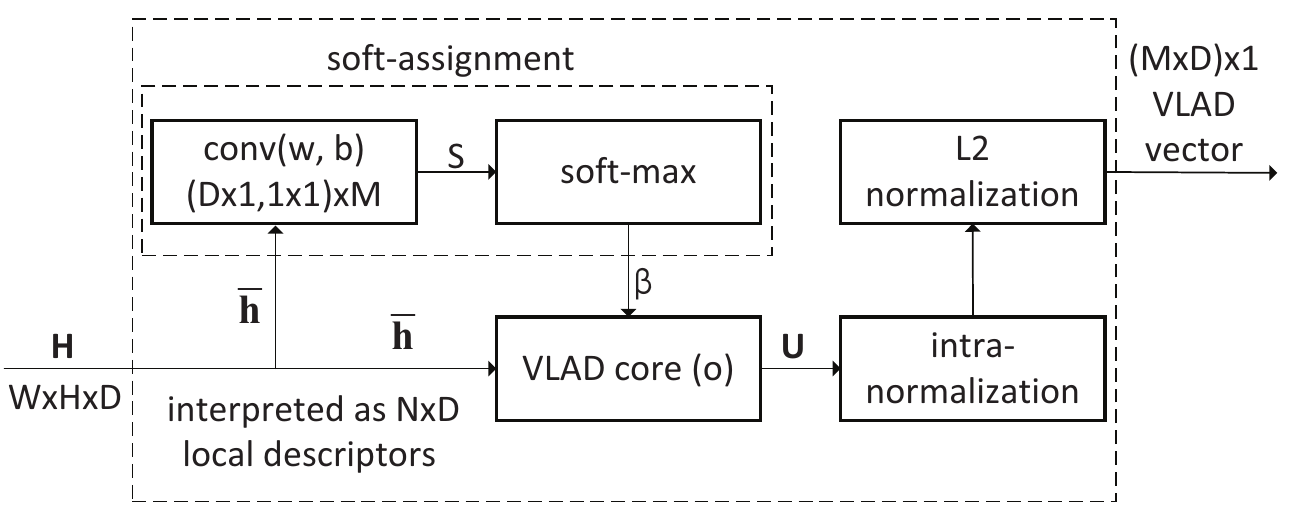}
  \caption{ Diagram of the NetVLAD pooling layer (from \cite{arandjelovic2016netvlad}).}
  \label{fig:NetVLAD_pooling}
\end{figure}

Specifically, as shown in Fig. \ref{fig:NetVLAD_pooling}, suppose that the input of the NetVLAD  layer is a three-dimensional tensor $\mathbf{H}^{(W \times H) \times D}$,  where  $W$, $H$ and $D$  depend on  the speech length, the dimensions of the spectrum  frequency bins, and the number of convolution kernels respectively. By  only retaining the third  dimension, $\mathbf{H}$ can be converted to  $N$ one-dimensional tensors, i.e. $\overline{\mathcal{H}}=\{\overline{\mathbf{h}}_{n} \in \mathbb{R}^D|n=1,2,\cdots,N \}$ where $N=W\times H$.
As shown in Fig. \ref{fig:NetVLAD_pooling}, the NetVLAD pooling layer consists of the following four steps \cite{arandjelovic2016netvlad}:
\begin{enumerate}[1)]
  \item Calculate a matrix $\mathbf{U} \in \mathbb{R}^{D \times M}$ from $\overline{\mathcal{H}}$ by:
        \begin{equation}\label{eq:NetVLAD}
            \mathbf{U}(:,m)=\sum_{n=1}^{N} \beta_{m}(\mathbf{\overline{h}}_n)(\mathbf{\overline{h}}_n-\mathbf{o}_m)
        \end{equation}
        where $M$ is the number of the chosen clusters $\mathcal{O}=\{\mathbf{o}_m \in \mathbb{R}^D|m=1,2,\cdots,M\}$, and $\beta_{m}(\mathbf{\overline{h}}_n)$ is an assignment weight calculated by:
        \begin{equation}\label{eq:NetVLAD1}
            \beta_{m}(\mathbf{\overline{h}}_n)=\frac{{\rm
            exp}\big(\mathbf{w}_m^T \mathbf{\overline{h}}_n+b_m \big)}{\sum_{m^\prime=1}^{M}{\rm exp}\big(\mathbf{w}_{m^\prime}^T \mathbf{\overline{h}}_n+b_{m^\prime}\big)}
        \end{equation}
        with $\{\mathbf{w}_m \}$, $\{b_m\}$ and $\{\mathbf{o}_m\}$ as the parameters of the network.

  \item Normalize $\mathbf{U}$ by $\ell_2$-norm column-wisely. This step is termed as the intra-normalization.
  \item Convert the normalized $\mathbf{U}$ into a vector:
\begin{equation}
  \mathbf{u}=[\mathbf{U}(:,1)^T,\mathbf{U}(:,2)^T,\cdots,\mathbf{U}(:,M)^T]^T
\end{equation}

  \item Normalize $\mathbf{u}$ by $\ell_2$-norm to generate an $M \times D$ dimensional output vector. This step is termed as the $\ell_2$-normalization.
\end{enumerate}

In addition, \cite{xie2019utterance} also applied a variant of NetVLAD, named GhostVLAD. The main difference between them is that some of the clusters in the GhostVLAD layer, named \textit{``ghost clusters''}, are not included in the final concatenation, and hence do not contribute to the final representation. When aggregating the frame-level features, the contribution of the noisy and undesirable sections of a speech segment to the normal VLAD clusters will be effectively down-weighted, since that larger weights are assigned to the ``ghost cluster''. See \cite{zhong2018ghostvlad} for the details.

\subsection{Learnable dictionary encoding pooling}
Motivated by GMM-UBM, {{\cite{cai2018exploring}}} proposed a learnable dictionary encoding (LDE) pooling layer which models the distribution of the frame-level features $\mathcal{H}$ by a dictionary. The dictionary learns a set of dictionary component centers $\mathcal{\overline{O}}=\big \{\mathbf{\overline{o}}_m \in \mathbb{R}^{d_2}|m =  1,2,\cdots, M \big\} $, and assigns weights to the frame-level features by:
\begin{equation}
  \overline{\beta}_{tm}=\frac{{\rm exp}(-\tau_m||\mathbf{h}_t-\mathbf{\overline{o}}_m||^2)}{\sum_{m^\prime=1}^{M}{\rm exp}(-\tau_{m^\prime}||\mathbf{h}_t-\mathbf{\overline{o}}_{m^\prime}||^2)}
\end{equation}
where the smoothing factor $\tau_m$ for each dictionary center $\mathbf{\overline{o}}_m$ is learnable. The aggregated output of the pooling layer with respect to the center $\mathbf{\overline{o}}_m$ is:
\begin{equation}
\label{eq:LDEP}
   \mathbf{u}_m=\frac{\sum_{t=1}^{T}\overline{\beta}_{tm}(\mathbf{h}_t-\overline{\mathbf{o}}_m)}{\sum_{t=1}^{T}\overline{\beta}_{tm}}
\end{equation}
In order to facilitate the derivation, \eqref{eq:LDEP} is simplified to:
\begin{equation}
   \mathbf{u}_m=\frac{\sum_{t=1}^{T}\overline{\beta}_{tm}(\mathbf{h}_t-\overline{\mathbf{o}}_m)}{T}
\end{equation}
Finally, the output of the pooling layer is $\mathbf{u}=[\mathbf{u}_1^T,\mathbf{u}_2^T,\cdots,\mathbf{u}_M^T]^T$.

\subsection{Spatial pyramid pooling}
\label{sec:spatial_pyramid_pooling}
In order to  handle variable-length utterances,  \cite{zhang2018text} incorporated a Spatial Pyramid Pooling operation into a CNN-based network, which can directly produce fixed-length feature vectors from variable-length utterances.

\begin{figure}[t]
  \centering
  \includegraphics[width=2.5in]{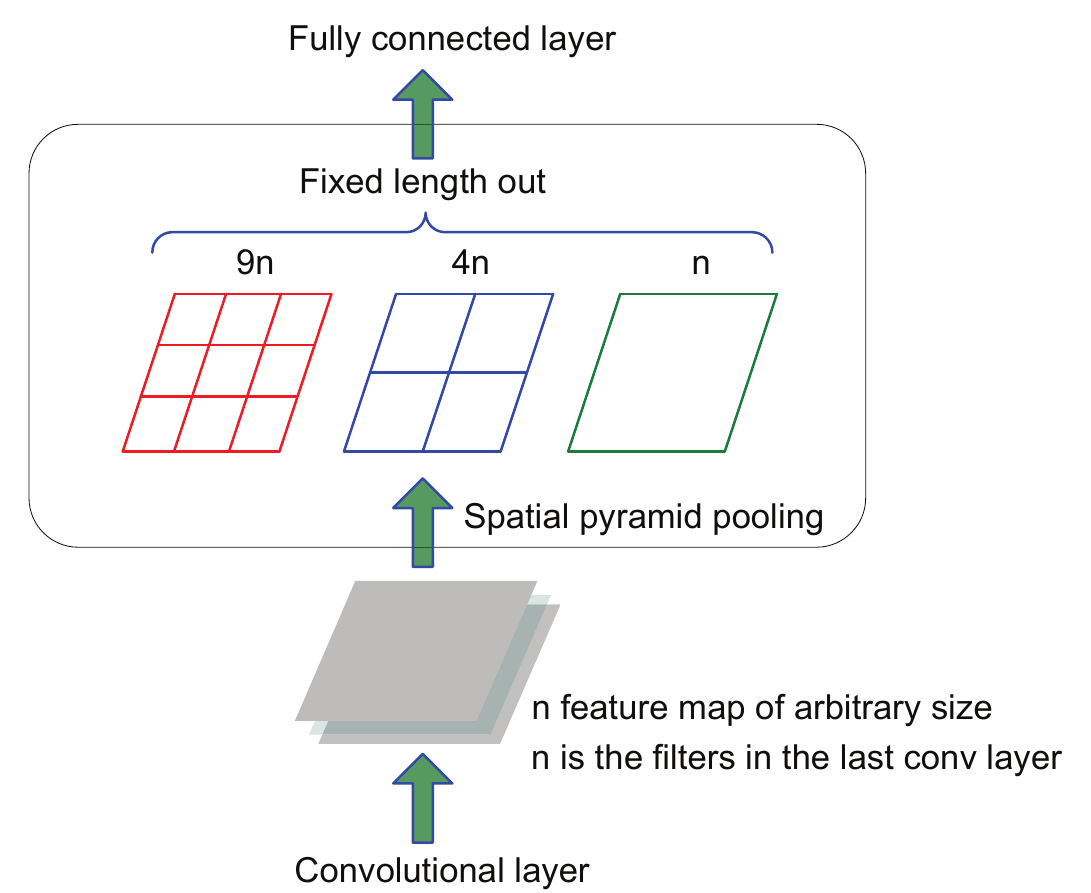}
  \caption{ Spatial pyramid pooling (from \cite{zhang2018text}).}
  \label{fig:pyramid_pooling}
\end{figure}

As shown in Fig. \ref{fig:pyramid_pooling}, the spatial pyramid pooling layer outputs a fixed length vector by first dividing the input feature maps into $1 \times 1$, $ 2 \times 2$, and $3 \times 3$ small patches and then performing average pooling  over these  patches.
An exceptional advantage of the spatial pyramid pooling layer is that it maintains spatial information of the last frame-level feature maps by making average pooling in each local small patches.

\cite{jung2019spatial} further extracted embeddings from the divided small patches via a parameter-sharing LDE layer instead of applying the averaging pooling on them.

\subsection{Other temporal pooling functions}
 There are many other successful pooling methods. For example, \cite{gao2018improved} proposed a  cross-convolutional-layer pooling method to capture the first-order statistics for modelling long-term speaker characteristics. \cite{travadi2019total} reported  a total variability model based pooling layer. \cite{heigold2016end} connected the last output of LSTM to the loss function for an utterance-level speaker representation.
 Apart from the single-scale aggregation methods in Sections \ref{sec:average_pooling} to \ref{sec:spatial_pyramid_pooling} which generate the pooling output from the last frame-level layer, multiscale aggregation methods have also been proposed \cite{tang2019deep, gao2019improving, seo2019shortcut, hajavi2019deep, Jung2020Improving} which utilize multiscale features from different frame-level layers to generate the pooling output.

\subsection{Discussion to the temporal pooling layers}
Because temporal pooling layers behave fundamentally different in different datasets, network structures, or loss functions, it is difficult to conclude which one is the best. To our knowledge, temporal pooling functions with learnable parameters achieved better (at least competitive) results than the simple pooling layers such as the average pooling and statistical pooling in most cases, with a weakness of higher computational complexity than the latter. Some examples are listed in Table \ref{tab:experiments_pooling}.
\renewcommand\arraystretch{1.5}
\begin{table}[t]
  \centering
  \footnotesize
  \caption{Experimental results of different temporal pooling functions selected from literature. Each row represents a comparison. The results across rows are not comparable.}
  \label{tab:experiments_pooling}
  \scalebox{0.8}{
  \begin{tabular}{m{2cm} m{1.1cm} m{2.5cm}<{\centering} m{0.8cm}<{\centering} m{0.8cm}<{\centering} m{1.2cm}<{\centering}}
     \hline
    \hline
    \multicolumn{2}{c}{Comparison methods}   & \multirow{2}*{Test dataset [condition]} &\multicolumn{3}{c}{EER}\\
    \cline{1-2}    \cline{4-6}
    Main models         & Baselines           &        & Main   & Baseline & Relative reduction  \\
    \hline
     Attention\cite{zhu2018self}      & Average    &SRE16 Cantonese    &   5.81\%    &  7.33\%     & \textbf{21\%}\\
     NetVLAD \cite{xie2019utterance}  & Average    &VoxCeleb1 test set &   3.57\%    &  10.48\%    & \textbf{66\%}\\
     GhostVLAD \cite{xie2019utterance}& Average    & VoxCeleb1 test set&   3.22\%    &  10.48\%    & \textbf{69\%}\\
     LDE \cite{villalba2020state}     & Statistics &  SITW             &   2.50\%    &  3.01\%     & \textbf{17\%}\\
    \hline
    \hline
  \end{tabular}}
\end{table}

\section{Deep embedding: Classification-based objective functions}
\label{sec:embedding_objectives}
The objective function largely determines the performance of a neural network. Deep-embedding-based speaker recognition systems usually adopt classification-based objective functions. Before reviewing the objective functions, we first summarize the deep-embedding-based speaker recognition as the following multi-class classification problem.

Let $\mathcal{X}=\{(\mathbf{x}_n, l_n)|n=1,2, \cdots, N\}$ denote the training samples in a mini-batch\footnote{DNN is often trained using a  mini-batch data in an iteration}, where $\mathbf{x}_n \in \mathbb{R}^{d_4}$ represents the input of the last fully connected layer, $ l_n \in \{1,2,\cdots,J\}$ is the class label of  $\mathbf{x}_n$ with $J$ as the number of speakers in the training set, and $N$ is the batch size.
In addition, $\mathbf{W}=[\mathbf{w}_1,\mathbf{w}_2,\cdots,\mathbf{w}_J]$ and $\mathbf{b}=[b_1,b_2,\cdots,b_J]$ denote the weight matrix and bias vector of the last fully connected layer respectively.

In this section, we comprehensively summarize the classification based objective functions. Without loss of generality, the ``loss function'',``cost function'' and ``objective function'' are equivalent in this article.

\subsection{The variants of softmax loss }\label{sec:softmax_variants}
As shown in Section \ref{sec:embedding}, both the d-vector and x-vector extractors take the minimum cross entropy as the objective function, and take softmax as the output layer. For short, we denote the objective function as the \textit{Softmax loss}\footnote{Following \cite{liu2017sphereface}, we define the softmax loss as a combination of the last fully connected layer, softmax function, and cross-entropy loss function.}.
{{For a multiclass classification problem, the cross-entropy error function over $\mathcal{X}$ can be calculated as:
\begin{equation}
\label{eq:CE}
  L = -\frac{1}{N} \sum_{n=1}^{N} {\sum_{j=1}^{J} t_{nj} {\rm log} p_{nj} }
\end{equation}
where  $[t_{n1}, t_{n2}, \cdots, t_{nJ}]$ is a one-hot vector encoded from the label $l_n$, in other words, $t_{nj}$ equals to 1 if and only if sample $\mathbf{x}_n$ belongs to class $j$. $p_{nj}$ is the posterior probability of $\mathbf{x}_n$ belonging to class $j$. It is produced from neural networks with the following Softmax function:
\begin{equation}
\label{eq:softmax_function}
  p_{nj}  =   \frac{{\rm exp}{(\mathbf{w}_{j}^T \mathbf{x}_n+b_{j}})}{\sum_{j=1}^{J}{\rm exp}{(\mathbf{w}_{j}^T \mathbf{x}_n+b_{j}})}
\end{equation}
Combining \eqref{eq:CE} and \eqref{eq:softmax_function} derives an equivalent form of the Softmax loss:
\begin{equation}
\label{eq:softmax}
   \mathcal{L}_{\rm S}= -\frac{1}{N}\sum_{n=1}^{N} {\rm log}  \frac{{\rm exp}{(\mathbf{w}_{l_n}^T \mathbf{x}_n+b_{l_n}})}{\sum_{j=1}^{J}{\rm exp}{(\mathbf{w}_{j}^T \mathbf{x}_n+b_{j}})}
\end{equation}
}}

Softmax loss is the most common objective function for deep embedding. However, from \eqref{eq:softmax}, one can see that  Softmax loss is only good at maximizing the between-class distance, but does not have an explicit constraint on minimizing the within-class variance. Therefore, the performance of deep embedding has much room of improvement. Here we present some representative variants of Softmax loss as follows.

\begin{itemize}
\item \textbf{Angular softmax (ASoftmax) loss} \cite{huang2018angular,novoselov2018deep,cai2018exploring}: Because the inner product between  $ \mathbf{w}_j$ and $\mathbf{x}_n$ in \eqref{eq:softmax} can be rewritten as:
\begin{equation}
  \mathbf{w}_j^T\mathbf{x}_n=\|\mathbf{w}_j\|\,\|\mathbf{x}_n\|{\rm cos}(\theta_{j,n})
\end{equation}
where $\theta_{j,n} (0 \leq \theta_{j,n} \leq \pi)$  denotes the angle between $ \mathbf{w}_j$ and $\mathbf{x}_n$, Softmax loss can be rewritten as:
\begin{equation}
\label{eq:softmax_variant}
    \mathcal{L}_{\rm S}= -\frac{1}{N}\sum_{n=1}^{N} {\rm log}  \frac{{\rm exp}{\big(\|\mathbf{w}_{l_n}\|\,\| \mathbf{x}_n\| {\rm cos}(\theta_{l_n,n})+b_{l_n}}\big)}{\sum_{j=1}^{J}{\rm exp}{\big(\|\mathbf{w}_{j}\|\, \|\mathbf{x}_n\|{\rm cos}(\theta_{j,n})+b_{j}}\big)}
\end{equation}
If we further set the bias terms to zero, normalize the weights at the forward propagation stage, and add a margin to the angle:
\begin{equation}
\label{eq:ASoftmax_margin}
  b_j=0, \quad \|\mathbf{w}_j\|=1, \quad \psi(\theta_{l_n,n})= (-1)^{a}{\rm cos}(m_1 \theta_{l_n,n})-2a
\end{equation}
then, we explicitly constrain the learned features to have a small intra-speaker variation, where $m_1 \geq 1$ is an integer margin hyperparameter, $ \theta_{l_n,n} \in [\frac{a \pi}{m_1}, \frac{(a+1)\pi}{m_1} ]$, and $a \in \{0,1,\cdots,m_1-1\}$. The intuition behind the angle function $\psi(\theta_{l_n,n})$ is illustrated in Fig. \ref{fig:ASoftmax_margin}.
Then, we obtain ASoftmax loss as follows:
\begin{equation}
  \label{eq:A-Softmax}
\begin{split}
  &\mathcal{L}_{\rm AS}= \\
  &-\frac{1}{N}\sum_{n=1}^{N} {\rm log}  \frac{{\rm exp}{\big(\| \mathbf{x}_n\| \psi(\theta_{l_n,n})})}{{\rm exp}{\big(\| \mathbf{x}_n\| \psi(\theta_{l_n,n})})+\sum_{j=1,j\neq l_n}^{J}{\rm exp}{\big(\|\mathbf{x}_n\|{\rm cos}(\theta_{j,n})}\big)}
\end{split}
\end{equation}
Note that, because $m_1$ is limited to a positive integer instead of a real number, the margin is not flexible enough.

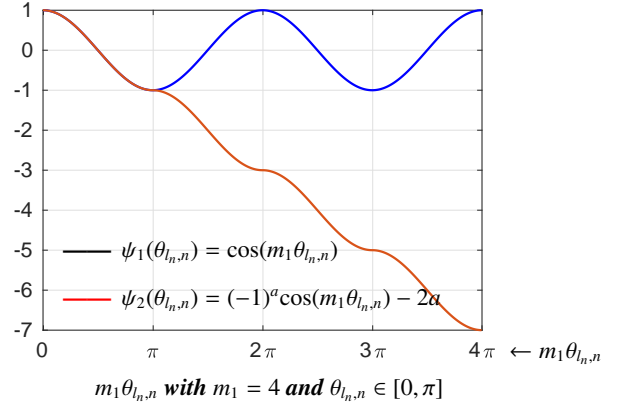
\begin{figure}[t]
  \centering
  \includegraphics[width=2.5in]{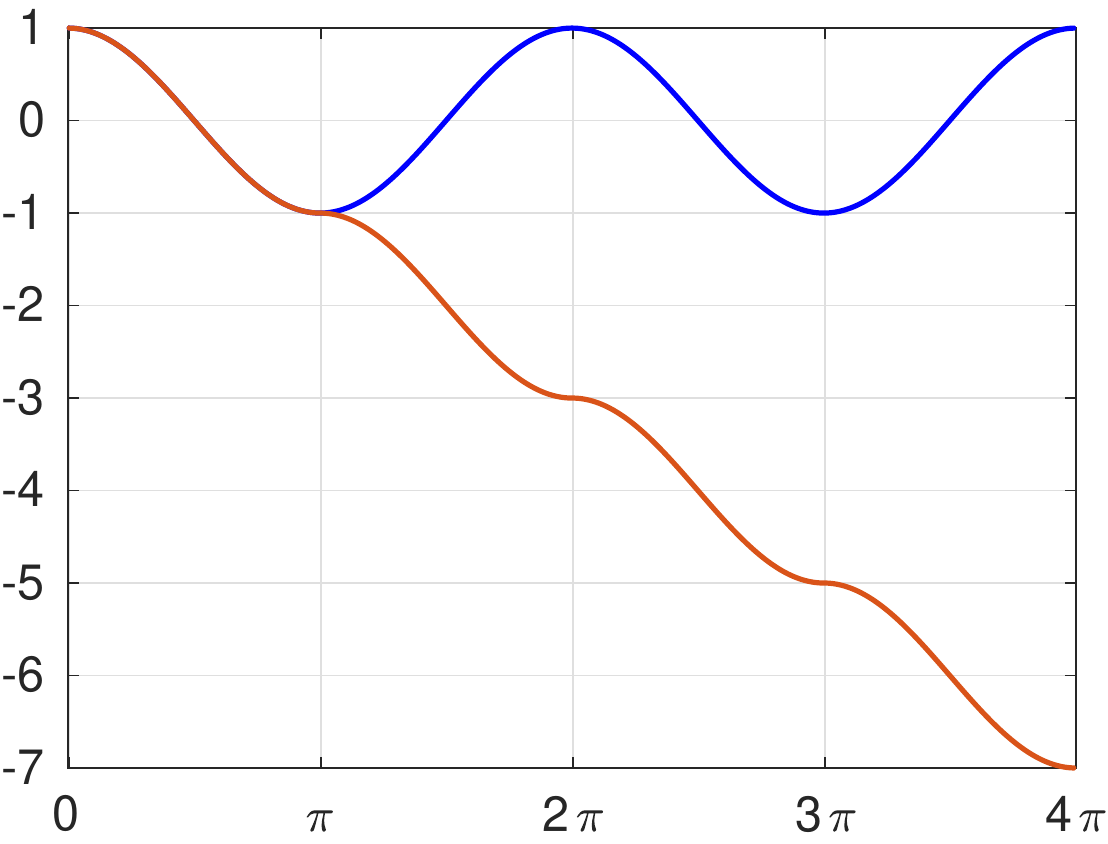}
   \put(-162,40){\small\bfseries{------ }}
   \put(-140,40){\small\bfseries\color{black}{$\psi_1(\theta_{l_n,n})= {\rm cos}(m_1 \theta_{l_n,n})$}}
   \put(-162,22){\small\bfseries\color{red}{------}}
   \put(-140,22){\small\bfseries\color{black}{$\psi_2(\theta_{l_n,n})= (-1)^{a}{\rm cos}(m_1 \theta_{l_n,n})-2a$}}
   \put(-150,-12){\small\bfseries\color{black}{$m_1 \theta_{l_n,n}$ \textit{with} $m_1=4$  \textit{and} $\theta_{l_n,n} \in [0, \pi]$}}
   \put(5,2){\small\bfseries\color{black}{$ \leftarrow m_1 \theta_{l_n,n}$ }}

   \caption{Illustration of the angle function of the ASoftmax loss. Obviously,  the angle $\theta_{l_n,n}$ in the training stage is in $[0,\pi]$. If we simply multiply an integer  margin $m_1$ to $\theta_{l_n,n}$, then the angle function $\psi_1(\cdot)$ is monotonic when $\theta_{l_n,n} \in [0,\frac{\pi}{m_1}]$ only. Therefore, in practice, $\psi_1(\cdot)$ is generalized to $\psi_2(\cdot)$  to ensure that the  angle function is monotonically decreasing when $\theta \in [0,\pi]$.}
   \label{fig:ASoftmax_margin}
\end{figure}

\item
\textbf{Additive margin softmax (AMSoftmax) loss} \cite{xie2019utterance,hajibabaei2018unified,yu2019ensemble}: It is a revision of ASoftmax loss by replacing $ \psi(\theta_{l_n,n})$  in \eqref{eq:A-Softmax} with $ ({\rm cos}(\theta_{l_n,n})-m_2)$, and normalizing $\|\mathbf{x}_n\|=1$:
\begin{small}
\begin{equation}\label{eq:AMSoftmax}
\begin{split}
  &\mathcal{L}_{\rm AMS} =   \\
  &-\frac{1}{N}\sum_{n=1}^{N} {\rm log}  \frac{{\rm exp}{\big(\tau ({\rm cos}(\theta_{l_n,n})-m_2)}\big)}{{\rm exp}{\big(\tau ({\rm cos}( \theta_{l_n,n})-m_2)\big)}+\sum_{j=1,j\neq l_n}^{J}{\rm exp}{\big(\tau ({\rm cos}(\theta_{j,n})})\big)}
\end{split}
\end{equation}
\end{small}
where $\tau$ is a scaling factor for preventing gradients too small during the training process \cite{xiang2019margin}. In addition, \cite{Zhou2020} also proposed a dynamic-additive margin softmax, where $m_2$ is replaced by a dynamic margin for each training sample.

\item
\textbf{Additive angular margin softmax (AAMSoftmax)} \cite{xiang2019margin,liu2019large}: It replaces $({\rm cos}(\theta_{l_n,n})-m_2)$  in \eqref{eq:AMSoftmax} by ${\rm cos}(\theta_{l_n,n}+m_3)$:
 \begin{small}
\begin{equation}\label{eq:AAMSoftmax}
\begin{split}
  &\mathcal{L}_{\rm AAMS} = \\
  &-\frac{1}{N}\sum_{n=1}^{N}{\rm log}  \frac{{\rm exp}{\big(\tau ({\rm cos}(\theta_{l_n,n}+m_3))}\big)}{{\rm exp}{\big(\tau ({\rm cos}( \theta_{l_n,n}+m_3))\big)}+\sum_{j=1,j\neq l_n}^{J}{\rm exp}{\big(\tau ({\rm cos}(\theta_{j,n})})\big)}
\end{split}
\end{equation}
\end{small}
\end{itemize}
To further improve the convergence speed and accuracy, \cite{Rybicka2020} recently proposed a parameter adaptation method which  adapts the scaling factor $\tau$ and margin $m_3$ at each iteration.

{
Compared to Softmax, both ASoftmax, AMSoftmax, and AAMSoftmax benefit from the following two aspects: first,
the learned features are angularly distributed, which matches with the cosine similarity scoring back-end; second, they introduce an angle, i.e. a cosine margin, to quantitatively control the decision boundary between training speakers for minimizing the within-class variance.
More information can be found in \cite{liu2017sphereface,wang2018additive,wang2018cosface,deng2019arcface}.

\subsection{Regularization for Softmax loss and its variants}
\label{sec:class_regular}
As illustrated in Section \ref{sec:softmax_variants}, the learned feature by the Softmax loss is not discriminative enough. To address this issue, an alterative way is to combine the Softmax loss with some regularizers \cite{liu2017sphereface}:
\begin{equation}
  \mathcal{L} = \mathcal{L}_{\rm S} + \lambda \mathcal{L}_{\rm Regular}
\end{equation}
where $\lambda$ is a hyperparameter for balancing the Softmax loss $\mathcal{L}_{\rm S}$ and the regularizer $\mathcal{L}_{\rm Regular}$. Besides, the regularizer is also applicable to other Softmax variants.

Because the embedding layer that produces the embedding speaker features is not always the last hidden layer, e.g. the x-vector in Fig.\ref{fig:x-vector}, the regularizer was sometimes added to the embedding layer. For clarity, we define the output of the embedding layer as $\mathcal{E}=\{(\mathbf{e}_n,l_n)|n=1,2,\cdots,N\}$ , where $\mathbf{e}_n \in \mathbb{R}^{d_5}$. Here we introduce some regularizers as follows:

\begin{itemize}
\item
\textbf{Center loss} \cite{cai2018exploring,li2018deep,wang2019discriminative}:
It is a typical regularizer for Softmax loss. It explicitly minimizes the within-class variance by:
\begin{equation}\label{eq:center_loss}
  \mathcal{L}_{\rm C}=\frac{1}{2}\sum_{n=1}^{N}\|\mathbf{e}_n-\mathbf{c}_{l_{n}}\|^2
\end{equation}
where $\mathbf{c}_{l_n} \in  \mathbb{R}^{d_5}$ denotes the $l_n$th class center of the elements in $\mathcal{E}$. At each training iteration, the centers are updated as follows \cite{li2018deep}:
\begin{equation}\label{eq:center1}
  \mathbf{c}_j^{t+1}=\mathbf{c}_j^t-\epsilon \Delta\mathbf{c}_j^t
\end{equation}
\begin{equation}\label{eq:center2}
  \Delta\mathbf{c}_j=\frac{\sum_{n=1}^{N}\delta(l_n=j)\cdot(\mathbf{c}_j-\mathbf{e}_n)}{1+\sum_{n=1}^{N}\delta(l_n=j)}
\end{equation}
where  $\epsilon \in [0,1]$ controls the learning rate of the centers, the superscript ``$t$'' represents the number of iterations, and $\delta(\cdot)$ is an indicator function. If the condition of the indicator function, i.e. $l_n=j$, is satisfied, then $\delta=1$; otherwise, $\delta=0$.

The center loss is usually combined with Softmax loss:
\begin{equation}
  \mathcal{L} = \mathcal{L}_{\rm S}+\lambda\mathcal{L}_{\rm C}
\end{equation}
See \cite{wen2016discriminative} for more  information about the  center loss.

\item
\textbf{Ring loss}  \cite{liu2019large}: It restricts $\|\mathbf{e}_n\|$ to be close to a target value $R$ for AMSoftmax loss:
\begin{equation}\label{eq:Ring loss}
  \mathcal{L}=\mathcal{L}_{\rm AMS}+\lambda \times \frac{1}{N}\sum_{n=1}^{N}(\|\mathbf{e}_n\|-R)^2
\end{equation}
where  the target norm $R$ is optimized during the network training. Equation \eqref{eq:Ring loss} essentially applies a normalization constraint to the features.

\item
\textbf{Minimum hyperspherical energy criterion} \cite{liu2019large}:
It enforces the weights of the output layer to distribute evenly on a hypersphere:
\begin{equation}\label{eq:MHE}
  \mathcal{L}=\mathcal{L}_{\rm AMS}+\frac{\lambda}{N(J-1)}\sum_{n=1}^{N}\sum_{j=1,j \neq l_n}^{J} h(\|\hat{\mathbf{w}}_{l_n}-\hat{\mathbf{w}}_j\|)
\end{equation}
where  $ \hat{\mathbf{w}}_{l_n}$ and   $\hat{\mathbf{w}}_j$ are the
$\ell_2$-normalized $\mathbf{w}_{l_n}$ and $\mathbf{w}_j$ respectively, and $h(z)=\frac{1}{z^2}$ is a decreasing function. Intuitively, the minimum hyperspherical energy based regularizer enlarges the inter-class separability.

\item
\textbf{Gaussian prior} \cite{li2019gaussian}:
To reduce information leak, \cite{li2019gaussian} introduced a Gaussian prior to the output of the embedding layer, which results in the following objective function:
\begin{equation}
\label{eq:reglar_Gaussian}
  \mathcal{L}=\mathcal{L}_{\rm S}+\lambda \sum_{j}\sum_{\mathbf{e}_n \in\varepsilon(j)}\|\mathbf{e}_n-\mathbf{w}_j\|
\end{equation}
where $\varepsilon(j)$  is the set of the utterances belonging to the $j$th speaker, $\mathbf{e}_n$ represents  the x-vector,  $\mathbf{w}_j$ represents the parameters of the last layer corresponding to the output unit of speaker $j$.

\item
\textbf{Triplet loss} \cite{jati2019multi}: Because Softmax loss does not explicitly reduce intra-class variance, triplet loss was introduced to directly bring samples from the same class closer than the samples from different classes. Formally, the  triplet loss weighted Softmax loss is written as:
\begin{equation}
\label{eq:cross_trip}
  \mathcal{L}=\lambda \mathcal{L}_{\rm  S}+(1-\lambda)\mathcal{L}_{\rm Triplet}
\end{equation}
where $\mathcal{L}_{\rm Triplet}$ denotes the triplet loss, which will be introduced in Section \ref{sec:end2end_loss}.

\end{itemize}
There are also many other regularization approaches. For example, \cite{yu2019ensemble} added a Hilbert-Schmidt independence criterion based constraint to the embedding layer  for regularizing AMSoftmax loss $\mathcal{L}_{\rm AMS}$. See \cite{yu2019ensemble} for the details.

\subsection{Multi-task learning  for deep embedding}
\label{sec:multi_task}
Phonetic information is important in improving the performance of speaker recognition. As illustrated in Section \ref{sec:DNN/i-vector}, one way to  incorporate phonetic information into the i-vector-based systems is to employ an ASR acoustic model, e.g. the DNN-UBM/i-vector or the DNN-BNF/i-vector. As for the deep-embedding-based speaker recognition, the phonetic information was usually incorporated by multi-task learning.
For example, \cite{chen2015multi,liu2015deep} trained
a deep embedding network to discriminate the speaker identity and text phrases simultaneously. The training objective is to minimize:
\begin{equation}\label{eq:multi_task}
   {\rm CE}([\bm{l}_1,\bm{l}_2],[\bm{l}_1^\prime,\bm{l}_2^\prime]) = {\rm CE}_1(\bm{l}_1,\bm{l}_1^\prime) +
   {\rm CE}_2(\bm{l}_2,\bm{l}_2^\prime)
\end{equation}
where ${\rm CE}_1$ and ${\rm CE}_2$ are two cross-entropy criteria for speaker and text phrase respectively. $\bm{l}_1$ and $\bm{l}_2$ indicate the true labels for speakers and text individually, and  $\bm{l}_1^\prime$ and $\bm{l}_2^\prime$ are the two outputs of the network respectively. Some similar ideas can also be found in \cite{dey2018dnn}.

Although the text content may be a  harmful source to text-independent speaker recognition, some positive results were observed with the multi-task learning. {{The authors in \cite{liu2018speaker}}} added phonetic information to the frame-level layers of the x-vector extractor with an auxiliary ASR acoustic model by multi-task learning. {{The authors in \cite{wang2019usage}}} conjectured that the phonetic information is helpful for frame-level feature learning, however, it is useless in utterance-level speaker embeddings. They experimentally verified their assumptions by  multitask learning and adversarial training, where the phonetic information was used as positive and negative effects respectively.

Besides the phonetic information mining, some multi-task learning approaches intend to improve the performance of the auxiliary and main tasks together. For example, \cite{tang2016collaborative} proposed a collaborative learning approach based on multi-task recurrent neural model to improve the performance of both speech and speaker recognition. \cite{yao2018snr} proposed a multitask DNN structure to denoise i-vectors and classify speakers simultaneously. Considering that the acoustic and speaker domains are complementary, \cite{Jung2020MultTask} recently proposed a multi-task network that performs keyword spotting  and speaker verification simultaneously to fully utilize the interrelated domain information.

\subsection{Discussion to the  classification-based loss functions}
Because speaker verification is an open set recognition task, the deep embedding space produced from a training dataset with a limited number of speakers is required to generalize well to unseen test speakers. Therefore, it is the speaker discriminative ability of the embedding rather than the classification accuracy that is important, which accounts for the motivation why many classification-based loss functions are designed to minimize the within-class variance by adding a margin or a regularizer into the Softmax loss.

From the experimental results in literature, one can concluded that the design of loss functions is very important to performance. At present, nearly all stat-of-the-art deep embedding systems replaced the traditional Softmax by its variants, especially AMSoftmax and AAMSoftmax. In addition, the Softmax loss, its variants and regularizers are not mutually exclusive. For instance, the regularization terms \eqref{eq:Ring loss} and \eqref{eq:MHE} were originally added to AMSoftmax \cite{liu2019large}.

\section{End-to-end speaker verification:  Verification-based objective functions}
\label{sec:end2end_loss}
An emerging direction of speaker recognition is end-to-end speaker verification. It is able to produce the similarity score of a pair of utterances in a test trial directly. The main difference between deep embedding and end-to-end speaker verification is the objective function. Therefore, in this section, we mainly review the verification-based loss functions, and skip the other components that are similar to deep embedding, e.g. the network structures or temporal pooling layers.

Here we emphasize that the borderline between the classification-based deep embedding and verification-based end-to-end speaker verification is unclear in literature. Some work also called the end-to-end speaker verification systems as deep embedding extractors. The main reason for this confusion is that, although the end-to-end speaker verification systems have different objective functions and training strategies from the deep embedding extractors, they need to extract utterance-level speaker embeddings from the hidden layers as the input of some independent back-ends, e.g. PLDA, in the test stage, so as to achieve the state-of-the-art performance. Despite the confusion usage of the terms in literature, here we clearly regard the speaker verification systems whose loss functions yield similarity scores from training trials as end-to-end speaker verification.

In this section, we focus on summarizing verification-based objective functions, each of which needs to address the following three core issues:
\begin{itemize}
  \item How to design a \textbf{training loss} that pushes DNN towards our desired direction: As shown in Fig. \ref{fig:veri-iden-diar}, speaker verification can be viewed as a binary classification problem of whether a pair of utterances are from the same speaker. A natural solution to this problem is to train a binary classifier in an end-to-end fashion from a large number of manually constructed pairs of training utterances, i.e. training trials. The training loss of the binary classifier largely determines the effectiveness of the classifier.
  \item How to define a \textbf{similarity metric} between a pair of utterances: The similarity of a pair of utterances is calculated from the embeddings of the utterances at the output layer where a proper similarity metric for evaluating the similarity between the embeddings boosts the performance.
  \item How to \textbf{select and construct training trials} from an exponentially large number of training trials: Because the number of all possible training trials is at least the square of the number of training utterances, and also because many of the training trials are less informative, we need to select or even construct some informative training trials instead of using all training trials.
\end{itemize}

\subsection{Pairwise loss}
\label{sec:pairwise_loss}

Pairwise loss is a kind of training loss of the end-to-end speaker verification where each training trial contributes to the accumulation of the training objective value independently.
Suppose there is a set of pairwise training trials as $\mathcal{X}_{\rm pair}=\{(\mathbf{x}_n^e,\mathbf{x}_n^t; l_n)|n=1,2,\cdots,N\}$ where $\mathbf{x}_n^e$ and $\mathbf{x}_n^t$ denote a pair of speaker embedding features at the output layer, and $l_n \in \{0, 1 \}$ is the ground-truth  label. If $\mathbf{x}_n^e$ and $\mathbf{x}_n^t$ belong to the same speaker, then $l_n=1$; otherwise, $l_n=0$.

\textit{Binary cross-entropy loss} is the most common pairwise loss \cite{heigold2016end,rahman2018attention,zhang2016end,snyder2016deep,zhang2019seq2seq,rohdin2018end}:
\begin{equation}
\label{eq:binary_cross_entropy}
  \mathcal{L}_{\rm BCE}=-  \sum_{n=1}^{N} {\bigg [}l_n {\rm ln}\Big(p(\mathbf{x}_{n}^e,\mathbf{x}_{n}^t)\Big)+\eta (1-l_n) {\rm ln}\Big(1-p(\mathbf{x}_{n}^e,\mathbf{x}_{n}^t)\Big){\bigg ]}
\end{equation}
where $\eta$ is a balance factor between positive ($l_n=1$) and negative ($l_n=0$) trials, and $p(\mathbf{x}^e_n,\mathbf{x}^t_n)$ denotes the acceptance probability, i.e. the probability of $\mathbf{x}^e_n$ and $\mathbf{x}^t_n$  belonging to the same speaker. The reason why there needs a balance factor is that the number of the negative trials is usually much larger than that of positive trials. The difference between the variants of the binary cross-entropy loss is on the calculation method of $p(\mathbf{x}^e_n,\mathbf{x}^t_n)$ which is summarized as follows:
\begin{itemize}
 \item In \cite{heigold2016end,rahman2018attention}, {{the authors}} applied sigmoid function to cosine similarity:
    \begin{equation}
    \label{eq:end-to-end-sigmod-cosine}
    \begin{split}
        & p(\mathbf{x}^e_n,\mathbf{x}^t_n) =\frac{1}{1+{\rm exp}{\big(-wS(\mathbf{x}^e_n, \mathbf{x}^t_n)-b}\big)}     \\
        & S(\mathbf{x}^e_n,\mathbf{x}^t_n) =\frac{\mathbf{x}^{e^T}_n \mathbf{x}^t_n}{\|\mathbf{x}^e_n\| \, \|\mathbf{x}^t_n\|}
    \end{split}
    \end{equation}
    where $w$ and $b$ are two learnable parameters, and $-b/w$ corresponds  to the verification threshold. Some similar idea can also be found in \cite{zhang2016end}.

 \item In \cite{snyder2016deep}, {{the authors}} further introduced a PLDA-like similarity metric:
    \begin{equation}
    \label{eq:end-to-end-sigmod-PLDA}
    \begin{split}
        &p(\mathbf{x}^e_n,\mathbf{x}^t_n)= \frac{1}{1+{\rm exp}{\big(-S(\mathbf{x}^e_n, \mathbf{x}^t_n))}}     \\
        & S(\mathbf{x}^e_n, \mathbf{x}^t_n) =  {(\mathbf{x}^e_n)}^T\mathbf{x}^t_n-{(\mathbf{x}^e_n)}^T\mathbf{S}\mathbf{x}^e_n-{(\mathbf{x}^t_n)}^T\mathbf{S}\mathbf{x}^{t}_n+b
    \end{split}
    \end{equation}
    where $\mathbf{S}$ and $b$ are learnable parameters. Another similar PLDA-based similarity metric was proposed in \cite{rohdin2018end}.

\item The third calculation method is to learn a score from a joint vector ${\mathbf{x}^{e,t}_n}$ by \cite{zhang2019seq2seq}:
\begin{equation}
    \label{eq:end-to-end-sigmod-PLDA}
    \begin{split}
        &p(\mathbf{x}^e_n,\mathbf{x}^t_n)= \frac{1}{1+{\rm exp}{\big(-s^{e,t}_n)}}     \\
        & s^{e,t}_n = S(\mathbf{x}^{e,t}_n)
    \end{split}
    \end{equation}
    where $s^{e,t}_n$ is a scalar produced from a fully-connected feedforward neural network $S(\cdot)$, and  ${\mathbf{x}^{e,t}_n}$ is a joint vector of ${\mathbf{x}^e_n}$ and  ${\mathbf{x}^t_n}$ produced by a sequence-to-sequence attention mechanism \cite{zhang2019seq2seq}. Similar ideas can also be found in \cite{heo2017joint}.
\end{itemize}

The training trials of the aforementioned end-to-end speaker verification are constructed from two utterances.
To reduce the variability of the training trials, some work \cite{heigold2016end,rahman2018attention,zhang2016end} obtains the embedding of the enrollment speech $\mathbf{x}_n^e$ from an average of a small amount of utterances.

\textit{Contrastive loss} \cite{chung2018voxceleb2,yu2019ensemble} is another  commonly used  pairwise loss:
\begin{equation}
  \mathcal{L}_{\rm C} = \frac{1}{2N} \sum_{n=1}^{N}\Big(l_n d_n^2+(1-l_n){\rm max}(\rho-d_n,0)^2 \Big)
\end{equation}
where $d_n$ denotes the Euclidean distance between $\mathbf{x}_n^e$ and $\mathbf{x}_n^t$,  and $\rho$ is a manually-defined margin.
Unfortunately, training an end-to-end network with the contrastive loss is notoriously difficult. In order to avoid bad local minima in the early training stage, \cite{chung2018voxceleb2} proposed to first pre-train a speaker embedding system using Softmax loss, and then fine-tune the system with the contrastive loss.
 \cite{wan2018generalized} proposed a generalization of the contrastive loss as follows:
\begin{equation}
\label{eq:GE2E-contrast}
  L_{\rm GC}=\sum_{n=1}^{N}\Big( l_n \big(1-p(\mathbf{x}^e_n,\mathbf{x}^t_n)\big)+ (1-l_n)\mathop{\mathop{\rm max}_{\mathbf{x}^{e}_{n} \in \{\mathbf{c}_{j^\prime}\}_{j^\prime=1}^{J^\prime}}}p(\mathbf{x}^e_n,\mathbf{x}^t_n) {\Big )}
\end{equation}
where  $p(\mathbf{x}^e_n,\mathbf{x}^t_n)$ is the same  as  \eqref{eq:end-to-end-sigmod-cosine}, and   $\mathbf{c}_{j^\prime}$ with $j^\prime=1,2,\ldots,J^\prime$ is the speaker centroid of the $j$th speaker in a mini-batch which is obtained by averaging the utterances that belong to the $j^\prime$th speaker.

Besides the above two common training losses, some other loss functions are as follows.
In \cite{gao2019improving}, the authors  proposed  a \textit{discriminant analysis loss} $\mathcal{L}_{\rm DALoss}$ to learn  discriminative embeddings:
\begin{equation}\label{eq:DALoss}
  \mathcal{L}_{\rm DALoss} = \eta_1 \mathcal{L}_{\rm intra} + \eta_2 \mathcal{L}_{\rm inter}
\end{equation}
where $\eta_1$ and $\eta_2$ are the weights of the two loss items, and $\mathcal{L}_{\rm intra}$ and $\mathcal{L}_{\rm inter}$ are described as follows. $\mathcal{L}_{\rm intra}$ represents the intra-speaker variabilities which  is defined as:
\begin{equation}
  \mathcal{L}_{\rm intra}=\sum_{j^\prime=1}^{J^\prime}\frac{C^{j^\prime}}{\sum_{k=1}^{C^{j^\prime}}\frac{1}{d_k(\mathbf{x}_{n_1}^{j^\prime},\mathbf{x}_{n_2}^{j^\prime})}}
\end{equation}
where $j^\prime=1,2,\cdots,J^\prime$ denotes the index of the training speaker in each mini-batch, and $d_k(\mathbf{x}_{n_1}^{j^\prime},\mathbf{x}_{n_2}^{j^\prime})$ denotes the $k$th largest squared Euclidean distance between the embeddings of the $j^\prime$th speaker. The overall cost is the mean of the first $C^{j^\prime}$th largest distances within each speaker. $\mathcal{L}_{\rm inter}$ represents the inter-speaker variabilities:
\begin{equation}
  \mathcal{L}_{\rm inter} = {\rm max}(0,\zeta-{\rm min}(d(\widetilde{\mathbf{x}}^{j^\prime_1},\widetilde{\mathbf{x}}^{j^\prime_2})))
\end{equation}
where $\mathbf{\widetilde{x}}^{j_1^\prime}$ and $\mathbf{\widetilde{x}}^{j_2^\prime}$ denote the centers of the feature vectors of the $j_1^\prime$th and $j_2^\prime$th speakers respectively with $j_1^\prime \neq j_2^\prime \in \{1,2,\cdots,J^\prime \}$, $d(\cdot)$ denotes the distance (e.g. the squared Euclidean distance), and $\zeta$ denotes a margin. Thus, minimizing $ \mathcal{L}_{\rm inter} $ is equivalent to maximizing the distances between the centers to be larger than the minimum margin $\zeta$.

In \cite{mingote2019optimization}, the authors proposed to minimize both the empirical false alarm rate $P_{\rm fa}$ and miss detection rate $P_{\rm miss}$:
\begin{equation}
  \mathcal{L}=\eta_1 \cdot  {P}_{\rm fa}(\xi) + \eta_2  \cdot {P}_{\rm miss}(\xi)
\end{equation}
\begin{equation}
\label{eq:p_fa}
  P_{\rm miss}(\xi) = \frac{ \sum_{n=1}^{N} l_n \ \delta \Big (S(\mathbf{x}^e_n,\mathbf{x}^t_n) < \xi \Big)}{\sum_{n=1}^{N} \delta(l_n=1)}
\end{equation}
\begin{equation}
\label{eq:p_miss}
  P_{\rm fa}(\xi) = \frac{\sum_{n=1}^{N} (1-l_n) \  \delta \Big (S(\mathbf{x}^e_n,\mathbf{x}^t_n) > \xi \Big)}{\sum_{n=1}^{N} \delta(l_n=0)}
\end{equation}
where $\delta(\cdot)$ denotes an indicator function, $\xi$ denotes a decision threshold which is optimized with the neural network, and $\eta_1$ and $\eta_2$ are two tunable hyperparameters. The score $S(\mathbf{x}^e_n,\mathbf{x}^t_n)$ is obtained from the output linear layer of the neural network, where the number of units of the output layer equals to the number of the speakers in the training data. Specifically, it uses a batch of input vectors $\{\mathbf{x}_i\}_{i=1}^{I}$ and the parameters $\{\mathbf{w}_j,b_j\}_{j=1}^{J}$ of the output linear layer to construct training trials in a mini-batch:
\begin{equation}
  S(\mathbf{x}^e_n,\mathbf{x}^t_n)={(\mathbf{x}^{e}_{n})}^T \mathbf{x}^t_n +b_n^e,
\end{equation}
where $(\mathbf{x}^{e}_{n},b_n^e) \in \{\mathbf{w}_j,b_j\}_{j=1}^{J}$ and $\mathbf{x}^{t}_{n} \in \{\mathbf{x}_i\}_{i=1}^{I}$.  To make \eqref{eq:p_fa} and \eqref{eq:p_miss} differentiable, the indicator function $\delta(z>0)$ is relaxed to a sigmoid function $\sigma(z)=1 / (1+{\rm exp}{(-z)})$.

\subsection{Triplet loss}
\label{sec:triplet}
Triplet loss is a kind of training loss that each training sample that contributes to the accumulation of the training objective value independently is constructed from three utterances.
A triplet training sample consists of three utterances, including an anchor utterance, a positive utterance that is produced from the same speaker as the anchor utterance, and a negative utterance from a different speaker. Suppose the speaker features of a training sample produced from the top hidden layer are $\mathbf{x}^a$ (anchor), $\mathbf{x}^p$ (positive), and $\mathbf{x}^n$ (negative), respectively. We denote the training set as $\mathcal{X}_{\rm trip} = \{ (\mathbf{x}_n^a,\mathbf{x}_n^p,\mathbf{x}_n^n)|n=1,2,\cdots,N \}$.

Triplet loss designs a margin-based loss to push the positive utterance $\mathbf{x}_n^p$ closer to the anchor $\mathbf{x}_n^a$ than the negative utterance $\mathbf{x}_n^n$ in a trial as shown in Fig. \ref{fig:Triplet_Loss}.
For any training sample in $\mathcal{X}_{\rm trip}$, we require:
\begin{equation}
\label{eq:triplet}
  s_n^{an} - s_n^{ap}+\zeta \le 0
\end{equation}
where, without loss of generality, $s_n^{an}$ denotes the cosine similarity between $\mathbf{x}^a$ and $\mathbf{x}^n$, $s_n^{ap}$ denotes the cosine similarity between $\mathbf{x}^a$ and $\mathbf{x}^p$, and $\zeta \in \mathbb{R}^+$ is a manually-defined \textit{safety} margin between positive and negative pairs. Note that $s_n^{an}$ and $s_n^{ap}$ could be the scores of any similarity measurement instead of merely the cosine similarity. Given $\eqref{eq:triplet}$, the triplet loss is defined as:
\begin{equation}
\label{eq:triplet_loss}
  \mathcal{L}_{\rm trip}=\sum_{n=1}^{N} {\rm max}(0, s_n^{an} - s_n^{ap}+\zeta)
\end{equation}

Cosine similarity \cite{li2017deep} and squared Euclidean distance \cite{zhang2017end,huang2018joint,bredin2017tristounet} are the most common similarity metric for the triplet loss. Before calculating the similarities, each speaker embedding in the training samples needs to be length-normalized. It is easy to prove that the two similarity metrics are equivalent \cite{zhang2018text,bai2020speaker} after the length normalization.
Besides the two similarity metrics, {{the authors in}} \cite{dey2018end} proposed several distance functions to explore phonetic information for text-dependent speaker verification. They first compute the Euclidean distance between any pair of frame-level hidden representations of the two input utterances $\mathcal{H}_1=\{\mathbf{h}_{1,t_1}|t_1=1,2,\cdots,T_1\}$ and $\mathcal{H}_2=\{\mathbf{h}_{2,t_2}|t_2=1,2,\cdots,T_2 \}$ via $\mathbf{D}=\{d(\mathbf{h}_{1,t_1},\mathbf{h}_{2,t_2})|t_1=1,2,\cdots,T_1, t_2=1,2,\cdots,T_2\} \in \mathbb{R}^{T_1,T_2}$.  Then, they integrate the $T_1 \times T_2$ frame-level Euclidean distances into an utterance-level similarity score of $\mathcal{H}_1$ and $\mathcal{H}_2$ by, e.g. the attention mechanism.

\begin{figure}[t]
  \centering
  \includegraphics[width=3in]{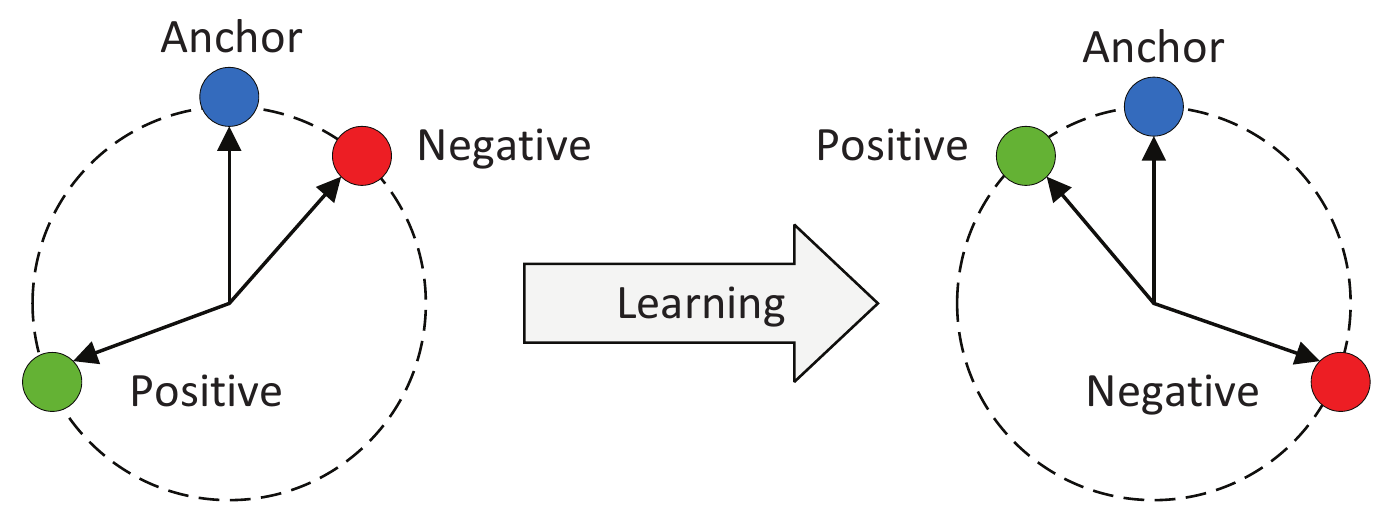}
  \caption{Triplet loss based on cosine similarity (from \cite{li2017deep}).}
  \label{fig:Triplet_Loss}
\end{figure}

Given a training set, we can see that the number of all possible triplet training samples is cubically larger than the number of training utterances. It is  neither efficient nor effective to enumerate all possible triplets \cite{bredin2017tristounet}, and only those that violate the constraint of $s_n^{an} - s_n^{ap}+\zeta \le 0$ contributes to the training process. Therefore, how to select informative triplet training samples is fundamental to the effectiveness of the model training.
In practice, the \textit{``hard negative''}  sampling strategy is popular  \cite{bredin2017tristounet}. It consists of the following two steps at each epoch:
\begin{enumerate}[1)]
  \item  Randomly sample $m$ utterances from each of the $M$ speakers of the training set, which constructs $Mm(m-1)/2$ anchor-positive pairs.
  \item For each of the anchor-positive pairs, randomly choose one negative utterance that satisfies $s_n^{an} - s_n^{ap} + \zeta > 0$  from the $(M-1)m$ negative candidates.
\end{enumerate}
Several variants of the \textit{``hard negative''}  sampling were proposed as well. For example,  \cite{zhang2017end} changed the first step by randomly selecting a small number of speakers from the speaker pool instead of from all speakers. In \cite{huang2018joint}, {{the authors}} divided training speakers into different groups and constructed each triplet training sample from a single group. Besides the \textit{``hard negative''}  sampling, the ``semi-hard'' negative sample selection \cite{schroff2015facenet,jati2019multi}
and softmax pre-training \cite{li2017deep} are all used to stabilize the training process of the triplet loss.

\subsection{Quadruplet loss}
Quadruplet loss is a kind of training loss for end-to-end speaker verification where each training sample that contributes to the accumulation of the training objective value independently is constructed from four utterances. Suppose there are
a  positive pairwise training set $\mathcal{X}_{\rm same} = \{ (\mathbf{x}_{n_1}^e,\mathbf{x}_{n_1}^t)|n_1=1,2,\cdots,N_1 \}$ and a negative pairwise training set $\mathcal{X}_{\rm diff}  = \{ (\mathbf{x}_{n_2}^e,\mathbf{x}_{n_2}^t)|n_2=1,2,\cdots, N_2\}$ respectively,  where $\mathbf{x}_{n_1}^e$ and $\mathbf{x}_{n_1}^t$ are from the same speaker while $\mathbf{x}_{n_2}^e$ and $\mathbf{x}_{n_2}^t$ are from different speakers. We have $\mathcal{X}_{\rm same} \cup \mathcal{X}_{\rm diff} = \mathcal{X}_{\rm pair}$. Currently, the quadruplet loss is {{formulated}} as the maximization of the partial interested area under the ROC curve (pAUC) \cite{bai2020partial}.

\begin{figure}[t]
  \centering
  \includegraphics[width=2in]{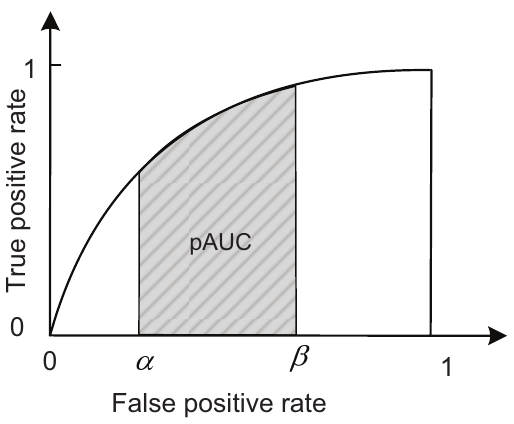}
  \caption{Illustration of the ROC curve, AUC, and pAUC (from \cite{bai2020partial}).}
  \label{fig:pAUC_Loss}
\end{figure}

The maximization of pAUC is to maximize an interested gray area of Fig. \ref{fig:pAUC_Loss} that is defined by two hyperparameters $\alpha$ and $\beta$. It has the following three steps:
\begin{enumerate}[1)~]
  \item Rank the similarity scores of all pairwise trials in $\mathcal{X}_{\rm diff}$ in descending order, and selecting  those
         elements that rank between the $(\lceil N_2 \times \alpha \rceil+1)$-th to $\lfloor N_2 \times \beta \rfloor$-th positions to construct $\mathcal{X}_{\rm diff}^\prime = \{(\mathbf{x}_{n_3}^e,\mathbf{x}_{n_3}^t)|n_3= 1,2,\cdots,N_3\} $ where $N_3 = \lfloor N_2 \times \beta \rfloor - (\lceil N_2 \times \alpha \rceil+1)$.
  \item Calculate pAUC on $\mathcal{X}_{\rm same}$ and $\mathcal{X}_{\rm diff}^\prime$:
   \begin{equation}
   \label{eq:pAUC}
     {\rm pAUC} =1-\frac{1}{N_1N_3} \sum_{n_1=1}^{N_1}\sum_{n_3=1}^{N_3}\Big (\delta(s_{n_1}<s_{n_3})+\frac{1}{2}\delta(s_{n_1}=s_{n_3}) \Big)
   \end{equation}
    where $\delta(\cdot)$ denotes the indicator function, and $s_{n_1}$ and $s_{n_3}$ denote the cosine similarity of the pairwise trials in $\mathcal{X}_{\rm same}$ and $\mathcal{X}_{\rm diff}^\prime$ respectively.
  \item  Relax the indicator function by the hinge loss which reformulates \eqref{eq:pAUC} to:
        \begin{equation}
        \label{eq:pAUC_relax}
         \mathcal{L}_{\rm pAUC}=\frac{1}{N_1N_3} \sum_{n_1=1}^{N_1}\sum_{n_3=1}^{N_3} {\rm max}\Big (0, \zeta-\big(
         s_{n_1}-s_{n_3})\big)\Big)^2
        \end{equation}
        where $\zeta$ is a margin hyperparameter.
\end{enumerate}
It can be seen clearly that \eqref{eq:pAUC_relax} is a quadruplet loss, since that $(s_{n_1}-s_{n_3})$ is calculated from four utterances.

\cite{bai2020partial} proposed two training sample construction methods. The first one is named random sampling. For a mini-batch, it first randomly chooses a mini-batch number of speakers, then  randomly selects two utterances for each  of the selected speaker, and finally generates the training trials of the batch by pairing all the selected utterances. The second one is named class-center learning. Before training, it first assigns a class center to each speaker in the training set. Then, for each training iteration, it generates the training trials of a mini-batch by pairing each of the class centers with each of the utterances in the batch, where the class centers are updated together with the DNN parameters.

The pAUC based loss has several advantages: (i) it directly optimizes the detection error tradeoff (DET) curve which is the major evaluation metric of speaker verification \cite{bai2020speaker}; (ii) it naturally overcomes the  class-imbalanced problem; (iii) it is able to select difficult quadruplet  training samples by setting $\alpha=0$ and $\beta$ to a small value, e.g. 0.01; (iv) triplet training samples is a subset of quadruplet training samples when given the same training utterances \cite{bai2020speaker}.
{{ Actually the pAUC should be calculated on the entire dataset, however, due to the limited computation resource,   \eqref{eq:pAUC} is  an empirical approximation to it within a mini-batch. Therefore,  a large batch size is usually  set to reduce the approximation error as much as possible. Fortunately, experimental results demonstrate that a  good approximation can be obtained with a batch size of no larger than 512.
}}

\subsection{ Prototypical network loss}

In \cite{wang2019centroid}, the \textit{prototypical network loss}  \cite{snell2017prototypical}, which was originally proposed for few-shot learning,  was applied to speaker embedding models.  Suppose that a  mini-batch contains a support set of $N$ labeled samples $\mathcal{S}=\{(\mathbf{x}_n,l_n)|n=1, 2,\cdots, N\}$ where $ l_n \in \{1,2,\cdots,J\}$ is the label of the sample $\mathbf{x}_n$, and  $\mathcal{S}_j$ denotes the set of all samples of class $j$. Then, the \textit{prototype} of each class is the mean vector of the  support points belonging to the class:
\begin{equation}
\label{eq:PNL-centroid}
  \mathbf{c}_j = \frac{1}{|\mathcal{S}_j|} \sum_{(\mathbf{x}_n,l_n)\in \mathcal{S}_j}^{} \mathbf{x}_n, \quad j=1,2,\cdots,J
\end{equation}
Given a query set $\mathcal{Q}=\{(\mathbf{x}_q,l_q)|q=1, 2,\cdots, Q\}$ with $l_q \in \{1,2,\cdots,J\}$, the prototypical network loss  classifies each query point $\mathbf{x}_q$ against $J$ prototypes $\{\mathbf{c}_j|j=1,2,\cdots,J\}$ via a softmax function:
\begin{equation}
  \mathcal{L}_{\rm PNL} = - \sum_{(\mathbf{x}_q,l_q) \in Q }^{}{\rm log}\frac{{\rm exp}\Big(-d\big(\mathbf{x}_q,\mathbf{c}_{l_q} \big) \Big)}{\sum_{j\prime=1}^{J}{\rm exp}\Big(-d\big(\mathbf{x}_q, \mathbf{c}_{j\prime}  \big) \Big)}
\end{equation}
where $d(\cdot)$ denotes the squared Euclidean distance.

For each mini-batch, \cite{wang2019centroid} first randomly selects a number of speakers from the training speaker pool, and then randomly chooses a support set and a query set for each of the selected speakers, where the samples of the support set and query set do not overlap.
Similar works were also conducted in \cite{anand2019few,Chung2020In, Kye2020}.

Before \cite{wang2019centroid}, {{\cite{wan2018generalized}}} proposed a generalized end-to-end loss based on the softmax function, which shares a similar idea with the prototypical network loss except that it uses a single set as both the support and query sets. In addition, \cite{Wei2020} recently  proposed an AM-Centroid loss which replaced the weights of the AAMSoftmax loss function with speaker centroids proposed in \cite{wan2018generalized}. This loss function aims to overcome the weakness of the AAMSoftmax loss based deep networks whose number of parameters at the output layer grows linearly with the number of training speakers.

\subsection{Other end-to-end loss functions}
Some loss functions cannot be categorized to the above categories. For example, given learnable speaker bases $\{\mathbf{w}_j\}_j^{J}$ and a mini-batch of utterances $\{(\mathbf{x}_n,l_n)\}_n^{N}$ where $l_n \in \{1,2,\cdots,J\}$, \cite{Heo2019} proposed a between-class variation based loss $\mathcal{L}_{\rm BC}$,
\begin{equation}
  \mathcal{L}_{\rm BC}= \sum_{j_2=1}^{J}\sum_{j_1=1, j_1 \neq j_2}^{J} \frac{\mathbf{w}_{j_1}^T\mathbf{w}_{j_2}}{\|\mathbf{w}_{j_1}\| \ \|\mathbf{w}_{j_2}\|}
\end{equation}
 and a hard negative mining loss $\mathcal{L}_{\rm H}$,
\begin{equation}
  \mathcal{L}_{\rm H} =  \sum_{n=1}^{N} \sum_{\mathbf{w}_h \in {\rm Hard}_n}^{} \!\! {\rm log}\Big(1+{\rm exp}\big(S(\mathbf{w}_h,\mathbf{x}_n)-S(\mathbf{w}_{l_n},\mathbf{x}_n)\big)\Big)
\end{equation}
where $\mathbf{x}_n$ denotes the \textit{n}th utterance, $\mathbf{w}_{l_n}$ denotes the basis that $\mathbf{x}_n$ belongs to, $S(\cdot)$ is the cosine similarity, and ${\rm Hard}_n$ is a set of so-called hard negative speaker bases of $\mathbf{x}_n$ which correspond to the top $H$ largest values in $\{S(\mathbf{w}_{j},\mathbf{x}_n) | {j\neq l_n}, j=1,2,\cdots,J\}$.

\subsection{Discussion to the verification-based loss functions}

The verification-based loss functions are fundamentally different from the classification-based loss functions in at least the following aspects.
First, speaker verification is essentially an open-set metric learning problem instead of a {{closed set}} classification problem. The verification-based losses are consistent with the test pipeline, which directly outputs verification scores.

Second, the output layers of the verification-based losses are very small and irrelevant to the number of training speakers, which is an important advantage of the verification-based methods over classification-based methods. Specifically, the number of parameters of a classification-based network at the output layer grows linearly with the increase of the number of training speakers, which make the network large-scale and easily overfit to the training data. For example, if a training set consists of 50000 speakers and if the top hidden layer of a classification network has 512 hidden units, then the output layer of the network contains 25.6 million parameters.  On the contrary, the verification-based systems do not suffer the aforementioned weakness. Aware of this issue, \cite{Wei2020} tried to solve the parameter explosion problem by drawing lessons from the  prototypical network loss.

The main weakness of the verification-based systems is that they are harder to train than the classification-based systems, since that they need to construct a large number of training trials and then select those that contributes significantly to the effectiveness of the systems, while the classification-based systems just classify each training utterance to its corresponding speaker. To overcome this weakness, many sample selection strategies for selecting highly-informative trials have been developed, such as the hard negative sampling in the triplet loss and the pAUC optimization in the quadruplet loss. Because the highly-informative trials are dynamically changing during the training process, the optimization process is not very stable and consistent. Some unstable examples include the triplet loss or pAUC maximization with the random sampling strategy. Fortunately, this weakness can be alleviated by constructing trials with speaker centroids via, e.g. the class-center learning \cite{bai2020partial}.

In respect of the performance, although the classification-based systems outperformed the verification-based systems once, recently results shown that the latter can achieve competitive performance with the former \cite{bai2020partial,Kye2020}.

A remark: the term ``end-to-end'' in this section intends to make a difference from the embedding systems in Section \ref{sec:embedding_objectives}. However, the term in many other speech processing tasks, which takes raw wave signals as the input and directly output decisions, is broader than the concept here. From the broader concept of ``end-to-end'', an end-to-end speaker verification system needs to further integrate additional procedures, including the voice activity detection, cepstral mean and variance normalization, into the network. It also has to prevent using additional back-ends, such as PLDA \cite{LinWeiwei2020}.

\section{Speaker diarization}
\label{sec:diarization}

In this section, we overview four kinds of speaker diarization technologies---stage-wise diarization, end-to-end diarization, online diarization, and multimodal diarization, where the stage-wise diarization has been studied for a long time, while the last three are emerging directions.

\subsection{Stage-wise speaker diarization}
\label{sec:stagewise}
\begin{figure*}[t]
  \centering
  \includegraphics[width=5in]{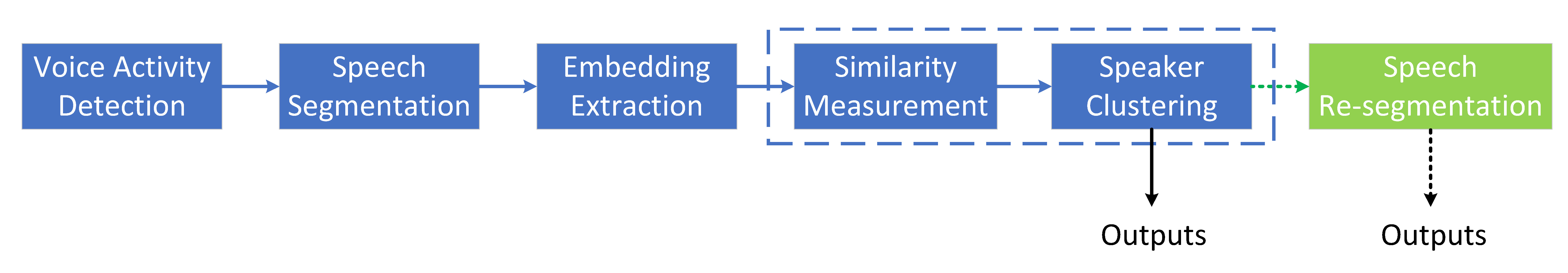}
  \caption{Diagram of the stage-wise speaker diarization, where speech re-segmentation is an optional module.}
  \label{fig:SpeakerDiarization_Diagram}
\end{figure*}

Stage-wise speaker diarization is composed of multiple independent modules. As shown in Fig. \ref{fig:SpeakerDiarization_Diagram}, most stage-wise speaker diarization systems consist of four  modules---voice activity detection, speech segmentation, speaker feature extraction, and speaker clustering. Some systems also have an optional re-segmentation module. This subsection briefly reviews deep learning based methods for each module. Voice activity detection detects speech in an audio recording and removes non-speech regions. Although it is an important module, it is usually studied independently. Therefore, we focus on reviewing the other modules.

\subsubsection{Speech segmentation}
 Speech segmentation splits speech into multiple speaker-homogeneous segments where each segment belongs to a single speaker. It usually can be categorized to two classes---\textit{uniform segmentation} and \textit{speaker change detection} (SCD). Uniform segmentation divides a long audio stream into short segments evenly by a  sliding window \cite{Lin2020DIHARD, garcia2017speaker,sell2018diarization,9053952,wang2018speaker}. For example, a 1.5 seconds sliding window with 0.75 second overlap is a common setting of uniform segmentation.

SCD partitions an audio recording according to the detected speaker change points, which results in non-uniform segments.
Generally, it first partitions an audio recording into small segments, then computes the similarity between the two adjacent speech segments in terms of the distance between their  representations, and finally decides whether the two adjacent segments are produced from the same speaker by thresholding the distance or finding a local extremum in the consecutive distance stream \cite{bredin2017tristounet}.
A common method for segmenting the audio recording into short segments is to use a sliding window \cite{bredin2017tristounet,wang2017speaker}. Recently, an ASR based segmentation \cite{9053280,sari2019pre} is also employed.

Conventional SCD algorithms usually adopt common handcrafted features, e.g. MFCC, as the acoustic representation \cite{chen1998speaker,siegler1997automatic}. An important advantage of the conventional methods is that only the step of tuning the threshold needs some experience, while the other parts do not need training \cite{yin2017speaker}.
Recently, the deep speaker embedding features are used as the representation of speech segments instead of conventional handcrafted features \cite{wang2017speaker,jati2018unsupervised,bredin2017tristounet,9053280,sari2019pre}.

To calculate the similarity between two adjacent segments, Euclidean distance \cite{bredin2017tristounet} and cosine similarity \cite{wang2017speaker} are two common similarity measurements. Some methods also feed two \cite{sari2019pre, jati2018unsupervised} or more \cite{9053280} consecutive embeddings together into a pre-trained DNN to predict the similarity between two adjacent segments. Recently, some work formulates SCD  as a sequence labeling task, which directly predicts if there is a change point in a speech segment \cite{yin2017speaker, zajic2017speaker, hruz2017convolutional}.

To summarize, on one side, the uniform segmentation is simple and works fine in many cases, which is the choice of many real-world diarization systems \cite{Lin2020DIHARD,sell2014speaker,9053280}; on the other side, the research on SCD is important not only for speaker diarization but also for many other applications, such as the closed captioning of broadcast television for hearing-impaired people \cite{9053280}.

\subsubsection{Speaker feature extraction}

The speaker feature extraction module in diarization shares similar technologies with speaker verification. Both of them map speech segments into speaker embeddings by i-vector \cite{sell2014speaker}, DNN-UBM/i-vector \cite{sell2015speaker},  x-vector \cite{sell2018diarization,diez2019bayesian,9054251,9053982}, or some other deep embedding extractors \cite{yella2015comparison, wang2018speaker, sun2019speaker}. See Sections \ref{sec:DNN/i-vector} to \ref{sec:end2end_loss} for the details. Here we only review some embedding methods that utilize additional information for speaker diarization. {{The authors in \cite{9054273}}} incorporated acoustic conditions, such as the distances between speakers and microphones in a meeting or the channel conditions of different speakers in a telephone conversation, into the speaker embeddings, given the fact that the acoustic conditions provide discriminative information for diarization. {{\cite{9054176}}} utilized a graph neural network to refine speaker embeddings, where the local structural information between speech segments is utilized as additional information.

\subsubsection{Speaker clustering}
Given the segment-level embedding features of an audio recording, speaker clustering aims to partition the speech segments into several groups, each of which belongs to a single speaker. It first defines a similarity measurement for evaluating the similarity of two segments, and then conducts clustering according to the similarity scores.
Popular similarity measurements include cosine similarity \cite{wang2018speaker} and PLDA-bases similarity \cite{sell2018diarization,sell2014speaker}. Recently, some deep-learning-based similarity measurements were also introduced, such as the LSTM-based scoring  \cite{lin2019lstm}, self-attentive similarity measurement strategies \cite{Lin2020SelfAttentive}, and joint training of speaker embedding and PLDA scoring \cite{garcia2017speaker}.
  Common clustering algorithms include  k-means \cite{wang2018speaker}, agglomerative hierarchical clustering \cite{sell2018diarization}, spectral clustering \cite{lin2019lstm,wang2018speaker},  Bayesian Hidden Markov Model  based clustering  \cite{diez2019bayesian,9054251,9053982} etc. Recently, Zhang \cite{zhang2018multilayer} proposed a non-neural-network deep model, named multilayer bootstrap networks, and applied it to speaker clustering \cite{zhang2016universal,li2018investigation}, which demonstrates competitive performance to the common speaker clustering algorithms.
However, these clustering algorithms are unsupervised, which is difficult to utilize manually-labeled training data, e.g. the time-stamped speaker  ground truth \cite{zhang2019fully}.

To address the problem, some work formulated speaker clustering as a semi-supervised learning problem \cite{milner2016dnn,yu2017active}.
Specifically, \cite{milner2016dnn} built a new DNN iteratively from a pre-trained speaker separation DNN for the speaker clustering of each audio file. In \cite{yu2017active}, {{the authors}} proposed an active-learning-based speaker clustering algorithm, which needs some involvement of human labor during the clustering process.

Some work formulated speaker clustering as a supervised learning problem \cite{zhang2019fully,li2019discriminative,9053477}. Specifically, \cite{zhang2019fully} and \cite{li2019discriminative} defined the speaker labels of training data according to their first appearance in the training data.
 As shown in Fig. \ref{fig:Diarization_labeling}, given a sequence of speech segments $\mathcal{E}=\{(\mathbf{e}_n,l_n)|n=1,2,\cdots,N\}$ where  $l_n \in \{\rm A,B,C,D,E,\cdots\}$ is the ground truth label of the $n$th segment with each capital letter representing a speaker, the training label of the sequence $\hat{l}_n \in \{1,2,3,4,5,\cdots\}$  is tagged with positive integers in the order of the speaker appearance in the sequence. For example, the training labels of the two sequences  $[\rm E,A,C,A,E,E,C]$ and $[\rm A,C,A,B,B,C,D,B,D]$ are tagged as $[\rm 1,2,3,2,1,1,3]$ and $[\rm 1,2,1,3,3,2,4,3,4]$ respectively.
{{ Under this labeling manner,}} \cite{zhang2019fully} trained a parameter-sharing RNN clustering model in a supervised way by multiple-instance-learning. They further integrated the RNN with a distance-dependent Chinese restaurant process to address the difficult problem of the unknown number of speakers. An improvement to \cite{zhang2019fully} was further presented in \cite{9053477}. Additionally, the clustering procedure was also modeled by a discriminative sequence-to-sequence neural network \cite{li2019discriminative}.

\begin{figure}[t]
  \centering
  \includegraphics[width=3.3in]{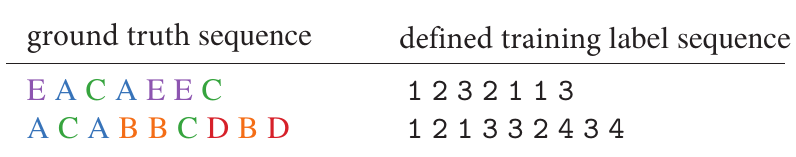}
  \caption{Two examples of the training label sequence definition of the supervised clustering in \cite{li2019discriminative} (from \cite{li2019discriminative}).}
  \label{fig:Diarization_labeling}
\end{figure}

\subsubsection{Speech re-segmentation}

Re-segmentation is an optional step after the speaker clustering. It refines speech boundaries between the speech segments. Variational-Bayesian refinement \cite{sell2015diarization,sell2018diarization,diez2018speaker} is the most famous conventional method. Recently, deep learning based re-segmentation methods were also developed. For example, following the successful application of the RNN to voice activity detection and SCD, \cite{yin2018neural} proposed to address re-segmentation with LSTM.

\subsubsection{Speech overlap detection}
 Most stage-wise speaker diarization systems simply assume that only one person is speaking at any time. In other words, they do not consider the speech overlap problem. However, speech overlap is one of the most important factors that hinder the diarization performance \cite{sell2018diarization,ryant2018first,ryant2019second,Lin2020DIHARD}, since it happens frequently in practice, e.g. a fast multi-speaker conversation. There have been several traditional studies on overlap detection for speaker diarization \cite{otterson2007efficient,boakye2008overlapped,huijbregts2009speech,yella2014overlapping}.
Recently, some deep learning based speech overlap detection methods were also proposed  \cite{9053096,9053760}. Specifically, \cite{9053096} first addressed the overlapped speech detection  as a sequence labeling problem by an LSTM-based architecture, and then assigned the detected overlap regions to two speakers by a  simple yet effective overlap-aware re-segmentation module. {{In \cite{9053760}, }}  a region proposal network {{was first used}} to detect overlapped speech, and then removed the highly overlapped segments in the post-processing stage. In addition to the above methods, end-to-end speaker diarization, which will be reviewed in the next section, is another way to deal with the speech overlap problem.

\subsection{End-to-end speaker diarization}

Because conventional clustering algorithms are unsupervised, it cannot minimize the diarization error rate directly \cite{fujita2020end} and is difficult to deal with the speech overlap problem \cite{fujita2020end}. Moreover, because each module of the stage-wise speaker diarization in Fig. \ref{fig:SpeakerDiarization_Diagram} is optimized independently, the performance is difficult to be boosted.
 Although several semi-supervised and supervised speaker clustering methods have recently been proposed, the potential of deep neural networks, which are mainly used to extract speaker embeddings in the stage-wise speaker diarization, has not been fully explored yet.
To address these problems, \cite{fujita2019end2,fujita2019end,fujita2020end} proposed an end-to-end  diarization method by  formulating speaker diarization as a  multi-label classification problem.

\begin{figure}[t]
  \centering
  \includegraphics[width=3in]{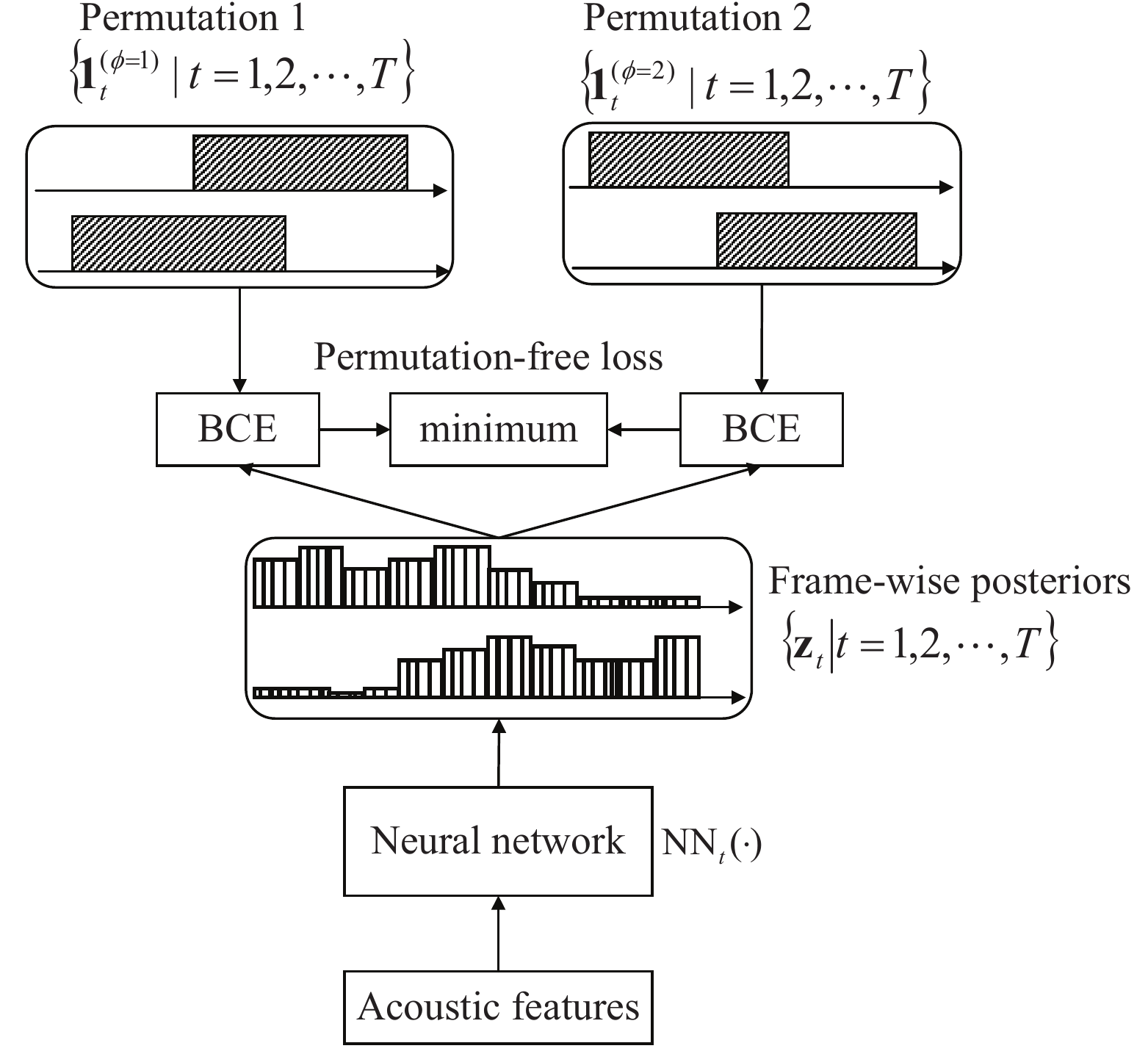}
  \caption{End-to-end neural diarization with the permutation-free loss for two-speaker diarization (from \cite{fujita2020end}).}
  \label{Fig:E2E_diarization_PIT}
\end{figure}

As shown in Fig. \ref{Fig:E2E_diarization_PIT}, given an acoustic feature sequence of an audio recording $\mathcal{Y}=\{\mathbf{y}_t \in \mathbb{R}^{d_1}|t=1,2,\cdots,T\}$, speaker diarization estimates the speaker label sequence $L=\{\bm{l}_t|t=1,2,\cdots,T\}$, where  $\bm{l}_t=\{l_{t,j}\in \{0,1\}|j=1,2,\cdots,J\}$ denotes a joint activity of a total number of $J$ speakers at time $t$. This multi-label formulation can represent speaker overlap regions properly. For example, $l_{t,j}=l_{t,j^\prime}=1$ with $j\neq j^\prime$ represents that the speech of the speakers $j$ and $j^\prime$ overlaps at time $t$.
With the assumption that each speaker is present independently, the frame-wise posteriors are estimated by a neural network as follows \cite{fujita2020end}:
\begin{equation}\label{eq:EEND_MultiLabel}
 \!\!  \mathbf{z}_t=[P(l_{t,1}|\mathcal{Y}),P(l_{t,2}|\mathcal{Y}),\cdots,P(l_{t,J}|\mathcal{Y})]={\rm NN}_t(\mathcal{Y})\in(0,1)^J
\end{equation}
where ${\rm NN}_t(\cdot)$ denotes a neural network, which was implemented by a BLSTM \cite{fujita2019end}, or a self-attention based neural network \cite{fujita2019end2}.

However, a difficult problem for the end-to-end speaker diarization is the speaker-label permutation ambiguity problem when aligning the ground truth label with the speech recording in preparing the training data.
For example, given an audio recording with three speakers A, B and C. If the ground-truth label of the recording is ``AAABBC'' , then the encoded labels ``111223'' and ``222113'' are equally correct, making the neural network hard to define a unique training label sequence \cite{lin2020optimal}.
To cope with this problem, as shown in Fig. \ref{Fig:E2E_diarization_PIT}, a neural network is trained to minimize the permutation-invariant training (PIT) loss  between the output $\mathbf{z}_t$  and the reference speaker label $\mathbf{1}_t \in \{0,1 \}^J$ \cite{fujita2019end2,fujita2019end,fujita2020end}:
\begin{equation}
    \mathcal{L}^{PF}=\frac{1}{TJ} \mathop{{\rm min}}_{\phi \in {\rm perm}(J)}\sum_{t}^{}{\rm BCE}(\mathbf{1}_t^\phi,\mathbf{z}_t)
\end{equation}
where ${\rm perm}(J)$ is the set of all possible permutations of the speaker identifiers $\{1,2\cdots, J\}$, and $\mathbf{1}_t^\phi$ is the $\phi$-th permutation of the ground-truth label sequence, and ${\rm BCE}(\cdot, \cdot)$ is the binary cross entropy between the label and the network output.

Under the multi-label classification framework \eqref{eq:EEND_MultiLabel}, the end-to-end diarization system is unable to deal with the test scenario where the number of speakers is larger than the maximum number of speakers in any of the training conversations. Here we bravely call this problem the \textit{fixed speaker capacity} issue for short.
 To this end, the end-to-end diarization framework is less flexible than the stage-wise methods which can handle any number of speakers in a test conversation \cite{horiguchi2020end}. To enlarge the speaker capacity, the number of the output nodes of the neural network \eqref{eq:EEND_MultiLabel} tends to be set large.
Unfortunately, the computational resource for training with the PIT loss will be exponentially increased along with the increase of the number of the output nodes. To deal with this contradiction,  \cite{lin2020optimal} proposed  an  optimal mapping loss, which directly computes the best match between the output speaker sequence and the ground-truth speaker sequence through a so-called Hungarian algorithm. It reduces the computational complexity to polynomial time, and meanwhile yields similar performance as the PIT loss. It should be noted that, the fixed {speaker capacity} issue remains unsolved in the optimal mapping loss.

\begin{figure}[t]
  \centering
  \includegraphics[width=3in]{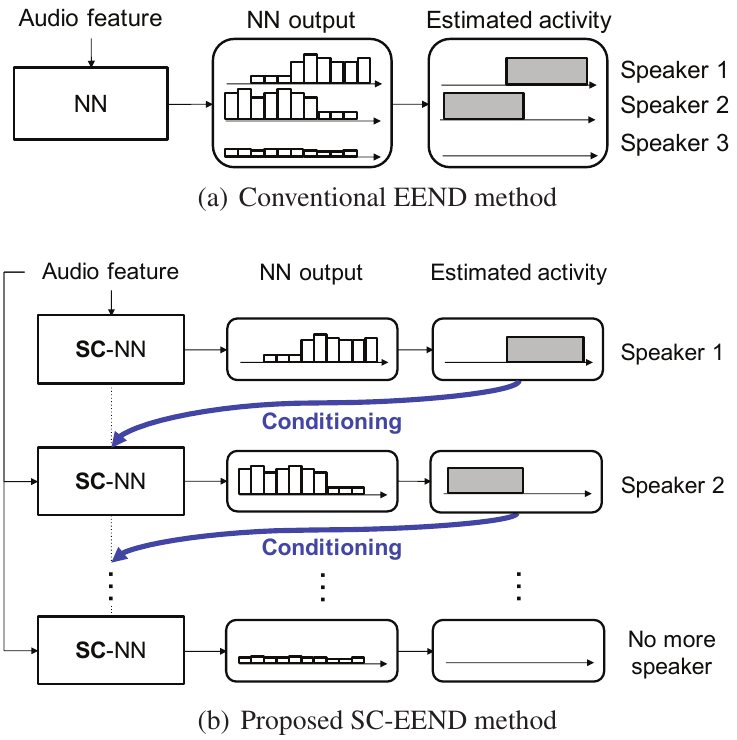}
  \caption{Diagrams of the conventional end-to-end neural diarization (EEND) method with fixed speaker numbers in \cite{fujita2020end}, and the  speaker-wise conditional end-to-end  neural diarization (SC-EEND) method (from \cite{fujita2020neural}).}
  \label{Fig:SC-EEND}
\end{figure}

To address the fixed speaker capacity issue, \cite{fujita2020neural}  proposed a  speaker-wise conditional end-to-end (SC-EEND) speaker diarization method.
 As shown in Fig. \ref{Fig:SC-EEND}, it uses an encoder-decoder architecture to decode each speaker's speech activity iteratively conditioned on the estimated historical speech activities. Although the SC-EEND method achieves {{increased}} speaker counting accuracy, it is difficult to handle more than four speakers \cite{fujita2020neural}, since it still has to deal with the PIT problem during decoding. Besides, they also proposed an encoder-decoder based attractor calculation  method, where a flexible number of speaker attractors are calculated from a speech embedding sequence \cite{horiguchi2020end}.  However, the speaker capacity of the method remains limited, due to the PIT loss which determines the assignment of the attractors to the training speakers.

\subsection{Online speaker diarization}
\label{sec:online_diarization}
Most state-of-the-art speaker diarization systems work in an offline manner. Online speaker diarization, which outputs the diarization result right after the audio segment arrives, is not an easy task, since that future information is unavailable when analyzing the current segment \cite{xue2020online}. In history, a number of online speaker diarization and speaker tracking solutions have been reported \cite{patino2018low,dimitriadis2017developing}. Here we focus on the deep learning based ones, which can be categorized to stage-wise online diarization and end-to-end online diarization methods.

In respect of the stage-wise online diarization, \cite{zhang2019fully} replaced the commonly used clustering module with a trainable unbounded interleaved-state RNN to make prediction in an online fashion{{, and it was   further improved  by introducing a \textit {sample mean loss} \cite{9053477}}}.
Additionally, a transformer-based discriminative neural clustering  model proposed in \cite{li2019discriminative}
can also perform online diarization. However, although it is possible to transfer  the online clustering to end-to-end speaker diarization, these methods still suffer from the assumption that only one speaker is present in a single segment \cite{xue2020online,lin2020optimal}.

In respect of the end-to-end online diarization,  \cite{xue2020online} extended the self-attention-based  end-to-end speaker diarization method in Fig. \ref{Fig:E2E_diarization_PIT} to an online version, with a speaker-tracing mechanism that is important for the success of the online diarization. Specifically, a straightforward online extension to the framework in Fig. \ref{Fig:E2E_diarization_PIT} is to perform diarization independently for each chunked recording. However, it is observed that the extension degrades the diarization error rate, due to the speaker permutation inconsistency across the chunk, especially for the short-duration chunks. To overcome this weakness, they applied a speaker-tracing buffer to record the speaker permutation information determined in previous chunks for a correct speaker-tracing in the following chunk. Because this method is limited to two-speaker diarization, more flexible speaker online end-to-end diarization methods need to be further explored  \cite{xue2020online}. In addition, \cite{von2019all} presented an all-neural approach to simultaneously conduct speaker counting, diarization, and source separation in a block-online manner, where a speaker-tracing mechanism is also employed to avoid the permutation inconsistency problem across time chunks.

\subsection{Multimodal speaker diarization}
Although speaker diarization is conventionally an audio-only task, the linguistic content carried by speech signals \cite{Flemotomos2020Linguistically} and the  speaker behaviors,  e.g., the movement of lips, recorded by videos  \cite{9054376} offer valuable supplementary cues to the detection of active speakers. To incorporate the aforementioned knowledge, \textit{multimodal speaker diarization} is emerging. Here we summarize some work as follows.

The first class is \textit{audio-linguistic} speaker diarization, \cite{park2018multimodal} integrated lexical cues and acoustic cues together by a gated recurrent unit-based sequence-to-sequence model, which improves the diarization performance by exploring linguistic variability deeply. The effectiveness of using both the linguistic and acoustic cues for diarization has been manifested further in structured scenarios \cite{el2019joint,Flemotomos2020Linguistically} where the speakers are assumed to produce distinguishable linguistic patterns. For instance, a teacher is likely to speak in a more didactic style while a student tends to be more inquisitive; a doctor is likely to inquire on symptoms and prescribe while a patient describe symptoms, etc. Another emerging direction is \textit{audio-visual} speaker diarization. In \cite{9054376}, the authors proposed a self-supervised audio-video synchronization learning method for the scenario where there lacks massive labeled data. The authors in \cite{chung2019said} proposed an iterative audio-visual approach which first enrolls speaker models using audio-visual correspondence, and then combines the enrolled models together with the visual information to determine the active speaker. In addition, microphone arrays provide important spatial information as well. For example, \cite{9053122} recently combined d-vectors with the spatial information produced from beamforming for the multimodal speaker diarization.

\section{Robust speaker recognition}
\label{sec:robust}
\renewcommand\arraystretch{1.5}
\begin{table*}[t]
  \centering
  \footnotesize
  \caption{Summary of adversarial-training-based domain adaptation methods.}
  \label{tab:adversarial_adaptation}
  \scalebox{1}{
  \begin{tabular}{m{0.2 cm} m{ 4.5cm}  m{6.5cm}  m{3.5cm} }
    \hline
    \hline
          &\textbf{Variable factors}  &   \textbf{Instantiation}             & \textbf{References}\\
    \hline
     1.   & Source and target mappings $M_s$ and $M_t$&  Shared parameters $M_s=M_t=M$, or unshared parameters $M_s \neq M_t$.           & \cite{wang2018unsupervised,9053323,9053905,fang2019channel};  \cite{xia2019cross}   \\
     2.   & Domain discriminator loss $\mathcal{L}_{{\rm adv}D}$ &  Binary cross-entropy, multi-class cross-entropy, or Wasserstein distance.& \cite{wang2018unsupervised}; \cite{zhou2019training}; \cite{rohdin2019speaker}\\
     3.   & Adversarial loss $\mathcal{L}_{{\rm adv}M}$ &  The gradient reversal layer, or the GAN loss function. & \cite{wang2018unsupervised,9053323, bhattacharya2019adapting};  \cite{bhattacharya2019generative,xia2019cross}    \\
     4.   & Input feature $\mathbf{X}_s$ and $\mathbf{X}_t$ &  Utterance-level speaker features (e.g. i-vector and x-vector), or frame-level acoustic features (e.g. MFCC and  F-bank). & \cite{wang2018unsupervised,tu2019variational} ;  \cite{9053323,bhattacharya2019adapting} \\
     5.   & Adaptation target &  Channel invariant, language invariant, phoneme invariant, noise robust, or short utterance compensation. & \cite{wang2018unsupervised};\cite{bhattacharya2019adapting};\cite{9053871};\cite{zhou2019training};\cite{zhang2018vector} \\
    \hline
    \hline
  \end{tabular}}
\end{table*}

Along with the rapid progress of speaker recognition, the frontier turns to ``\textit{recognition in the wild}''  \cite{nagrani2017voxceleb,mclaren2016speakers,ryant2018first}, where lots of domain mismatch and noisy problems arise. To overcome these difficulties, many domain adaptation and noise reduction methods were proposed. In this section, we comprehensively review these robust speaker recognition methods, including domain adaptation in Section \ref{subsec:da}, speech enhancement preprocessing in Section \ref{subsec:se}, and data augmentation techniques in Section \ref{sec:data_augmentation}.

\subsection{Domain adaptation}\label{subsec:da}

Over the past few years, speaker recognition has achieved great success due to the application of deep learning and large amount of labeled speech data. However, because collecting and annotating data for every new application is extremely expensive and time-consuming, sufficient training data may not be always available. For example, although large-scale labeled English databases are publicly available, Cantonese databases may be scarce \cite{sadjadi20172016}. Hence, it is needed to improve the performance of low-resource speaker recognition by using the large amount of auxiliary data. However, there are many distribution mismatch or domain shift problems between the low-resource data and auxiliary data, including different languages, phonemes, recording equipments, etc., which hinder the effectiveness of the auxiliary data. Fortunately, the mismatch problem can be alleviated by domain adaptation techniques.

Without loss of generality, the domain of interest is called the \textit{target} domain, and the domain with sufficient labeled training data is called the \textit{source} domain. The data distributions of the target domain and source domain are denoted as $p_t(\mathbf{x}, y)$ and $p_s(\mathbf{x}, y)$ respectively, where $p_s(\mathbf{x}, y) \neq p_t(\mathbf{x}, y)$.
Domain adaptation uses large amount of labeled data in the source domain to solve the problems in the target domain. If the training data in the target domain is manually labeled, then the domain adaptation is supervised; otherwise, it is unsupervised.
This paper focuses on the unsupervised domain adaptation, since it is more common and technically more challenging than the supervised domain adaptation in speaker recognition.

Domain adaptation has long been a significant topic in speaker recognition. It received much attention after the domain adaptation challenge in 2013\footnote{https://www.clsp.jhu.edu/workshops/13-workshop/speaker-and-language-recognition/}. Over the past years, lots of  supervised \cite{garcia2014supervised,wang2016domain,wang2020generalized} and unsupervised \cite{kanagasundaram2015improving,misra2018maximum,alam2018speaker,glembek2014domain,garcia2014unsupervised,lee2019coral,villalba2014unsupervised,wang2020generalized,shum2014unsupervised,bousquet2019robustness}  shallow domain adaptation methods based on the well known i-vector/PLDA pipline have been developed.
Among them, the adaptation is usually  accomplished at the back-end, including the methods of compensating the  domain mismatch in the i-vector space by an independent module before LDA and PLDA \cite{aronowitz2014inter, kanagasundaram2015improving,misra2018maximum,alam2018speaker}, and conducting domain adaptation at LDA \cite{glembek2014domain,mclaren2011source} or PLDA \cite{garcia2014supervised,garcia2014unsupervised,wang2016domain,lee2019coral,villalba2014unsupervised,wang2020generalized,shum2014unsupervised}. Although the methods are quite effective, here we do not discuss their details, as this article focuses on deep learning based ones.

Recently, many deep learning based domain adaptation methods were proposed. Following the categorization of the domain adaptation techniques \cite{wang2018deep,tzeng2017adversarial}, this paper categorize the deep-learning-based domain adaptation in speaker recognition into the following three classes:
\begin{itemize}
  \item \textbf{Adversarial-training-based domain adaptation:} It seeks to minimize an approximate domain discrepancy distance through an adversarial objective with a domain discriminator.
  \item \textbf{Reconstruction-based domain adaptation:} It assumes that the data reconstruction of the source or target samples can be helpful for improving the performance of domain adaptation.
  \item \textbf{Discrepancy-based domain adaptation:} It aligns the statistical distribution shift between the source and target domains using some mechanisms.
\end{itemize}

\subsubsection{Adversarial-training-based domain adaptation} \label{sec:adversarial-training-based}
\begin{figure}[t]
  \centering
  \includegraphics[width=3.3in]{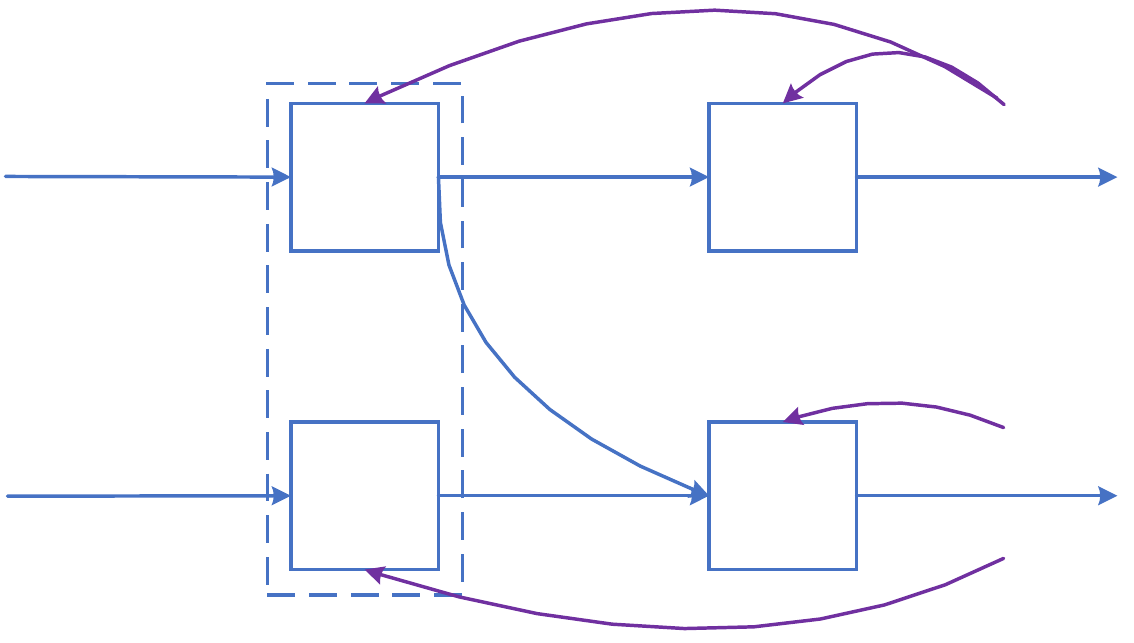}
  \put(-166,97){\small\bfseries\color{black}{$M_s$}}
  \put(-166,27){\small\bfseries\color{black}{$M_t$}}
  \put(-77,27){\small\bfseries\color{black}{$D$}}
  \put(-77,97){\small\bfseries\color{black}{$C$}}
  \put(-220,102){\small\bfseries\color{black}{$\mathbf{X}_s, Y_s$}}
  \put(-215,35){\small\bfseries\color{black}{$\mathbf{X}_t$}}
  \put(-242,62){\small\bfseries\color{black}{$p_s(\mathbf{x}, y) \!\! \neq \!\! p_t(\mathbf{x}, y)$}}
  \put(-112,62){\small\bfseries\color{black}{$ p(M_s(\mathbf{X}_s)) \!\! \approx \!\! p(M_t(\mathbf{X}_t))$}}
  \put(-35,102){\small\bfseries\color{purple}{$\mathcal{L}_{\rm cls}$}}
  \put(-133,102){\small\bfseries\color{black}{$M_s(\mathbf{X}_s)$}}
  \put(-133,35){\small\bfseries\color{black}{$M_t(\mathbf{X}_t)$}}
  \put(-35,35){\small\bfseries\color{purple}{$\mathcal{L}_{{\rm adv}_D}$}}
  \put(-35,22){\small\bfseries\color{purple}{$\mathcal{L}_{{\rm adv}_M}$}}
  \put(-125,-10){\small\bfseries\color{purple}{Gradient descent}}
  \put(-125,140){\small\bfseries\color{purple}{Gradient descent}}
  \put(-205,2){\small\bfseries\color{red}{ $M_s \neq M_t$ or $M_s = M_t$}}
  \caption{A unified framework of adversarial-training-based domain adaptation.}
  \label{fig:Adversarial-training}
\end{figure}

\begin{figure*}[t]
  \centering
  \includegraphics[width=5in]{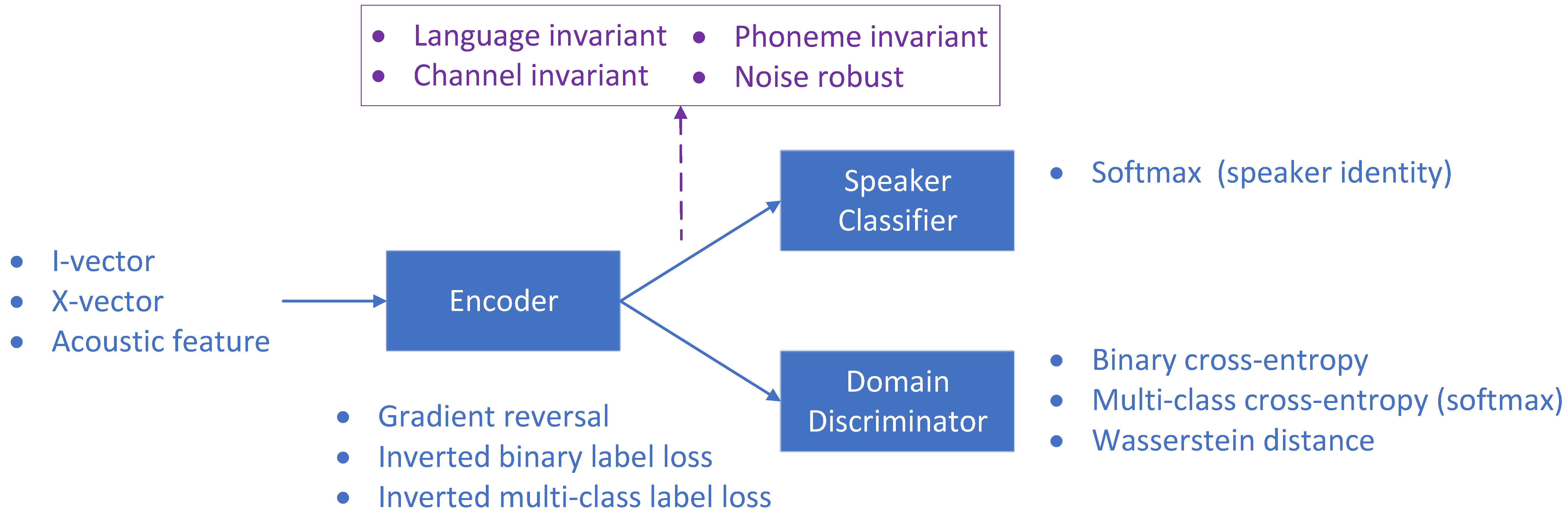}
  \put(-158,40){\small\bfseries\color{black}{$D$}}
  \put(-158,87){\small\bfseries\color{black}{$C$}}
  \put(-250,64){\small\bfseries\color{black}{$M$}}
  \caption{Domain-adversarial neural network (DANN) architecture with $M_s = M_t=M$.}
  \label{fig:Y-GAN-robust}
\end{figure*}

Before presenting the literature, we first build a unified framework of adversarial-training-based domain adaptation for speaker recognition {{by drawing lessons from \cite{tzeng2017adversarial}}}. As shown in Fig. \ref{fig:Adversarial-training}, the framework consists of:
\begin{itemize}
  \item $\mathbf{X}_s$ and $Y_s$ drawn from $p_s(\mathbf{x}, y)$ are the source features and speaker labels, and  $\mathbf{X}_t$   drawn from $p_t(\mathbf{x}, y)$  are the target features without labels;
  \item $M_s$ and $M_t$ are the feature mappings for the  source and target domains respectively;
  \item $C$ is a classifier for discriminating speakers in the source domain.
  \item $D$ is a domain discriminator for discriminating the source domain and target domain.

\end{itemize}
The adversarial adaptation methods aim at learning $M_s$ and $M_t$ for minimizing the distance between the empirical source and target data distributions in the feature space, i.e. making $ p(M_s(\mathbf{X}_s))\approx p(M_t(\mathbf{X}_t))$. After the adaptation,  the recognition models trained on $M_s(\mathbf{X}_s)$ can be directly applied to the target domain.

In the training stage, $M_s$ and $C$ are jointly trained using the standard supervised loss:
\begin{equation}
\label{eq:adv_class}
\begin{split}
   (\hat{M_s} , \hat{C})&= {\rm arg} \mathop{{\rm min}}_{M_s,C} \mathcal{L}_{\rm cls}(\mathbf{X}_s,Y_s;M_s,C) \\
    &= -\mathbb{E}_{(\mathbf{x}_s,y_s) \sim (\mathbf{X}_s,Y_s)} \sum_{j=1}^{J}\mathbb{I}_{[j=y_s]}{\rm log}[C(M_s(\mathbf{x}_s))]
\end{split}
\end{equation}
where $J$ is the number of speakers, and $\mathcal{L}_{\rm cls}$ is either the  standard softmax or its variants described in Section \ref{sec:embedding_objectives}, which ensures that the outputs of $M_s$ are speaker-discriminative.
To minimize the difference between the source and target representations, the adversarial adaptation methods conduct the following two steps alternatively for $D$ and $M_t$:
\begin{equation}
  \label{eq:adv_D}
     \hat{D} ={\rm arg} \mathop{{\rm min}}_{D} \mathcal{L}_{{\rm adv}_D}(\mathbf{X}_s,\mathbf{X}_t; \hat{M_s},\hat{M_t},D)
  \end{equation}
      \begin{equation}
      \label{eq:adv_Map}
       \hat{M_t} ={\rm arg}  \mathop{{\rm min}}_{M_t} \mathcal{L}_{{\rm adv}_M}(\mathbf{X}_s,\mathbf{X}_t; \hat{M_s},M_t,\hat{D})
     \end{equation}
where the symbols with hat, such as $\hat{M_s}$, denote that they are fixed during the alternative optimization.

By alternatively minimizing \eqref{eq:adv_D} and \eqref{eq:adv_Map}, $D$ and $M_t$ play an adversarial game: $D$ is optimized to predict the domain labels of $M_s(\mathbf{X}_s)$ and $M_t(\mathbf{X}_t)$, while $M_t$ is trained to make the prediction as incorrect as possible. When the training process converges, we have $ p(M_s(\mathbf{X}_s))\approx p(M_t(\mathbf{X}_t))$ since that $D$ is unable to discriminate $M_s(\mathbf{X}_s)$ and $M_t(\mathbf{X}_t)$. In addition, because the output of $M_s$ is speaker-discriminative which is ensured by the speaker classifier $C$, the output of $M_t$ is supposed to be speaker-discriminative too. For clarity, the manual label requirement in \eqref{eq:adv_class}, \eqref{eq:adv_D}, and \eqref{eq:adv_Map} are summarized in Table \ref{tab:adv_labels}.
\renewcommand\arraystretch{1.2}
\begin{table}[t]
  \footnotesize
  \centering
  \caption{The manual label requirement in the loss functions \eqref{eq:adv_class}, \eqref{eq:adv_D}, and \eqref{eq:adv_Map}. The domain labels can be any factors that are needed to be mitigated, including the types of channels, languages, phonemes, noise, etc.}
  \label{tab:adv_labels}
  \begin{tabular}{m{1cm}<{\centering}| m{1.8cm}<{\centering} m{1.8cm}<{\centering} m{1.8cm}<{\centering}}
  \hline
  \hline
   Losses & \textbf{Target} domain speaker labels & \textbf{Source} domain speaker labels & \textbf{Domain} labels\\
  \hline
    $\mathcal{L}_{\rm cls}$     &\checkmark    & \xmark   &\xmark      \\
    $\mathcal{L}_{{\rm adv}_D}$ &\xmark        & \xmark   &\checkmark  \\
   $\mathcal{L}_{{\rm adv}_M}$  &\xmark        & \xmark   &\checkmark  \\
  \hline
  \hline
  \end{tabular}
\end{table}
Different implementations of the framework in Fig. \ref{fig:Adversarial-training} are summarized in Table \ref{tab:adversarial_adaptation} which will be reviewed in detail as follows.

One of the most popular domain-adversarial neural network (DANN) architectures in literature is shown in Fig. \ref{fig:Y-GAN-robust} \cite{ganin2015unsupervised}. It consists of an encoder $M$, a speaker classifier $C$, and a domain discriminator $D$.
It follows the framework in Fig. \ref{fig:Adversarial-training} with a constraint $M_s=M_t=M$. Its training alternates the following steps:
\begin{enumerate}[1)~]
  \item The encoder is optimized by merging \eqref{eq:adv_class} and \eqref{eq:adv_Map}:
   $\hat{M} = {\rm arg} \mathop{{\rm min}}_{M} {\large[}\mathcal{L}_{\rm cls}(\mathbf{X}_s,Y_s;M,\hat{C}) + \lambda \mathcal{L}_{{\rm adv}_M}(\mathbf{X}_s,\mathbf{X}_t; M,\hat{D}){\large]}$, where $\lambda$ is a balance factor.

  \item The speaker classifier is obtained by minimizing \eqref{eq:adv_class}: $\hat{C} = {\rm arg} \mathop{{\rm min}}_{C} \mathcal{L}_{\rm cls}(\mathbf{X}_s,Y_s;\hat{M},C) $.

  \item The domain discriminator is obtained by minimizing \eqref{eq:adv_D}: $\hat{D} ={\rm arg} \mathop{{\rm min}}_{D} \mathcal{L}_{{\rm adv}_D}(\mathbf{X}_s,\mathbf{X}_t; \hat{M},D) $.
\end{enumerate}
After training, the encoder is responsible for extracting domain-invariant and speaker-discriminative speaker features, while the speaker classifier and domain discriminator will be discarded.

The domain-adversarial architecture has been used to explore channel-invariant \cite{wang2018unsupervised,9053323,9053905,fang2019channel}, language-invariant \cite{bhattacharya2019adapting,bhattacharya2019generative,rohdin2019speaker,tu2019variational,9053735}, phoneme-invariant \cite{9053871,wang2019usage}, noise-robust \cite{zhou2019training,meng2019adversarial,9054601} speaker features, etc. Here we present them briefly as follows.

\begin{figure*}[t]
  \centering
  \includegraphics[width=5in]{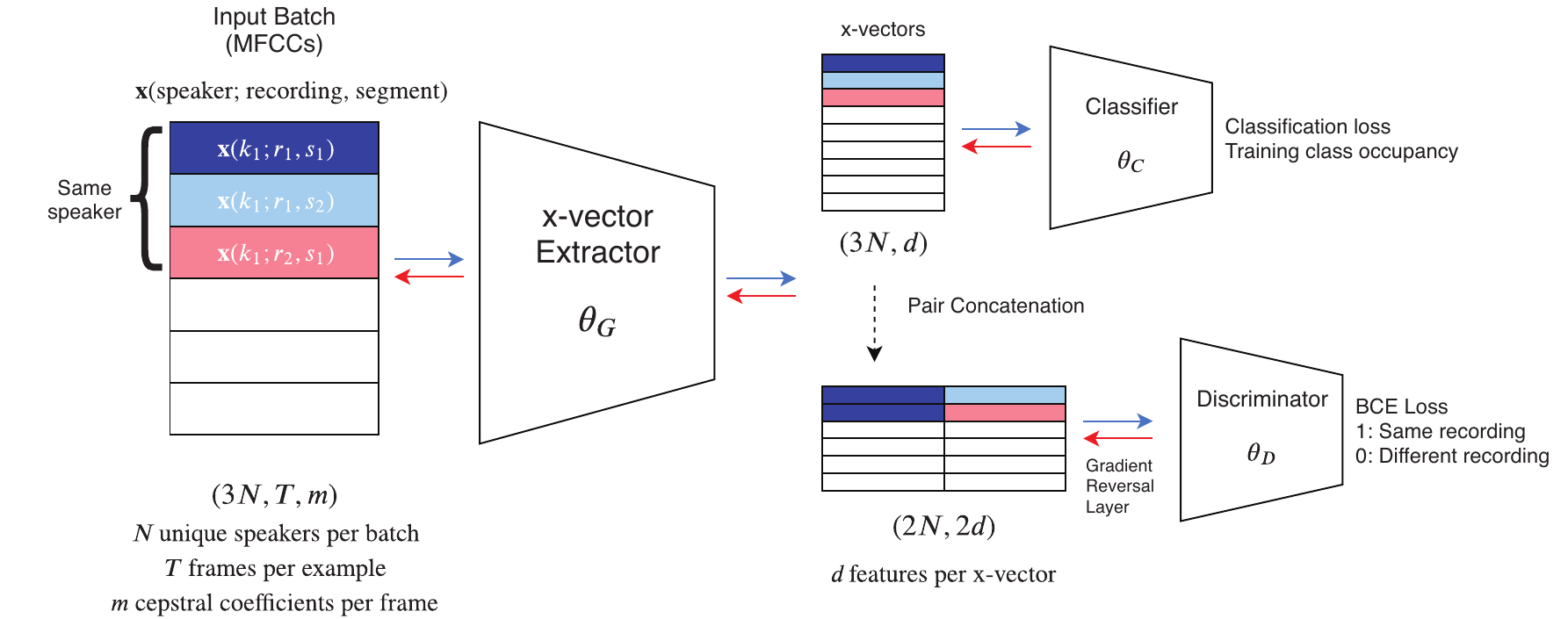}
  \caption{ Channel-invariant domain adaptation (from \cite{9053323}). The classifier is trained in the same way as the ordinary x-vector. The discriminator is trained on concatenated pairs of within-speaker pairs. The blue arrows represent the forward propagation. The red arrows represent the backward propagation of gradients.}
  \label{fig:Channel-invariant}
\end{figure*}

\cite{wang2018unsupervised} applied DANN to learn a channel invariant feature extractor from the i-vector subspace.
It takes the binary cross-entropy loss as the loss function of $\mathcal{L}_{{\rm adv}_D}$ to train the discriminator $D$. It adds a gradient reversal layer between the encoder and the discriminator to realize a minimax game $\mathcal{L}_{{\rm adv}_M}= - \mathcal{L}_{{\rm adv}_D}$.
\cite{9053323} developed a similar framework with Fig. \ref{fig:Y-GAN-robust} on an x-vector extractor. The framework produces speaker features that are invariant to the granularity of the recording channels. Instead of predicting the concrete domain labels, its domain discriminator predicts whether a pair of speaker embeddings that comes from the same speaker belong to the same recording in a Siamese fashion, see Fig. \ref{fig:Channel-invariant} for the above process. There is also a gradient reversal layer between then x-vector extractor and the domain discriminator. \cite{9053905} also proposed a similar work  to suppress the channel variability.

In addition to the channel mismatch problem, language mismatch is another challenge in speaker recognition. In \cite{bhattacharya2019adapting},  {{the authors}} applied the domain-adversarial architecture directly to acoustic features for a language invariant feature extractor, where the binary cross-entropy loss and a gradient reversal layer are applied to \eqref{eq:adv_D} and \eqref{eq:adv_Map} respectively. In \cite{bhattacharya2019generative}, they further
replaced the gradient reversal layer with a generative adversarial network (GAN) loss, a.k.a. inverted-label loss. Specifically,
 $\mathcal{L}_{{\rm adv}_D}$  still adopts the binary cross-entropy loss, while $\mathcal{L}_{{\rm adv}_M}$ fools the domain discriminator by inverting  the domain labels instead of using a gradient reversal layer, i.e.:
\begin{equation}
\begin{split}
   \mathcal{L}&_{{\rm adv}_D}(\mathbf{X}_s,\mathbf{X}_t;\hat{M},D) =  \\
    & -\mathbb{E}_{\mathbf{x}_s \sim \mathbf{X}_s}[{\rm log}(D(\hat{M}(\mathbf{x}_s)))]-\mathbb{E}_{\mathbf{x}_t \sim \mathbf{X}_t}[{\rm log}(1-D(\hat{M}(\mathbf{x}_t)))]
\end{split}
\end{equation}
\begin{equation}
\label{eq:GAN_loss}
\begin{split}
   \mathcal{L}&_{{\rm adv}_M}(\mathbf{X}_s,\mathbf{X}_t;M,\hat{D}) =  \\
    & -\mathbb{E}_{\mathbf{x}_s \sim \mathbf{X}_s}[{\rm log}(1-\hat{D}(M(\mathbf{x}_s)))]-\mathbb{E}_{\mathbf{x}_t \sim \mathbf{X}_t}[{\rm log}(\hat{D}(M(\mathbf{x}_t)))]
\end{split}
\end{equation}
where $\mathcal{L}_{{\rm adv}_D}$ tags the data from the source domain as ``1'' and the data from the target domain as ``0'', and $\mathcal{L}_{{\rm adv}_M}$ takes the opposite domain labels. This objective has the same fixed-point properties as the gradient reversal layer but provides stronger gradients to the encoder than the latter \cite{bhattacharya2019generative, tzeng2017adversarial}.
\begin{figure}[t]
  \centering
  \includegraphics[width=3in]{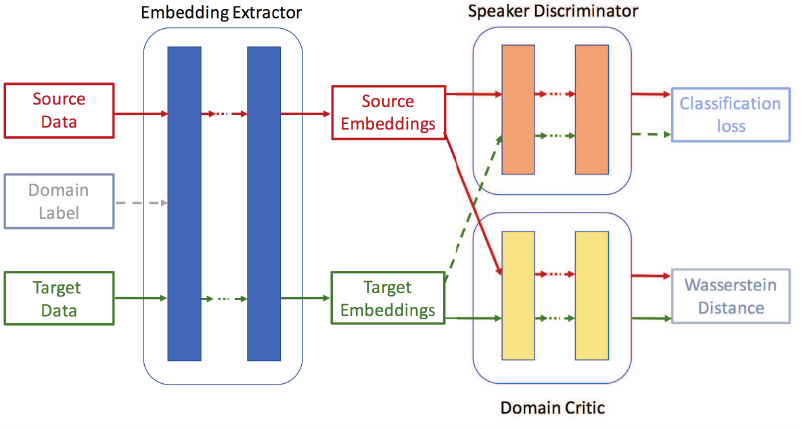}
  \caption{End-to-end adversarial language adaptation (from \cite{rohdin2019speaker}).}
  \label{fig:DAT_E2E_Wass_distance_adaptation}
\end{figure}
{{The authors in \cite{rohdin2019speaker}}} trained a language-invariant embedding extractor in an end-to-end fashion, where the embedding extractor is a standard TDNN based x-vector extractor (Fig. \ref{fig:DAT_E2E_Wass_distance_adaptation}). They utilized the discriminator to estimate the empirical Wasserstein distance between the source and target samples, and optimized the feature extractor network to minimize the distance in an adversarial manner.
\begin{figure}[t]
  \centering
  \includegraphics[width=2.5in]{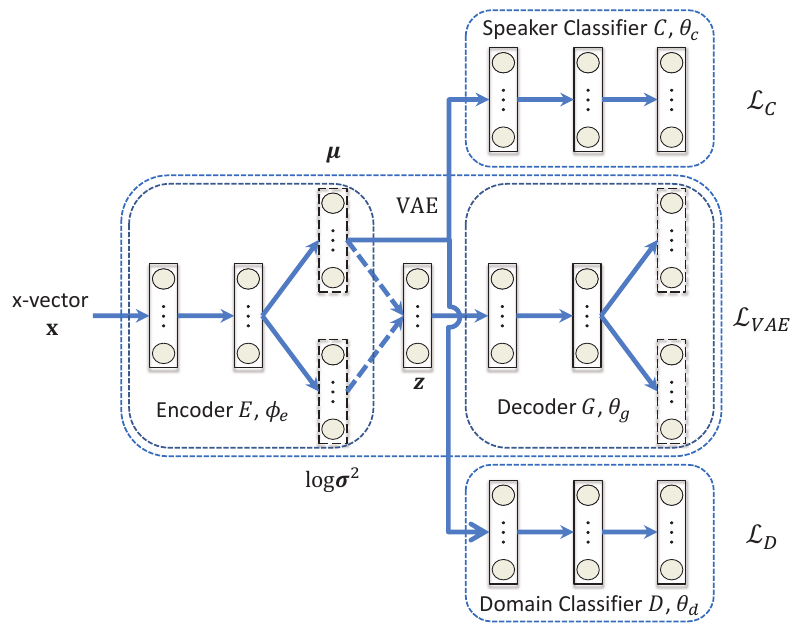}
  \caption{Variational domain adversarial neural network (from \cite{tu2019variational}).}
  \label{fig:VAE_regularizer_adaptation}
\end{figure}
\cite{tu2019variational} added an additional variational autoencoder (VAE) branch to the standard DANN structure of Fig. \ref{fig:Y-GAN-robust} for a language-invariant feature extractor as shown in Fig. \ref{fig:VAE_regularizer_adaptation}, where the state-of-the-art x-vector was used as the input. The VAE branch performs like a variational regularization which constrains the learned features to be Gaussian. As we know, Gaussian distribution is essential for the effectiveness of the standard PLDA backend. In \cite{9053735}, they further replaced the VAE with an information-maximized VAE, which not only retains the variational regularization but also inclines to preserve more speaker discriminative information than the VAE.

For text-independent speaker recognition, phonetic information is sometimes harmful, given the fact that it is difficult to ensure shared phonetic coverage across short enrollment and test utterances. However, for long recordings, phonetic information and even more higher level cues, such as word usages, actually differentiates speakers.
Quest for such high-level speaker features was an active area of research in about two decades ago. Recently, some studies also investigated its effect on text-independent speaker recognition in the era of deep learning. For example, \cite{wang2019usage} applied multi-task learning on frame-level layers to enhance the phonetic information in the frame-level features, and used the  adversarial training on segment-level layers to learn phoneme-independent representations. Finally, both operations result in improved performance.  \cite{9053871} concluded that phonetic information should be suppressed in text-independent speaker recognition working with frame-wise or extremely short utterances.

Additive noise is one of the most serious interferences of speaker recognition. To explore a noise-robust feature extractor, many works resorted to DANN.
For instance, \cite{zhou2019training} used a multi-class cross entropy loss as the loss function of a noise discriminator to train a noise-condition-invariant feature extractor. \cite{meng2019adversarial} applied the adversarial training to learn a feature extractor that are invariant to two kinds of conditions---different environments and different SNRs, where the different environments are represented as a categorical variable and the range of SNRs is formulated as a continuous variable.

\cite{9054601} applied an unsupervised adversarial invariance (UAI) architecture to disentangle speaker-discriminative  information.
As shown in Fig. \ref{fig:UAI-domain-adaptation}, the encoder generates two latent representations $\mathbf{h}_1$ and $\mathbf{h}_2$ from the
x-vector $\mathbf{x}$, where $\mathbf{h}_1$ only contains  the speaker-discriminative information, and $\mathbf{h}_2$ contains all other information of $\mathbf{x}$. This was implemented by optimizing the encoder,  predictor and decoder together, where the predictor aims to predict speaker labels $\hat{\mathbf{y}}$  from  $\mathbf{h}_1$, and the decoder aims to recover $\hat{\mathbf{x}}$ from a concatenation of $\mathbf{h}_2$ and a noise corrupted version of $\mathbf{h}_1$, denoted as $\mathbf{h}_1^\prime$. In order to further encourage  the ``disentanglement'', a minimax game between the disentanglers and the encoder was conducted, where the disentanglers tries to reconstructs the two latent representations from each other,  while the encoder is optimized against the disentanglers. An important merit of UAI over DANN is that the adversarial game does not need domain labels.

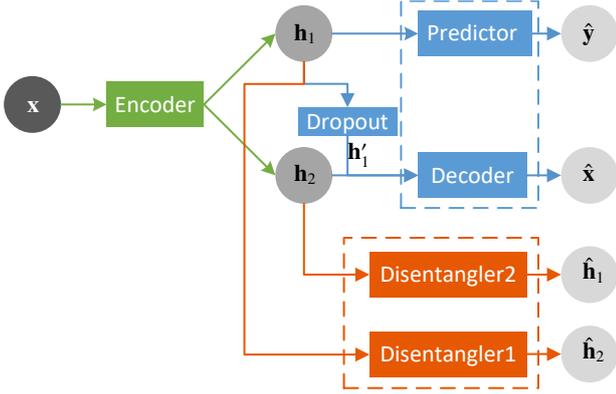
\begin{figure}[t]
  \centering
  \includegraphics[width=3.2in]{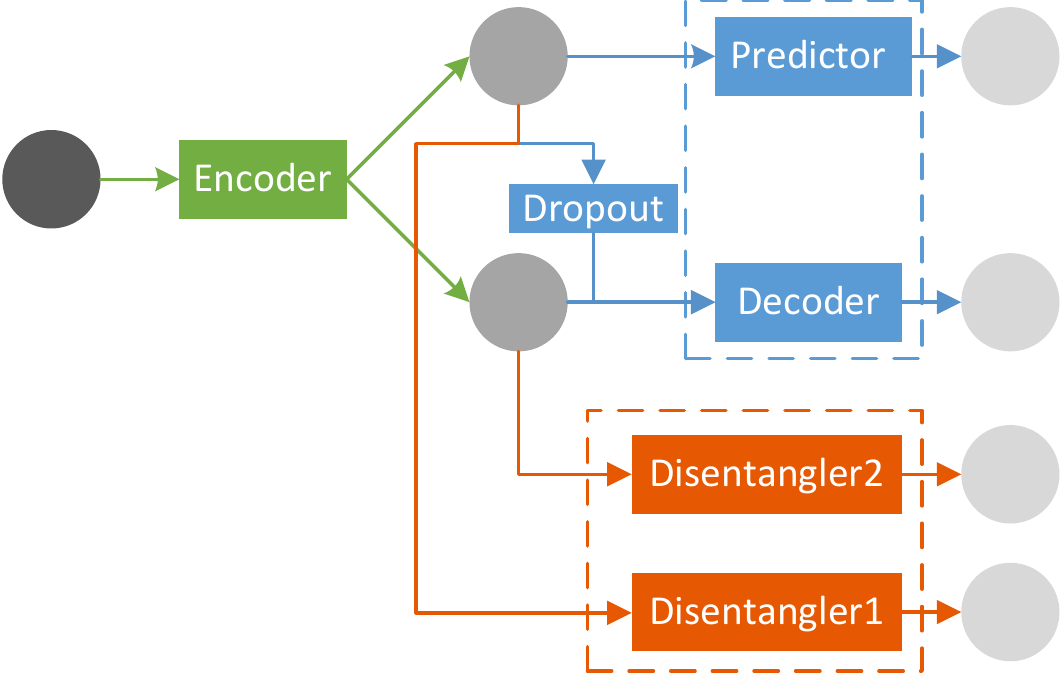}
  \put(-222,106){\small\bfseries\color{white}{$\mathbf{x}$}}
  \put(-122,133){\small\bfseries\color{black}{$\mathbf{h}_1$}}
  \put(-122,80){\small\bfseries\color{black}{$\mathbf{h}_2$}}
  \put(-102,88){\small\bfseries\color{black}{$\mathbf{h}_1^\prime$}}
  \put(-14,80){\small\bfseries\color{black}{$\hat{\mathbf{x}}$}}
  \put(-14,133){\small\bfseries\color{black}{$\hat{\mathbf{y}}$}}
  \put(-14,43){\small\bfseries\color{black}{$\hat{\mathbf{h}}_1$}}
  \put(-14,13){\small\bfseries\color{black}{$\hat{\mathbf{h}}_2$}}
  \caption{A brief diagram of unsupervised adversarial invariance.}
  \label{fig:UAI-domain-adaptation}
\end{figure}

In addition to DANN where $M_s=M_t=M$, another kind of adversarial-training-based domain adaptation methods trains an encoder for each domain, i.e. $M_s \neq M_t$.
Adversarial discriminative domain adaptation (ADDA) shown in Fig. \ref{fig:GAN-adaptation} is such a representative approach \cite{tzeng2017adversarial}.
In \cite{xia2019cross}, {{the authors}} applied ADDA to learn an asymmetric mapping that adapts the target domain encoder to the source domain encoder, where the two domains are in different languages.
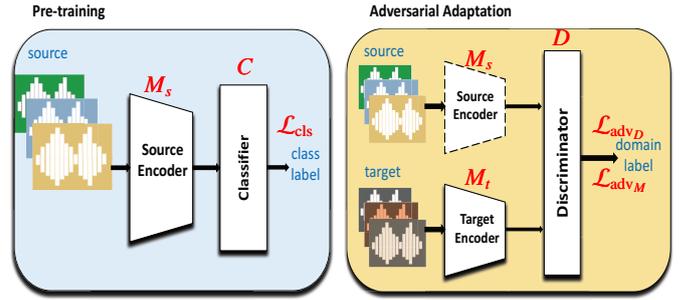
\begin{figure}[t]
  \centering
  \includegraphics[width=3.5in]{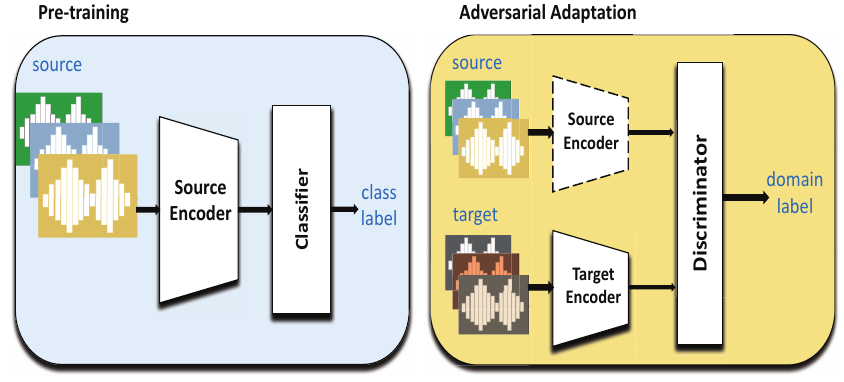}
  \put(-165,87){\small\bfseries\color{red}{$C$}}
  \put(-165,87){\small\bfseries\color{red}{$C$}}
  \put(-200,80){\small\bfseries\color{red}{$M_s$}}
  \put(-200,80){\small\bfseries\color{red}{$M_s$}}
  \put(-150,65){\small\bfseries\color{red}{$\mathcal{L}_{\rm cls}$}}
  \put(-150,65){\small\bfseries\color{red}{$\mathcal{L}_{\rm cls}$}}
  \put(-47,98){\small\bfseries\color{red}{$D$}}
  \put(-47,98){\small\bfseries\color{red}{$D$}}
  \put(-80,90){\small\bfseries\color{red}{$M_s$}}
  \put(-80,90){\small\bfseries\color{red}{$M_s$}}
  \put(-80,45){\small\bfseries\color{red}{$M_t$}}
  \put(-80,45){\small\bfseries\color{red}{$M_t$}}
  \put(-32,65){\small\bfseries\color{red}{$\mathcal{L}_{{\rm adv}_D}$}}
  \put(-32,65){\small\bfseries\color{red}{$\mathcal{L}_{{\rm adv}_D}$}}
  \put(-32,45){\small\bfseries\color{red}{$\mathcal{L}_{{\rm adv}_M}$}}
  \put(-32,45){\small\bfseries\color{red}{$\mathcal{L}_{{\rm adv}_M}$}}
  \caption{Adversarial discriminative domain adaptation (ADDA) approach (from \cite{xia2019cross}). Note that the source encoder is fixed during the adversarial adaptation stage.}
  \label{fig:GAN-adaptation}
\end{figure}
From the view of the framework in Fig. \ref{fig:Adversarial-training}, ADDA is trained by the following two successive steps:
\begin{enumerate}[1)~]
  \item Pre-train a source domain encoder $M_s$ and a speaker classifier $C$ with the labeled source data  by \eqref{eq:adv_class}.
  \item Fix the source domain encoder $M_s$, and perform adversarial training on the target encoder $M_t$ and domain discriminator $D$ by alternatively minimizing $\mathcal{L}_{{\rm adv}_D}$ and $\mathcal{L}_{{\rm adv}_M}$ via \eqref{eq:adv_D} and \eqref{eq:adv_Map} respectively. The domain discriminator $D$ minimizes the binary cross-entropy loss, i.e. $ \mathcal{L}_{{\rm adv}_D}= -\mathbb{E}_{\mathbf{x}_s \sim \mathbf{X}_s}[{\rm log}(D(\hat{M}_s(\mathbf{x}_s)))]-\mathbb{E}_{\mathbf{x}_t \sim \mathbf{X}_t}[{\rm log}(1-D(\hat{M}_t(\mathbf{x}_t)))] $. The target encoder $M_t$ minimizes an inverted label loss, i.e. $\mathcal{L}_{{\rm adv}_M}=-\mathbb{E}_{\mathbf{x}_t \sim \mathbf{X}_t}[{\rm log}(\hat{D}(M_t(\mathbf{x}_t)))]$\footnote{Different from \eqref{eq:GAN_loss}, the constant part $-\mathbb{E}_{\mathbf{x}_s \sim \mathbf{X}_s}[{\rm log}(1-\hat{D}(\hat{M}_s(\mathbf{x}_s)))]$  was removed from $\mathcal{L}_{{\rm adv}_M}$, given that $M_s$ is fixed.}.
\end{enumerate}
In the test stage, the data from the target domain is first mapped to the shared feature space by the target domain encoder, and then classified by a back-end classifier trained in the source domain, e.g. PLDA.

In order to compensate the unreliability of short utterances, \cite{zhang2018vector} proposed to compensate short utterances by long utterances using GAN. Specifically, it uses a generator to generate compensated i-vectors from short-utterance i-vectors, and uses a discriminator to determine whether an i-vector is generated by the generator or extracted from a long utterance. Similarly,  \cite{9054036} addressed the problem by adversely learning a mapping function that maps short embedding features to enhanced embedding features.

Despite of the success of the adversarial-training-based domain adaptation, its training is not easy in practice. For example, DANN learns a common encoder for both domains, which may make the optimization poorly conditioned, since a single encoder has to handle features from two separate domains \cite{tzeng2017adversarial}. Although ADDA is able to learn domain specific encoders, its target domain has no labels. As a result, without shared weights between the encoders, it may quickly fall into a degenerate solution if not properly initialized \cite{tzeng2017adversarial}. To remedy this weakness, a pre-trained source encoder is used to initialize the target encoder, leaving the source encoder fixed during the adversarial training \cite{tzeng2017adversarial,xia2019cross}. Besides, many training difficulties observed in related areas, such as image processing, have been encountered in speaker recognition as well, though seldom discussed in depth.

\subsubsection{Reconstruction-based domain adaptation}
In the machine learning community, CycleGAN, which was originally proposed for image-to-image translation \cite{zhu2017unpaired}, is one of the most common reconstruction-based domain adaptation methods \cite{wang2018deep}. Recently, it was introduced to speaker recognition \cite{nidadavolu2019cycle, nidadavolu2019low, 9053823}.
\begin{figure}[t]
  \centering
  \includegraphics[width=3.5in]{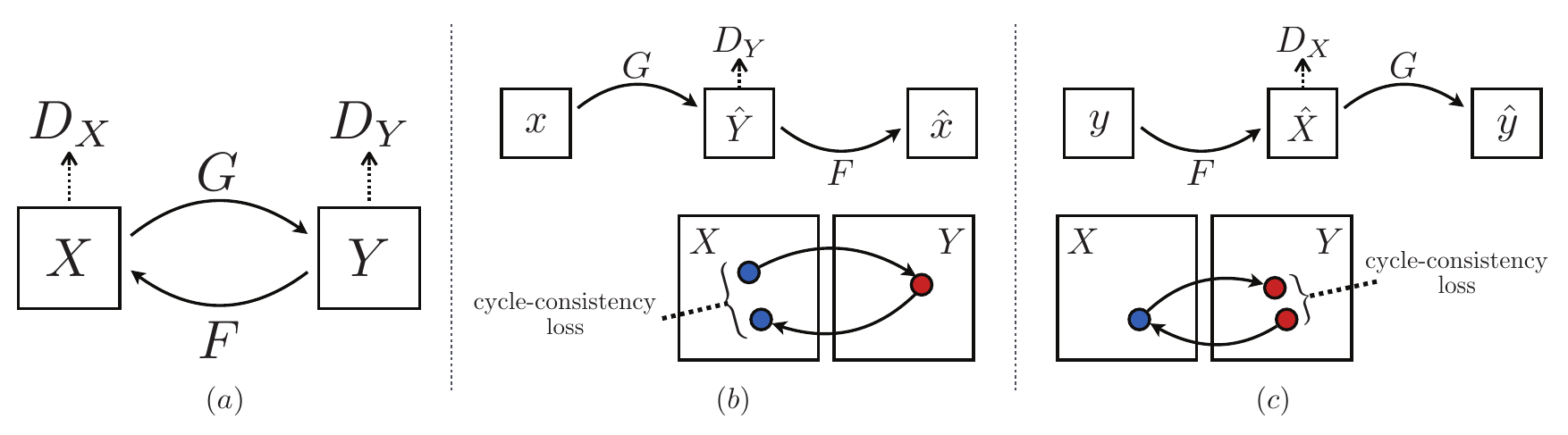}
  \caption{ The cycle GAN architecture (from \cite{zhu2017unpaired}).}
  \label{fig:Cycle-GAN}
\end{figure}
As shown in Fig. \ref{fig:Cycle-GAN}, CycleGAN comprises two generators and two discriminators. The generator $G$ transforms the feature $X$ in a domain to the feature $Y$ in another domain, producing an approximation of $Y$, i.e. $\hat{Y}=G(X)$. The discriminator $D_Y$, which aligns with  $G$, discriminates between $\hat{Y}$ and $Y$. The other generator-discriminator pair, i.e. $F$ and $D_X$, is intended to transfer features from $Y$ to $X$. The generators and discriminators are trained using a cycle consistency loss and a combination of two adversarial losses, where the cycle consistency loss measures how well the original input is reconstructed after a sequence of two generators, i.e. $F(G(X)) \approx X$ or $G(F(Y)) \approx Y $. Because of the adversarial and cycle reconstruction mechanisms of CycleGAN, it has an outstanding advantage that its training needs neither speaker labels nor paired data between the source and target domains.

{{At the acoustic feature level, \cite{nidadavolu2019cycle} explored a domain adaptation approach }} by learning feature mappings between a microphone domain and a telephone domain  using CycleGAN. It maps the acoustic features from the target domain (microphone) back to the source domain (telephone), and conducts speaker recognition using the system trained in the source domain.
In \cite{nidadavolu2019low}, they further investigated the effectiveness of CycleGAN in low resource scenarios where the target domain only has limited amount of data. They found that, the adaptation system trained on limited amount of target domain data performs slightly better than the adaptation system trained on a larger amount of target domain data, when some noise was added to the data.
In \cite{9053823}, they developed a CycleGAN-based feature enhancement approach in the log-filter bank space to improve the performance of speaker verification in noisy and reverberant environments.

Besides CycleGAN, the encoder-decoder structure is another popular reconstruction-based domain adaptation method in machine learning \cite{wang2018deep}. As for speaker recognition, \cite{shon2017autoencoder} combined an autoencoder with a denoising autoencoder to adapt resource-rich source domain data to the target domain.

\subsubsection{Discrepancy-based domain adaptation}

Discrepancy-based domain adaptation aligns the statistical distribution shift between the source and target domains by using some statistic criteria, including maximum mean discrepancy (MMD), correlation alignment (CORAL), Kullback-Leibler divergence, etc. {{For speaker verification, those criteria has been widely studied in shallow domain adaptation models \cite{alam2018speaker,lee2019coral}. Recently, some of them were also introduced to the deep learning based adaptation approaches. Specifically, \cite{lin2018reducing}}} added  a MMD based loss to the reconstruction loss  of an autoencoder to  train a domain-invariant encoder for multi-source adaptation of i-vectors.
In \cite{lin2018multisource,lin2019semi}, they further proposed a nuisance-attribute autoencoder based on MMD. In \cite{9054134}, they proposed a multi-level deep neural network adaptation method using MMD and consistency regularization.

\subsection{Speech enhancement and de-reverberation preprocessing}\label{subsec:se}
Speech is  always  distorted  by  noise and reverberation in real-world scenarios. A natural choice of coping with these distortions for speaker recognition is to add a speech enhancement or de-reverberation preprocessing module. Recently, deep learning based speech enhancement and de-reverberation techniques \cite{wang2018supervised} have been applied to speaker recognition, which can be roughly divided into three categories, i.e. masking-based \cite{zhao2014robust,kolboek2016speech,chang2017robust,zhao2019robust,shon2019voiceid}, mapping-based \cite{plchot2016audio,novotny2018on,novotny2019analysis,oo2016dnn,sun2018speaker}, and GAN-based \cite{michelsanti2017conditional} techniques.
{{It should be note that this section only reviews some general concepts and provides useful cues without considering the details of the speech enhancement techniques, since it is out the scope of this paper. More details can be found in speech enhancement related papers \cite{wang2018supervised}.}}

Masking-based speech enhancement has received a lot of attention and shown impressive performance in speech quality and speech intelligibility. It uses a DNN to estimate a time-frequency mask of noisy speech, and then uses the mask to recover the corresponding clean speech. In  \cite{zhao2014robust,kolboek2016speech,chang2017robust}, the authors applied masking-based  speech enhancement techniques {{as an independently noise reduction module}} for speaker recognition. In \cite{zhao2019robust}, {{the authors}} jointly optimized speech separation and speaker verification networks together. {{Besides,}} \cite{shon2019voiceid} designed a \textit{VoiceID loss} which jointly optimize a pre-trained speaker embedding system and a speech enhancement network where the speaker embedding system is fixed during the joint optimization, as shown in Fig. \ref{fig:VoiceID_Loss}.
 \begin{figure*}[t]
  \centering
  \includegraphics[width=5 in]{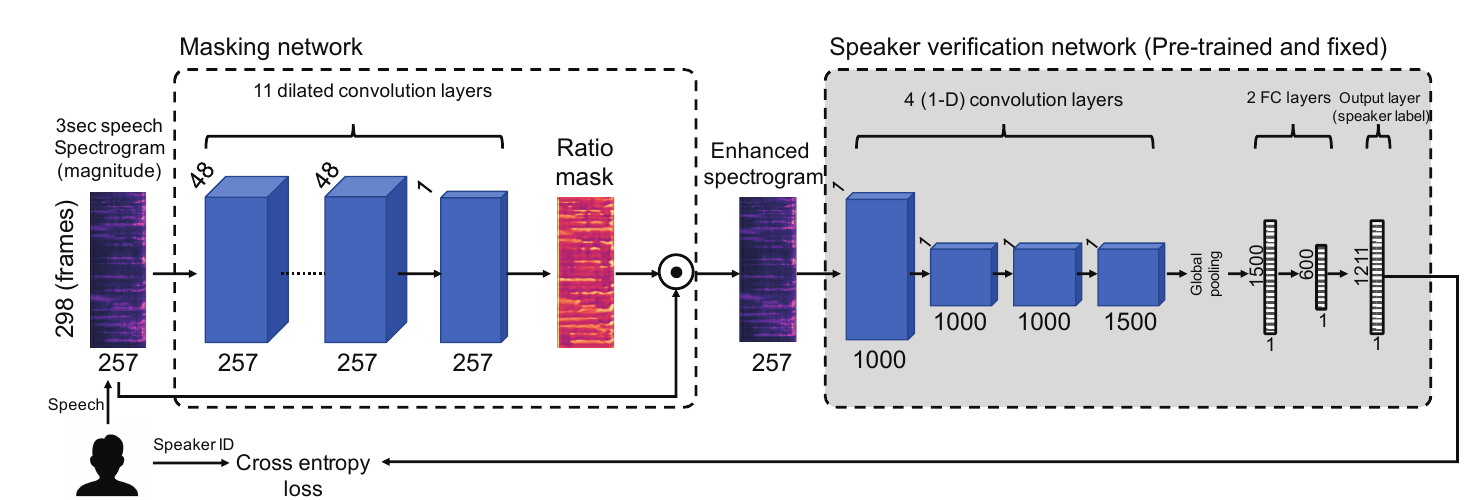}
  \caption{Flow chart of the VoiceID loss (from \cite{shon2019voiceid}).}
  \label{fig:VoiceID_Loss}
\end{figure*}

DNN-based autoencoder, which maps noisy speech directly to its clean counterpart, is another speech enhancement method for speaker recognition. {{\cite{plchot2016audio}}} trained an autoencoder to map the log magnitude spectrum of noisy speech to its clean counterpart by minimizing the mean squared error, which demonstrated its effectiveness on the text-dependent GMM-MAP  and  text-independent i-vector systems.
The authors of \cite{novotny2018on} explored a similar noise reduction method with \cite{plchot2016audio}, and applied it to an x-vector system \cite{novotny2019analysis}. In \cite{oo2016dnn}, the authors enhanced both the amplitude feature and the phase feature, where they used MFCC as the amplitude feature, and modified group delay cepstral coefficients as the phase feature. They concluded that simultaneous enhancing the amplitude and phase features is more effective than enhancing their individual components alone. Apart from the simplest feedforward DNN, more complicated LSTM based speech enhancement methods were also explored  \cite{sun2018speaker}.

Besides the masking and mapping based speech enhancement, GAN-based speech enhancement methods for speaker recognition were also developed. In specific, {{\cite{michelsanti2017conditional}}} used conditional GANs to learn a mapping from a noisy spectrum to its enhanced counterpart. The conditional GAN consists of a generator and a discriminator which are trained in an adversarial manner. The generator enhances the noisy spectrum; the discriminator aims to distinguish the enhanced spectrum from their clean counterpart using the noisy spectrum as a condition. In addition, {{\cite{yu2017adversarial}}} proposed a noise robust bottleneck feature extraction method based on adversarial training.

In addition to the above speech enhancement methods for speaker recognition, speech de-reverberation has also been studied in speaker recognition \cite{guzewich2017improving,movsner2018dereverberation,Mošner2018}. Recently, far-field and multi-channel speaker recognition also attracted much attention \cite{qin2019far,taherian2019deep,cai2019multi,taherian2020robust}.

\subsection{Data augmentation for robust speaker recognition}
\label{sec:data_augmentation}
Large-scale multi-condition training is an effective way to improve the generalization of speaker recognition in noisy environments. Particularly, we have observed that the performance of deep speaker embedding systems  appear to be highly dependent on the amount of training data. One way to prepare large-scale noisy training data is \textit{data augmentation}. In \cite{snyder2018x}, {{the authors}} employed additive noises and reverberation to the original training data for the data augmentation of x-vectors, which has shown to be very effective. {{\cite{zhu2019mixup}}} applied a mixup learning strategy  to improve the generalization of x-vector extractors.
To improve the performance of speaker verification for children with limited data, \cite{9053891} made speed and pitch perturbation  as well as voice conversion to increase the amount of training data. {{Besides, }}  \cite{9053481} validated the effectiveness of \textit{spectral augmentation}, which was originally proposed for speech recognition, for deep speaker embeddings.

\subsection{Other robust methods}
Apart from the aforementioned methods, some other robust speaker recognition methods are as follows. For example, \cite{9053110} optimized a deep feature loss for feature-domain enhancement of x-vector extractors. \cite{9053407} designed a new loss function for noise-robust speaker recognition. \cite{kim2019deep} proposed an orthogonal vector pooling strategy to remove unwanted factors. There are also many robust back-ends for speaker verification \cite{ghahabi2017deep,bhattacharya2016modelling,yang2017applying,mahto2017vector,guo2018deep}.

\section{Datasets}
\label{sec:data_tool}

\renewcommand\arraystretch{1.2}
\begin{table*}[t]
  \centering
{{
  \caption{Popular databases and challenges for speaker recognition. The term ``wild'' denotes that the audio is acquired across unconstrained conditions. The term ``quite'' denotes that the data were recorded indoor under a typical office environment. The term ``SV'',``SI'' and ``SR'' are the abbreviations of ``speaker verification'', ``speaker identification'' and ``speaker recognition'' respectively.   } \label{tab:challenge-dataset}
\scalebox{0.655}{
  \begin{tabular}{p{3.4cm}  p{1cm} p{3cm}  p{2.2cm}  p{2cm}  p{3.5cm}  p{3cm}  p{3cm}   p{2.8cm}}
  \hline
  \hline
  Dataset & Year& Condition & Language  & Sample rate& Application & Speakers & Data amount & Annotation method \\
  \hline
  NIST SRE \cite{greenberg2020two} &  1996 $\sim$ 2020 &  Clean, noisy &  Multilingual  &--- & Text-independent SR & --- & ---     &   Hand annotated \\

  VoxCeleb1 \cite{nagrani2017voxceleb}  & 2017& Multi-media (wild) &Mostly English &16kHZ &\makecell[l]{Text-independent SV \\ and SI} & 1,251 (690 males) &\makecell[l]{153,516 utterances, \\ 352 hours}  &  Automated pipeline     \\

  VoxCeleb2 \cite{chung2018voxceleb2}  & 2018 &  Multi-media (wild) & \makecell[l]{Multilingual \\ Mostly English}&--- & \makecell[l]{Text-independent SV \\ and SI}&  6,112 (3761 males) &\makecell[l]{1,128,246 utterances, \\ 2,442 hours}  &  Automated pipeline     \\

  SITW \cite{mclaren2016speakers} &  2016 &   Multi-media (wild) &  ---      &  16kHZ  & \makecell[l]{Text-independent single \\ and multi-speaker  SV} &  299 (203 males)  & 2,800 utterances   &  Hand annotated  \\

  RSR2015  \cite{larcher2014text} &  2015& \makecell[l]{Smart-phones and \\tablets (quite)}  &  English &16 kHZ&Text-dependent SV& 300 (157 males)  &\makecell[l]{196,844 files, \\  151 hours}&  Hand annotated       \\

  RedDots \cite{lee2015reddots} &  2015  & \makecell[l]{Mobile devices \\(through internet)} &   English      &---& \makecell[l]{Fixed phrase, free speech \\ and text-prompted SV}& 45  & ---    &  Manual, automatic, or semi-automatic    \\

  VOICES  \cite{richey2018voices,nandwana2019voices} &  2018    & \makecell[l]{Far-field microphones \\(noisy room)} & English &48 kHZ& \makecell[l]{Text-independent SV \\and SI}& 300  & \makecell[l]{374,688 files, \\ 1440 hours}   &   Hand annotated     \\

  Librispeech\cite{panayotov2015librispeech}& 2015  & ---& English & 16 kHz &  ASR and SR &Over 9,000 & 1000 hours   & Manually annotated  \\

  CN-CELEB\cite{9054017} &  2019  &  Multi-media (wild)&   Chinese  & --- &Text-independent SV &1000  & \makecell[l]{130,109 utterances, \\ 274 hours}    &  Automated pipeline  with human check      \\

  BookTubeSpeech \cite{9053258}& 2020  &   Multi-media       &    ---      & ---  &Text-independent SV& 8450  & ---  &    Automatic pipeline       \\

  Hi-MIA  \cite{9054423} &  2020  & \makecell[l]{Microphone arrays \\ (rooms, far-field)}& \makecell[l]{Chinese,\\ English} &16kHZ, 44.1kHz  & Text-dependent SV  &  340 (175 male) & \makecell[l]{More than 3,936,003 \\utterance,  1,561 hours}  &    Hand annotated     \\

  FFSVC 2020 \cite{qin2020ffsvc} &  2020   &  Close-talk cellphone, far-field microphone arrays (far-field)&    Mandarin    &  16kHZ, 48kHZ & \makecell[l]{Text-dependent and \\ text-independent SV}  &  --- & ---    &   Hand annotated    \\

  DIHARD1 \cite{ryant2018first} &   2018   &  \makecell[l]{Single channel (wild)}& English (most), Mandarin  & 16kHZ& Speaker diarization& --- &  40 hours & Hand annotated          \\

  DIHARD2 \cite{ryant2019secondDIHARD}& 2019   & \makecell[l]{Single channel and \\ multichannel (wild)} & English (most), Mandarin    &16kHZ& Speaker diarization & --- & \makecell[l]{503 files, \\ 339.95 hours}       & Hand annotated            \\

  AMI \cite{carletta2005ami} & 2005 & Multi-modal&  English  & --- &  Speaker diarization & ---  & 100 hours   &  Hand annotated           \\
  \hline
  \hline
  \end{tabular}
}
}}
\end{table*}

In this section, we make an overview to existing challenges, and datasets for speaker recognition.

Table \ref{tab:challenge-dataset} summarizes the brief information of most common and some recently developed datasets. The recording methods of the databases are briefly introduced as follows.
\begin{itemize}
  \item NIST SRE:  The National Institute of Standards and Technology (NIST) of America has successfully conducted 15 Speaker Recognition Evaluations (SREs) in the past 20 years.  It is the largest and most popular challenge in speaker recognition. More information can be found in \cite{greenberg2020two,gonzalez2014evaluating}.
  \item  VoxCeleb1,2: The VoxCeleb dataset was collected by a fully automated pipeline based on computer vision techniques from open-source media. The pipeline obtains videos from YouTube, performs active speaker verification using a two-stream synchronization CNN, and confirms the identity of the speaker using CNN based facial recognition \cite{nagrani2017voxceleb,chung2018voxceleb2}.
  \item  SITW: The speakers in the wild (SITW) database contains hand-annotated speech samples from open-source media for the purpose of benchmarking text-independent speaker recognition technology on single and multi-speaker audio acquired across unconstrained conditions \cite{mclaren2016speakers}.
  \item  RSR2015: The RSR2015 database were recorded indoor under a typical office environment with six mobile devices (five smart-phones and one tablet) \cite{larcher2014text}.
  \item RedDots: The RedDots project used a mobile app as the recording front-end. Speakers recorded their voices offline and later on uploaded the recordings to an Apache web server when Internet connection was available \cite{lee2015reddots}.

  \item VOICES: The voices obscured in complex environmental settings (VOICES) corpus was recorded in furnished rooms with background noise played in conjunction with foreground speech selected from the LibriSpeech corpus. The displayed  noises includes  television, music, or overlapping speech from multiple speakers (referred to as babble) \cite{richey2018voices,nandwana2019voices}.

  \item LibriSpeech: While intended created for speech recognition rather than verification or diarization, LibriSpeech does include labels of speaker identities and is thus useful for speaker recognition \cite{panayotov2015librispeech,9053258}.
  \item CN-CELEB: CN-CELEB was collected following a two-stage strategy: firstly {{the authors}} used an automated pipeline to extract potential segments of the Person of Interest from ``bilibili.com'', and then they applied a human check to remove incorrect segments \cite{9054017}.

  \item BookTubeSpeech: The BookTubeSpeech was collected by an automatic pipeline from BookTube videos \cite{9053258}.

  \item Hi-MIA: Hi-MIA database was designed for far-field scenarios. Recordings were captured by multiple microphone arrays located in different directions and distance to the speaker and a high-fidelity close-talking microphone \cite{9054423}.

   \item FFSVC 2020: The far-field speaker verification challenge 2020 (FFSVC20) is designed to boost the speaker verification research with special focus on far-field distributed microphone arrays under noisy conditions in real scenarios \cite{qin2020interspeech,qin2020ffsvc}.

  \item DIHARD1,2:  The DIHARD challenge  intended to improve the robustness of diarization systems to variation in recording equipment, noise conditions, and conversational domain \cite{ryant2018first,ryant2019secondDIHARD}.

  \item AMI: The AMI meeting corpus is a multi-modal data set consisting of 100 hours of meeting recordings, and it was recorded using a wide range of devices including close-talking and far-field microphones, individual and room-view video cameras, projection, a whiteboard, and individual pens, all of which produce output signals that are synchronized with each other \cite{carletta2005ami}.
\end{itemize}

\section{Conclusions and discussions}
\label{sec:discussion_conclusion}

This paper has provided a comprehensive overview of the deep learning based speaker recognition. We have analysed the relationship between different subtasks, including speaker verification, identification, and diarization, and summarized some common difficulties. Based on the analysis, we summarized the subtasks from three widely studied core issues---speaker feature extraction, speaker diarization, and robust speaker recognition. For speaker feature extraction, we reviewed two kinds of hybrid structures, which are DNN-UBM/i-vector and DNN-BNF/i-vector. In addition, the overview of the  state-of-the-art deep speaker embedding were made in respect of four key components, which are the inputs, network structures, temporal pooling strategies, and loss functions respectively. Particularly, we reviewed the loss functions of the end-to-end speaker verification for feature learning from the perspective of different training sample construction methods.
For speaker diarization, we reviewed stage-wise diarization, supervised end-to-end diarization, online diarization, and multimodal diarization. For robust speaker recognition, we surveyed three kinds of deep learning based domain adaptation methods as well as several speech preprocessing methods, which deal with the  domain mismatch and back-ground noise respectively. Some popular and recently developed datasets were summarized as well.
To conclude, deep learning has boosted the performance of speaker recognition to a new high level. We make our best to summarize the recent rapid progress of the deep learning based speaker recognition, hopefully this provides a knowledge resource and further blooms the research community.

Although the deep learning based speaker recognition has achieved a great success, many issues remain to be addressed. Here we list some open problems from the perspectives of network training, loss functions, real-world diarization, and domain adaptation respectively. For the network training, most speaker feature extraction methods need handcraft acoustic features as the input, which may not be optimal. The state-of-the-art deep models have a large number of parameters, which are difficult to be applied to portable devices. The network training also needs large amounts of labeled training data and heavy computation resources.
For the loss functions, although so many loss functions have been proposed, there is lack of strong theoretical base for the success of the loss functions, nor theoretical guidance that could lead to better loss functions. Although the verification losses for the end-to-end speaker verification meet the verification process tightly, their potentials have not been fully developed yet.
For the real-world diarization, from the recent DIHARD challenges, one can see that speaker diarization is still a hard problem. In a real-world conversation, a speech recording may be contaminated by serious speech overlap and strong background noise. Some technically difficult problems, such as the unknown number of speakers and rapid speaker changes, also hinder the performance of speaker diarization in real-world applications severely.
Finally, although many domain adaptation algorithms have been proposed, especially those based on adversarial learning, they have not made landmark progress compared to traditional shallow adaptation methods, e.g. the PLDA based adaptation, which needs further efforts.

\section{Acknowledgement}
The authors are grateful to Prof. DeLiang Wang, the Co-Editor-in-Chief, and the anonymous reviewers for their valuable comments,
which helped greatly improve the quality of the paper. The authors would also like to thank our colleagues Meng-Zhen Li and Rui Wang for their helpful discussions.

%\section*{References}
\bibliographystyle{elsarticle-num}
\bibliography{mybib}

\end{document}